\documentclass[aps,prd,onecolumn,showpacs,amsmath,amssymb]{revtex4}

\usepackage[colorlinks, citecolor=blue,anchorcolor=red,menucolor=red, linkcolor=red,filecolor=red,runcolor=red,urlcolor=blue,frenchlinks=red]{hyperref}

\usepackage{epsfig,float}
\usepackage{graphicx}
\usepackage{subfigure}
\usepackage{dcolumn}
\usepackage{morefloats}
\usepackage{color}
\usepackage{slashed}
\usepackage{bbm}
\usepackage{bm}
\usepackage{mathtools}
\usepackage{booktabs}
\usepackage{multirow}
\usepackage{mathrsfs}
\usepackage{verbatim}
\usepackage{threeparttable}
\allowdisplaybreaks[4]

\newcommand{\be}{\begin{equation}}
\newcommand{\ee}{\end{equation}}
\newcommand{\ba}{\begin{eqnarray}}
\newcommand{\ea}{\end{eqnarray}}

\begin{document}


\title{The mass splitting among the isospin multiplets of light vector mesons }

\author{Peng-Yu Niu$^{1,2}$\footnote{{\it Email address:} niupy@ihep.ac.cn}, Bin Zhou$^{1,2}$\footnote{{\it Email address:} bzhou@ihep.ac.cn; Corresponding author}, and  Qiang Zhao$^{1,2,3}$\footnote{{\it Email address:} zhaoq@ihep.ac.cn; Corresponding author} }

\affiliation{$^1$ Institute of High Energy Physics and Theoretical Physics Center for Science Facilities,
        Chinese Academy of Sciences, Beijing 100049, China}
\affiliation{$^2$ School of Physical Sciences,
University of Chinese Academy of Sciences, Beijing 100049, China}
\affiliation{$^3$ Synergetic Innovation Center for Quantum Effects and Applications (SICQEA),
Hunan Normal University, Changsha 410081, China}

\begin{abstract}
By including the strong isospin symmetry breaking effects and the electromagnetic contributions between the pseudoscalar mesons, we calculate the phase shifts of the $P$-wave $\pi\pi$ and $K\pi$ scattering up to $\mathcal{O}(p^4)$ in the framework of the SU(3) chiral perturbation theory (ChPT) and coupled channel inverse amplitude method. We re-fit the low energy constants with the present meson-meson scattering data and derive the mass differences for the charged and neutral iso-multiplets of $\rho$ and $K^{*}$. Our results show that the mass difference between $\rho^\pm$ and $\rho^0$ is very small while the mass difference between the charged and neutral $K^*$ can reach a relatively large value of $m_{K^{*0}}-m_{K^{*+}}= 2.91^{+1.43}_{-1.41}$ MeV. This full one-loop ChPT calculation would shed some light towards a better understanding of the long-standing puzzle about the $K^*$ mass splitting.
\end{abstract}
\date{\today}

\maketitle
\section{introduction}

The mass splittings among iso-multiplets has been recognized as one of the predominant features of isospin breaking effects. They can generally be attributed to two mechanisms. One is due to the $u$-$d$ quark mass difference, and the other is due to the electromagnetic (EM) interactions that contribute to the charged and neutral multiplets differently. However, although the sources for isospin symmetry breaking are known, quantitative evaluations of the isospin symmetry breaking effects in specific processes sometimes are nontrivial. In this work we will study the mass splitting between the charged and neutral iso-multiplets of $\rho$ and $K^*$. We will focus more on the mass splitting between $K^{*\pm}$ and $K^{*0}$ caused by the isospin symmetry breaking since it is still full of controversies in both experiment and theory.

The masses of both $K^{*0}$ and $K^{*\pm}$ have been measured in experiment. From Particle Data Group (PDG) the averaged value for the neutral state is $895.55\pm0.20$ MeV \cite{Tanabashi:2018oca} of which the most recent high-precision measurements were from $D$ meson decays at BESIII~\cite{Ablikim:2015mjo}, BaBar~\cite{delAmoSanchez:2010fd} and CLEO~\cite{Bonvicini:2008jw}. In contrast, early measurements from $pp$ scatterings or meson-baryon scatterings contain large uncertainties as listed in PDG~\cite{Tanabashi:2018oca}. For the charged $K^*$ mass measurements two different averaged values are listed. One is extracted from hadron produced processes with a value of $891.66\pm0.26 $ MeV~\cite{Tanabashi:2018oca}, and the the other one is measured in $\tau$ decays with a value of $895.47\pm0.20\pm 0.74$ MeV~\cite{Epifanov:2007rf}. The puzzling observation is that the mass values of the charged $K^*$ extracted in these two typical processes turn out to be dramatically different. Namely, in one case it appears to have about 4 MeV mass difference compared to the neutral one while and in the other case it becomes almost equal to each other. The mass difference between these two groups of measurements, which is about 4 MeV, can be regarded as significant. In contrast, the mass difference provided by PDG using only hadroproduction data~\cite{AguilarBenitez:1972bw,Barash:1900zz} yields $m_{K^{*0}}-m_{K^{*\pm}}=6.7\pm 1.2$ MeV, which is even larger. These controversies suggest that a reliable measurement and less model-dependent calculations are both needed to clarify this mysterious situation about the $K^*$ isospin breaking effects.

Early calculations of the isospin breaking effects among SU(3) flavor multiplets were done in the quark model based on the SU(6) spin-flavor symmetry~\cite{Harari:1965,Rubinstein:1966zzb}. A consequent relation is $m_{K^0} - m_{K^\pm}=m_{K^{*0}}-m_{K^{*\pm}}\simeq 4$ MeV, which seems to be a reasonable perspective. However, later calculations using  QCD-based constituent quark model (CQM) led to rather different answers~\cite{DeRujula:1975qlm,Isgur:1978xj,Isgur:1978wd,Isgur:1979ed,Godfrey:1985sp,Godfrey:1985xj}. In particular, both relativistic and nonrelativistic CQM favor a smaller value of about $2$ MeV mass difference for $m_{K^{*0}}-m_{K^{*\pm}}$~\cite{Isgur:1979ed,Godfrey:1985sp}. In recent  years, chiral constituent quark model~\cite{Manohar:1983md} is also applied to the study of the $K^*(892)$ mass splitting and their results also favor a relatively small mass difference which is about $1.31\pm0.56$ MeV~\cite{Gao:1996sa,Gao:1998gr,Gao:2007xh} and significantly smaller than $m_{K^0} - m_{K^\pm}$.

In Refs.~\cite{Jenkins:1995vb,Bijnens:1997ni,Bijnens:1997rv} a heavy vector meson effective theory was developed in the framework of chiral perturbation theory (ChPT)~\cite{Weinberg:1978kz,Gasser:1983yg,Gasser:1984gg,Gasser:1984ux,Gasser:1984pr} by including the vector mesons as massive fields interacting with the light mesons. This allows an effective field theory approach for the isospin breaking effects for the vector mesons. With the chiral correction up to $\mathcal O^(p^4)$ and EM corrections, the $K^*$ mass splitting is found to be 4.5 MeV in Ref.~\cite{Bijnens:1997ni}. In order to go beyond the restricted regime for the ChPT, theoretical tools were developed in the literature to combine nonperturbative methods and the ChPT together~\cite{Dobado:1989qm,Dobado:1992ha,Dobado:1996ps,Oller:1997ti,Nieves:2000km,Nieves:1998hp,Nieves:1999bx,Bruns:2013tja}.
In Refs.~\cite{Oller:1997ng,Oller:1998hw} a coupled-channel unitary approach using the inverse amplitude method was developed that extended the energy region successfully up to 1.2 GeV. In this framework light resonances can be described as dynamically generated states via hadron-hadron interactions. The vector meson phase shifts and light scalar mesons can be accounted for in the limit of isospin symmetry. Later, a complete one-loop calculation of the meson-meson scattering amplitudes with the same method was presented in Ref.~\cite{GomezNicola:2001as}. A coherent description of the meson-meson scattering data up to $1.2$ GeV was achieved with impressive success. The light scalar and vector mesons can be dynamically generated as the pole structures of the scattering amplitudes. The success of Refs.~\cite{Oller:1997ng,Oller:1998hw,GomezNicola:2001as} have shown the important role played by the coupled-channel dynamics with unitarization.

In this work we adopt the method developed in Refs.~\cite{Oller:1997ng,Oller:1998hw,GomezNicola:2001as}, but extend it to accommodate the isospin breaking effects. In particular, by extending the method of Ref.~\cite{GomezNicola:2001as} to include the isospin breaking effects, we will calculate the phase shifts for the $P$-wave $\pi\pi$ and $K\pi$  scatterings up to order of $\mathcal O(p^4)$. Since the $u$-$d$ quark mass difference and EM interaction will also contribute to the mass matrix of the ChPT Lagrangian it would be interesting to inspect the outcomes from this successful method. By fitting the $P$-wave $\pi\pi$ and $K\pi$  phase shift data simultaneously, we can constrain the chiral parameters using the $(I, \ J)=(1, \ 1)$ data and determine the isospin symmetry breaking effects in the $(1/2, \ 1)$ channel. This can be regarded as a reliable estimate of the $K^*$ mass splitting from the ChPT.

This paper is organized as follows. In Sec. II  the ChPT Lagrangian and main feature from the full one-loop calculation is discussed. In Sec. III  unitary and partial waves will be briefly illustrated in the framework of the ChPT and coupled channel inverse amplitude method. In Sec. IV the  results and discussions will be presented. A brief summary will be given in Sec. V. For the convenience of following the detailed computation by readers we include the partial wave scattering amplitudes and some useful formulas in Appendix.

\section{The ChPT Formalism}

We first briefly introduce the ChPT Lagrangians adopted in this work.
The leading order ChPT Lagrangian for the pseudoscalar mesons has the same expression as that in the isospin symmetry limit~\cite{Oller:1997ng,Oller:1998hw,GomezNicola:2001as} and is written as the following:
\begin{eqnarray}\label{L2-lagrangian}
{\cal L}_{2}=\frac{F_{0}^{2}}{4}\mbox{Tr}\left[D_{\mu}U(D^{\mu}U)^{\dagger}\right]+\frac{F_{0}^{2}}{4}\mbox{Tr}\left [\chi U^{\dagger}+U\chi^{\dagger}\right],
\end{eqnarray}
where $F_0$ is the decay constant of pseudoscalar mesons and $U(x)$ is expressed as:
\begin{eqnarray}
U(x)=\exp(i\sqrt{2}\Phi(x)/F_0) \ ,
\end{eqnarray}
with $\Phi(x)$ denoting the pseudoscalar meson octet matrix:
\begin{eqnarray}
\Phi \equiv\left(\begin{array}{ccc}
\frac{1}{\sqrt{2}}\pi^{0}+\frac{1}{\sqrt{6}}\eta & \pi^{+} & K^{+}\\
\pi^{-} & -\frac{1}{\sqrt{2}}\pi^{0}+\frac{1}{\sqrt{6}}\eta & K^{0} \notag \\
K^{-} & \bar{K}^{0} & -\frac{2}{\sqrt{6}}\eta
\end{array}\right).
\end{eqnarray}

For the traditional ChPT the mass matrix $\chi$ is a diagonal matrix with the bare masses of $\pi$ and $K$ or the bare masses of the $u \ (d)$ and $s$ quark. In the limit of isospin symmetry the bare masses of the $u$ and $d$ quark are the same. With the isospin breaking considered, the mass matrix $\chi$ should be rewritten as:
\begin{eqnarray}
\chi=\left(\begin{array}{ccc}
m_{0K^+}^{2}-m_{0K^{0}}^{2}+m_{0\pi^{+}}^{2} & 0 & 0\\
&&\\
0 & m_{0K^{0}}^{2}-m_{0K^{+}}^{2}+m_{0\pi^{+}}^{2} & 0\\
&&\\
0 & 0 & m_{0 K^{+}}^{2}+m_{0K^{0}}^{2}-m_{0\pi^{+}}^{2}
\end{array}\right) \ ,
\end{eqnarray}
where $m_{0P}$ (with $P=\pi, K$) stands for the bare mass of the corresponding pseudoscalar meson. When taking the isospin symmetry limit, the mass matrix returns to the normal form~\cite{GomezNicola:2001as,Gasser:1984gg,Gasser:1984ux,Gasser:1984pr} with the expressions $m_{0\pi^+}=B_0(m_u+m_d),~m_{0K^+}=B_0(m_u+m_s)$ and $m_{0K^0}=B_0(m_d+m_s)$~\cite{Scherer:2012xha}. In addition, the covariant derivative is adopted in Eq.~(\ref{L2-lagrangian}) in order to include the EM contributions. Then the next leading order, $\mathcal O(p^{4})$, Lagrangian is written as:
\begin{eqnarray}
{\cal L}_{4}&&=L_{1}\mbox{Tr}\left[D_{\mu}U(D^{\mu}U)^{\dagger}\right]^{2}+
L_{2}\mbox{Tr}\left[D_{\mu}U(D_{\nu}U)^{\dagger}\right]\mbox{Tr}\left[D^{\mu}U(D^{\nu}U)^{\dagger}\right]
\notag \\
&&+L_{3}\mbox{Tr}\left[D_{\mu}U(D^{\mu}U)^{\dagger}D_{\nu}U(D^{\nu}U)^{\dagger}\right]
+L_{4}\mbox{Tr}\left[D_{\mu}U(D^{\mu}U)^{\dagger}\right]\mbox{Tr}\left[\chi U^{\dagger}+U\chi^{\dagger}\right] \notag \\
&&+L_{5}\mbox{Tr}\left[D_{\mu}U(D^{\mu}U)^{\dagger}(\chi U^{\dagger}+U\chi^{\dagger})\right]
+L_{6}\mbox{Tr}\left[\chi U^{\dagger}+U\chi^{\dagger}\right]^{2}
+L_{7}\mbox{Tr}\left[\chi U^{\dagger}-U\chi^{\dagger}\right]^{2} \notag\\
&&+L_{8}\mbox{Tr}\left[\chi U^{\dagger}\chi U^{\dagger}+U\chi^{\dagger}U\chi^{\dagger}\right]
-iL_{9}\mbox{Tr}\left[ f_{\mu\nu}^{R}D^{\mu}U(D^{\nu}U)^{\dagger}+f_{\mu\nu}^{L}(D^{\mu}U)^{\dagger}D^{\nu}U \right],
\end{eqnarray}
where the chiral parameters $L_i(i=1-9)$ are energy-scale dependent and generally written as $L_i=L_i^r(\mu)+\Gamma_i \lambda$ with $\mu$ the $\overline{MS}-1$ renormalization scale and $\lambda\equiv \frac{\mu^{d-4}}{16\pi^2}\left[ \frac{1}{d-4}-\frac12 (\log 4\pi -\gamma+1)\right]$. The ${\cal L}_{4}$ Lagrangian contributes to the $\mathcal O(p^{4})$ tree diagrams of the meson scatterings and can be used to calculate the meson mass corrections. The term of $L_9$ comes from the EM interaction and does not appear in Refs.~\cite{Oller:1997ng,Oller:1998hw,GomezNicola:2001as} due to the neglect of isospin breaking effects there.

\begin{figure}[ht]
  \centering
  \includegraphics[scale=0.2]{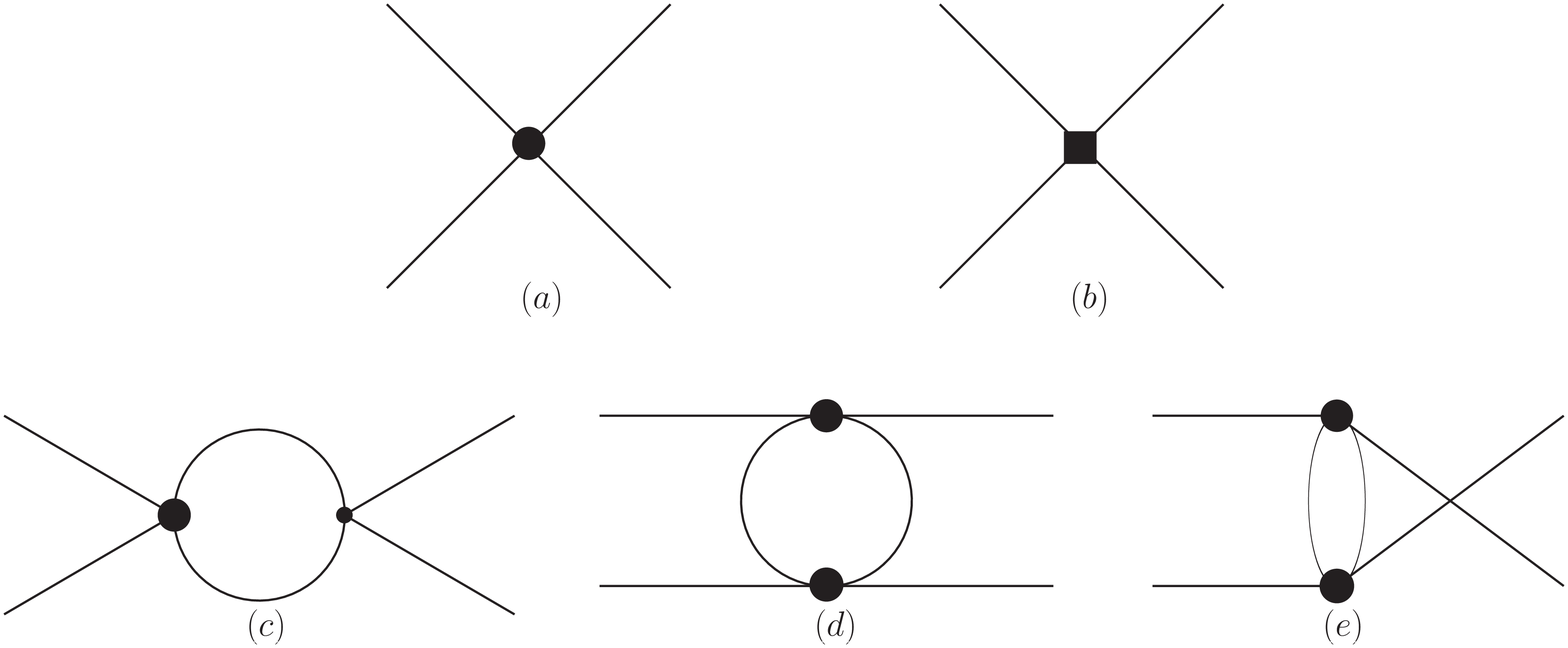}\\
  \includegraphics[scale=0.3]{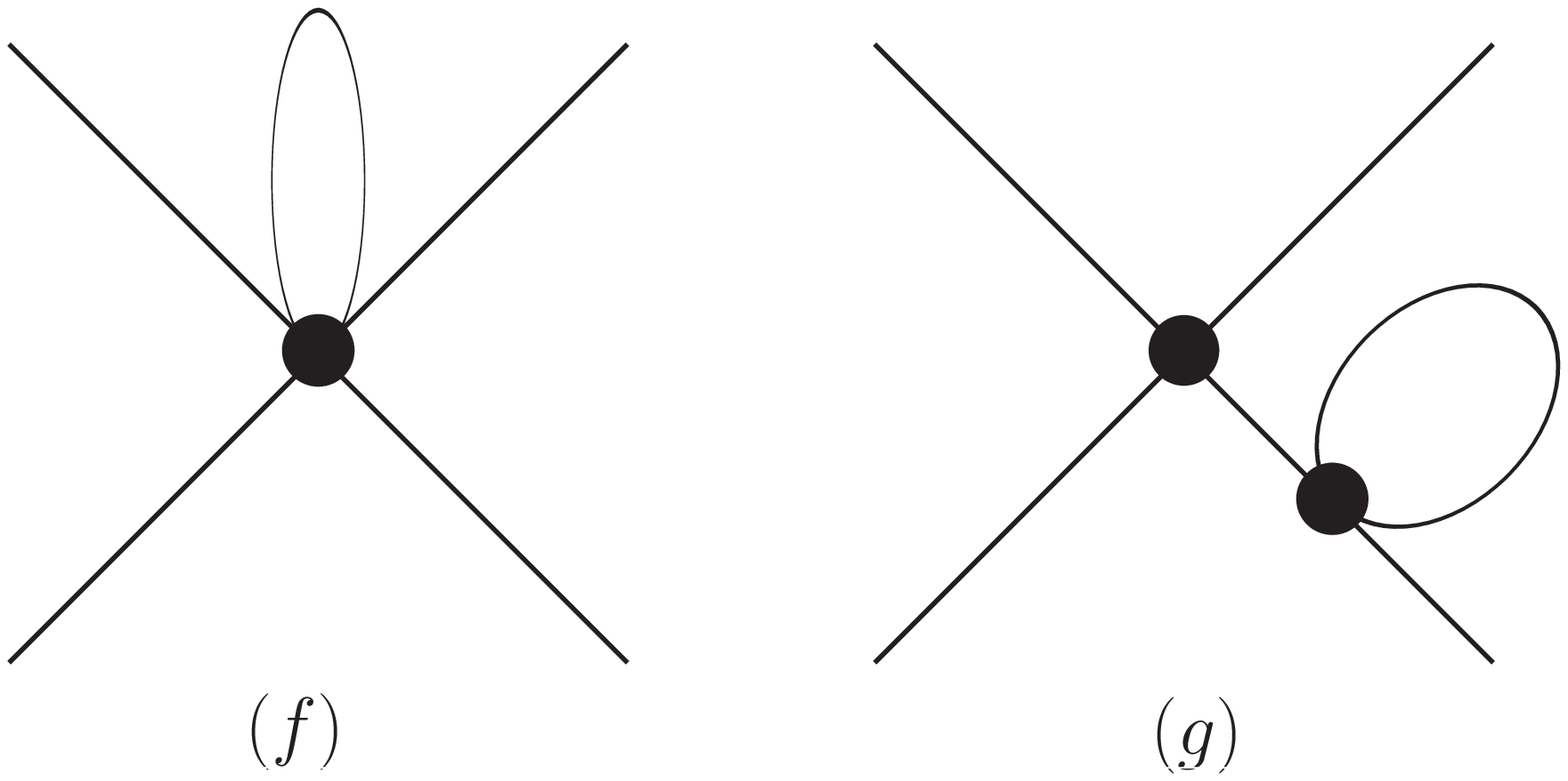}
  \caption{The Feynman diagrams of ``pure chiral" interaction that we consider for meson-meson scatting. The solid circle and solid square stand for the vertex of $\mathcal O (p^2)$ and $\mathcal O(p^4)$, respectively. (a) is the tree level of $\mathcal O(p^2)$ and (b) is the tree level of $\mathcal O(p^4)$. (c),(d),(e) are s,t,u channel. (f) is tadpole diagram and (g) is wave function normalization term.}
  \label{fig:treeloop}
\end{figure}

\begin{figure}[h]
\centering
\includegraphics[width=0.45\linewidth]{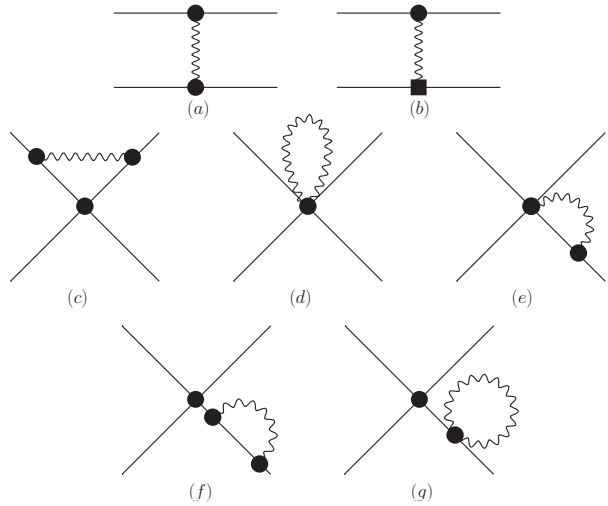}
\caption{The Feynman diagrams of EM interaction for meson-meson scattering. The solid circle and solid square stand for the vertex of order $\mathcal O (p^2)$ and $\mathcal O(p^4)$, respectively. (a) and (b) are the tree diagrams of order $\mathcal O(p^2)$ and $\mathcal O(p^4)$ and (c)-(g) are the one-loop diagrams which are neglected in our calculation.}
\label{fig:EM}
\end{figure}

Figures~\ref{fig:treeloop} and \ref{fig:EM} show all the Feynman diagrams for the meson-meson scattering up to $\mathcal O(p^4)$ including the ``pure chiral" interaction and EM interaction. We collect the amplitudes in several parts below:
\begin{eqnarray}
&&T_2=T^\text{tree}_2 +T^\text{EM}_2, \\
&&T_4=T^\text{tree}_4+T^\text{correct}_4+T^\text{stu}_4+T^\text{tadpole}_4 +T^\text{wf}_4+T^\text{EM}_4,
\end{eqnarray}
where $T^\text{tree}_2$ and $T^\text{tree}_4$ represent the $\mathcal{O} (p^2)$ and $\mathcal{O} (p^4)$ contributions from the tree diagrams of the ``pure chiral" interaction, respectively. It should be stressed that $T^\text{correct}_4$ is the contribution from the meson mass and decay constant correction of $T^\text{tree}_2$. The term $T^\text{stu}_4$ contains the contributions from the $s, \ t, \ u$ loops and $T^\text{tadpole}_4$ corresponds to the contribution of the tadpole diagram. The term $T^\text{wf}_4$ represents the contribution from wave function renormalization of the pseudoscalar mesons and is given by
 \begin{eqnarray}
T^\text{wf}_4=\frac{1}{2}\sum_{i=1,2,3,4} (Z_{p}^{(i)}-1)T^\text{tree}_2,
 \end{eqnarray}
where $Z_p$ is the wave function renormalization constant and the subscript $p$ represents the mesons of external lines. The superscript $i$ represents the sequence number of external lines. In our calculation, the renormalized masses of the pseudoscalar mesons have the following expressions:
\begin{eqnarray}
m_{\pi^{+}}^{2} &= & m_{0\pi^{+}}^{2}\left[1+\mu_{\pi^0}-\frac{\mu_{\eta}}{3}+\frac{8m_{0\pi^{+}}^{2}}{F_{0}^{2}}(2L_{6}+2L_{8}-L_{4}-L_{5}) +\frac{8m_{0K^{+}}^{2}}{F_{0}^{2}}(2L_{6}-L_{4})+\frac{8m_{0K^{0}}^{2}}{F_{0}^{2}}(2L_{6}-L_{4})\right] \notag, \\
m_{K^{0}}^{2} & = & m_{0K^{0}}^{2}\left[1+\frac{2 \mu_{\eta}}{3}+\frac{8m_{0\pi^{+}}^{2}}{F_{0}^{2}}(2L_{6}-L_{4})+\frac{8m_{0K^{+}}^{2}}{F_{0}^{2}}(2L_{6}-L_{4})+
\frac{8m_{0K^{0}}^{2}}{F_{0}^{2}}(2L_{6}+2L_{8}-L_{4}-L_{5})\right] \notag, \\
m_{K^{+}}^{2} & = & m_{0K^{+}}^{2}\left[1+\frac{2 \mu_{\eta}}{3}+\frac{8m_{0\pi^{+}}^{2}}{F_{0}^{2}}(2L_{6}-L_{4})+\frac{8m_{0K^{0}}^{2}}{F_{0}^{2}}(2L_{6}-L_{4})+
\frac{8m_{0K^{+}}^{2}}{F_{0}^{2}}(2L_{6}+2L_{8}-L_{4}-L_{5})\right],
\label{eq:massc}
\end{eqnarray}
where
 \begin{eqnarray}
\mu_i=\frac{m_p^2}{32\pi^2 F_0^2}\log\frac{m_p^2}{\mu^2},
 \end{eqnarray}
with
$p=\pi,K,\eta$. In our calculation the renormalization scale $\mu$ takes a value of 770 MeV. The wave function renormalization constants (only the finite part) of pseudoscalar mesons is written as:
\begin{eqnarray}
Z_{\pi^\pm}&=& 1+\frac{2}{3}(\mu_{\pi^\pm}+\mu_{\pi^0})+\frac{1}{3}(\mu_{K^\pm}+\mu_{K^0})
-\frac{8}{F_0^2} \left[ L_4^r( m_{0,K^\pm}^2+ m_{0,K^0}^2)+ (L_4^r+L_5^r)
m_{0\,\pi^\pm}^2\right],\nonumber\\
Z_{\pi^0}&=& 1+\frac{4}{3}\mu_{\pi^\pm}+\frac{1}{3}(\mu_{K^\pm}+\mu_{K^0})
-\frac{8}{F_0^2} \left[ L_4^r( m_{0,K^\pm}^2+ m_{0,K^0}^2)+ (L_4^r+L_5^r)
m_{0\,\pi^\pm}^2\right],\nonumber\\
Z_K&=& 1+\frac{1}{6}(2\mu_{\pi^\pm}+\mu_{\pi^0})+\frac{1}{3}(2\mu_{K^\pm}+\mu_{K^0})+\frac{1}{2}\mu_\eta
-\frac{8}{F_0^2} \left[ L_4^r( m_{0,K^\pm}^2+ m_{0,K^0}^2+m_{0\,\pi^\pm}^2)+ L_5^r m_{0\,K^\pm}^2\right],\nonumber\\
Z_\eta&=& 1+\mu_{K^\pm}+\mu_{K^0}-\frac{8}{3F_0^2}\left[\left(3L_4^r-L_5^r\right)m_{0,\pi^\pm}^2+
\left(3L_4^r+2L_5^r\right)(m_{0,K^\pm}^2+m_{0,K^0}^2)\right].
\label{eq:z}
\end{eqnarray}
The meson decay constants with corrections to one loop is written as:
 \begin{eqnarray}
f_\pi&=& F_0\left[1-2\mu_\pi-\mu_K+\frac{4 m_{0\,\pi}^2}{F_0^2}\left(L_4^r+L_5^r\right)+\frac{8m_{0\,K}^2}{F_0^2}L_4^r\right], \nonumber\\
f_K&=&F_0\left[1-\frac{3\mu_\pi}{4}-\frac{3\mu_K}{2}-\frac{3\mu_\eta}{4}+\frac{4 m_{0\,\pi}^2}{F_0^2}L_4^r+ \frac{4m_{0\,K}^2}{F_0^2} \left(2L_4^r+L_5^r\right) \right], \nonumber\\
f_\eta&=&F_0\left[1-3\mu_K+ \frac{4L_4^r}{F_0^2}\left(m_{0\,\pi}^2+2m_{0\,K}^2\right)+\frac{4m_{0\,\eta}^2}{F_0^2}L_5^r\right].
 \label{eq:f}
\end{eqnarray}
Considering the difference of decay constants between charged and neutral mesons is about $1\%$, we set $f_{\pi^0}=f_{\pi^\pm}$ and $f_{K^0}=f_{K^+}$.

We have also considered the isospin breaking effects for mass renormalizations and wave function renormalizations of the neural and charged mesons, respectively. Eq.~\ref{eq:massc} and Eq.~\ref{eq:z} will turn into the usual expression~\cite{GomezNicola:2001as} in the limit of isospin symmetry. Although we show all the Feynman diagrams of EM interaction up to $\mathcal O(p^4)$ in Fig.~\ref{fig:EM} (Note that the leading order of EM contribution is $\mathcal O(\alpha)$), we only calculate the amplitudes of tree diagrams of $\mathcal O(p^2)$ noted by $T_2^{\text{EM}}$ and $\mathcal O(p^4)$ noted by $T_4^{\text{EM}}$. This is because that the EM coupling is small. Higher order EM corrections are negligible and can be considered as next next leading order. Our final results are expressed explicitly in terms of the physical masses and physical values for the decay constants of the pseudoscalar mesons. To ensure the perturbative unitarity, we express $f_k$ and $f_\eta$ by $f_\pi$ using the Eq.~\ref{eq:f}. In the calculation we have adopted $f_\pi=92.4$ MeV as Ref.~\cite{GomezNicola:2001as}.

\section{Unitary and Partial waves}

\subsection {Unitary}
The $S$ matrix is unitary i.e. $SS^\dagger=1$. In the case of two dimension coupled channel, $S$ can be organized as $2\times2$ matrix
\begin{eqnarray}
S=\left[
  \begin{array}{cc}
    \eta e^{2 i \delta_1} & i\sqrt{1-\eta^2}e^{i(\delta_1+\delta_2)} \\
    i\sqrt{1-\eta^2}e^{i(\delta_1+\delta_2)} & \eta e^{2 i \delta_2} \\
  \end{array}
\right],
\end{eqnarray}
where $\delta_i(i=1,2)$ are the phase shifts which we will focus on and $\eta$ is the inelasticity. The $T$-matrix elements $T^{IJ}_{ij}$ are related to $S$ matrix elements through the equation
\begin{eqnarray}
S=\left[
  \begin{array}{cc}
    1 & 0 \\
    0 & 1 \\
  \end{array}
\right] +
\left[
  \begin{array}{cc}
    2i \sigma_1 T_{11}               & 2i\sqrt{\sigma_1\sigma_2} T_{12} \\
    2i\sqrt{\sigma_1\sigma_2} T_{21} & 2i\sigma_2 T_{22} \\
  \end{array}
\right],
\end{eqnarray}
where $T_{12}=T_{21}$ and $\sigma_i$ is the phase space of that state at $\sqrt {s}$ and given by
\begin{equation}
\sigma_n=-\frac{p_n}{8\pi\sqrt{s}}\theta\left(s-(m_{1n}+m_{2n})^2\right),
\end{equation}
where $p_n$ is the on-shell center of mass (c.m.) momentum of the meson in the intermediate state $n$ and $m_{1n}$, $m_{2n}$ are the masses of the two mesons in the state $n$. Note that we omit the $I$ (isospin) and $J$ (partial wave) labels for the $T$ matrix just for convenience. Based on the unitary of $S$ matrix, the $T$ matrix satisfies
\begin{equation}
\label{eq:ts}
\text{Im}T=T \Sigma T^*,
\end{equation}
where $\Sigma$ is a real diagonal matrix and written as:
\begin{eqnarray}
\Sigma=\left[
  \begin{array}{cc}
    \sigma_1 & 0 \\
    0 & \sigma_2 \\
  \end{array}
\right].
\end{eqnarray}

Next, considering in ChPT that the $T$ matrix can be expanded in powers of $p^2$, i.e. $T\simeq T_2+T_4 +\dots$ and combining Eqs.~(\ref{eq:ts}) and (\ref{euq:Ts}) together, we have
\begin{eqnarray}
\label{equ:Im}
&\text{Im}[T_2+T_4+\dots]=[T_2+T_4+\dots]\cdot \Sigma \cdot [T_2+T_4+\dots], \notag \\
&\text{Im}T_2=0, \notag \\
&\text{Im}T_4=T_2 \cdot \Sigma \cdot T_2, \notag \\
&\dots .
\end{eqnarray}
Then, we could have
\begin{eqnarray}
\label{euq:Ts}
&T^{-1}\simeq T_2^{-1}\cdot[1+ T_4 \cdot T_2^{-1}+\dots]^{-1}\simeq T_2^{-1}\cdot[1- T_4 \cdot T_2^{-1}+\dots] \ ,
\end{eqnarray}
which leads to
\begin{eqnarray}
\label{equ:T2}
&T_2\cdot\text{Re} T^{-1}\cdot T_2 \simeq  T_2-\text{Re} T_4\dots.
\end{eqnarray}

Now we deduce the expression of the $T$ matrix in terms of $T_2$ and $T_4$. From Eq.~(\ref{eq:ts}) we can easily obtain
\begin{eqnarray}
\Sigma &&=T^{-1} \cdot \text{Im}T  \cdot T^{*-1} \notag \\
&&=\frac{1}{2i}T^{-1}\cdot ( T-T^* )\cdot T^{*-1} \notag \\
&&=-\text{Im}T^{-1}.
\end{eqnarray}
Then, we have
\begin{eqnarray}
\label{equ:Tmatrix1}
T&=\left[ \text{Re}T^{-1}-i \Sigma \right]^{-1} \ .
\end{eqnarray}
In order to avoid using $T_2^{-1}$ which may not be invertible, we modify Eq.~(\ref{equ:Tmatrix1}) by multiplying $T_2 T_2^{-1}$ to the left and $T_2^{-1}T_2$ to the right. Combining Eqs.~(\ref{equ:T2}) and (\ref{equ:Im}) together, the $T$ matrix can be written as
\begin{eqnarray}
T&=T_2\cdot (T_2 \cdot \text{Re}T^{-1}\cdot T_2-i  T_2\cdot \Sigma \cdot T_2  )^{-1}\cdot T_2 \notag \\
&=T_2 \cdot (T_2 -\text{Re}T_4 -i \text{Im} T_4 )^{-1} \cdot T_2
\end{eqnarray}

Finally, we obtain a simple expression
\begin{equation}
\label{equ:Tmatrix2}
T=T_2 \cdot [T_2-T_4]^{-1}\cdot T_2.
\end{equation}
This is the IAM formula for the coupled channel transitions. This two-dimension coupled channel formula can avoid some problems that occur in the single channel IAM when $T_2=0$ or $|T_2|=|T_4|$~\cite{Oller:1998hw}. On the other hand, the left-hand cuts, which can be ignored in the single channel IAM~\cite{Guerrero:1998ei,GomezNicola:2001as}, will be properly included. Equation~(\ref{equ:Tmatrix2}) requires the complete evaluation of $T_4$ which contains quite a number of processes being neglected in the isospin symmetry limit, and the calculation details are given in the next section.

\subsection{Coupled channel partial waves}

In order to dynamically generate the meson with definite isospin and spin, we must extract the definite isospin component from meson-meson scattering processes and do partial wave analysis. Note that the strong interaction is invariant under the isospin transformation in the isospin symmetry limit, i.e.
\begin{eqnarray}
\langle I',I_3'|T|I,I_3 \rangle=T^I\delta_{I'I}\delta_{I_3'I_3}.
\label{eq:iso}
\end{eqnarray}
If the isospin symmetry is broken, Eq.~(\ref{eq:iso}) should be re-written as:
\begin{eqnarray}
\langle I',I_3'|T|I,I_3 \rangle=T^{I,I_3}\delta_{I'I}\delta_{I_3'I_3} \ ,
\end{eqnarray}
which means that the processes with different $I_3$ will have different amplitudes for the same isospin. Thus, the number of independent processes will increase compared with the cases that isospin symmetry is conserved. On the other hand, the charged and neutral channel can be conveniently distinguished by $I_3$. All the independent processes we adopt are given as follows:

For $\rho^{+}$, the following processes can contribute:
\begin{itemize}
\item $\pi\pi \to \pi\pi$ scattering:
\begin{eqnarray}
T^{I=1,I_3=1}(s,t,u)  = T_{\pi^{+}\pi^{0}\rightarrow\pi^{+}\pi^{0}}(s,t,u)-T_{\pi^{+}\pi^{0}\rightarrow\pi^{+}\pi^{0}}(s,u,t),
\end{eqnarray}

\item $\pi \pi \to K K$/$K K \to \pi \pi$ scattering:
\begin{eqnarray}
T^{I=1,I_3=1}(s,t,u) = \sqrt{2}T_{\pi^{+}\pi^{0}\rightarrow K^+\bar{K}^{0}}(s,t,u),
\end{eqnarray}

\item $K K \to K K$ scattering:
\begin{eqnarray}
T^{I=1,I_3=1}(s,t,u)
  =  T_{K^{+}K^{-}\rightarrow K^{0}\bar{K}^{0}}(t,s,u).
\end{eqnarray}
\end{itemize}

For $\rho^{0}$, the following processes can contribute:
\begin{itemize}
\item $\pi\pi \to \pi\pi$ scattering:
\begin{eqnarray}
T^{I=1,I_3=0}(s,t,u)  = T_{\pi^{+}\pi^{-}\rightarrow\pi^{+}\pi^{-}}(s,t,u)-T_{\pi^{+}\pi^{-}\rightarrow\pi^{+}\pi^{-}}(s,u,t),
\end{eqnarray}
\item $\pi \pi \to K K$/$K K \to \pi \pi$ scattering:
\begin{eqnarray}
T^{I=1,I_3=0}(s,t,u)
  =  T_{\pi^{+}\pi^{-}\rightarrow k^{+}k^{-}}(s,t,u)-T_{\pi^{+}\pi^{-}\rightarrow k^{+}k^{-}}(s,u,t),
\end{eqnarray}
\item $K K \to K K$ scattering:
\begin{eqnarray}
T^{I=1,I_3=0}(s,t,u)  =  T_{K^{+}K^{-}\rightarrow K^{+}K^{-}}(s,t,u)+T_{K^{+}K^{-}\rightarrow K^{0}\bar{K}^{0}}(s,t,u).
\end{eqnarray}
\end{itemize}

For $K^{*+}$, the following processes can contribute:
\begin{itemize}
\item $K\pi \to K\pi$ scattering:
\begin{eqnarray}
T^{I=\frac{1}{2},I_3=\frac{1}{2}}(s,t,u)=2T_{K^{0}\pi^{+}\rightarrow K^{0}\pi^{+}}(s,t,u)-T_{K^{+}\pi^{0}\rightarrow K^{+}\pi^{0}}(s,t,u),
\end{eqnarray}
\item $K \pi \to K \eta$/$K \eta \to K \pi$ scattering:
\begin{eqnarray}
T^{I=\frac{1}{2},I_3=\frac{1}{2}}(s,t,u)  = \sqrt{3}T_{K^{+}\pi^{0}\rightarrow K^{+}\eta}(s,t,u),
\end{eqnarray}
\item $K \eta \to K \eta$ scattering: The process $K^+ \eta \to K^+ \eta$ is pure $I=1/2$ process.
\begin{eqnarray}
T^{I=\frac{1}{2},I_3=\frac{1}{2}}(s,t,u)  =  T_{K^{+}\eta\rightarrow K^{+}\eta}(s,t,u).
\end{eqnarray}
\end{itemize}

Similarly, for $K^{*0}$ with $I=\frac{1}{2},I_{3}=-\frac{1}{2}$, the following processes can contribute:
\begin{itemize}
\item $K\pi \to K\pi$ scattering:
\begin{eqnarray}
T^{I=\frac{1}{2},I_3=-\frac{1}{2}}(s,t,u)=2T_{\pi^{+}\pi^{-}\rightarrow K^{+}K^{-}}(t,s,u)-T_{K^{0}\pi^{0}\rightarrow K^{0}\pi^{0}}(s,t,u),
\end{eqnarray}
\item $K \pi \to K \eta$/$K \eta \to K \pi$ scattering:
\begin{eqnarray}
T^{I=\frac{1}{2},I_3=-\frac{1}{2}}(s,t,u) = \sqrt{3}T_{K^{0}\pi^{0}\rightarrow K^{0}\eta}(s,t,u),
\end{eqnarray}
\item $K \eta \to K \eta$ scattering:
\begin{eqnarray}
T^{I=\frac{1}{2},I_3=-\frac{1}{2}}(s,t,u)  =  T_{K^{0}\eta\rightarrow K^{0}\eta}(s,t,u).
\end{eqnarray}
\end{itemize}

In the above formulas cross symmetry is applied. The projection to each partial wave $J$ is done via the following decomposition:
\begin{equation}
\label{equ:partialwave}
T^{I,I_3,J}=\frac{1}{32N} \int ^1_{-1}P_J(\cos \theta)T^{I,I_3}(\theta) d (\cos \theta).
\end{equation}
Note that $N=2$ is for processes $\pi\pi \to \pi\pi$ and $\eta \eta \to \eta\eta$, and $N=1$ for other process. We emphasize that the mass differences of the pseudoscalar mesons are reserved as mentioned before in the final calculation and they reflect the effects from the isospin symmetry breaking. Taking $\pi^+ \pi^- \to \pi^0 \pi^0$ as an example, the tree level amplitude of $\mathcal O(p^2)$ is expressed as
\begin{eqnarray}\label{pipitopipi-1}
T_{\pi^+ \pi^- \to \pi^0 \pi^0}(s,t,u)=\frac{3s-m_{\pi^+}^2-2 m_{\pi^0}^2}{3f_\pi^2} \ ,
\end{eqnarray}
with the isospin symmetry breaking manifested by the pseudoscalar meson masses. In contrast, in the isospin symmetry limit the amplitude has a form of
\begin{eqnarray}
T_{\pi^+ \pi^- \to \pi^0 \pi^0}(s,t,u)=\frac{s-m_\pi^2}{f_\pi^2} \ ,
\end{eqnarray}
which is equivalent to Eq.~(\ref{pipitopipi-1}) by taking $m_{\pi^\pm}=m_{\pi^0}=m_{\pi}$, i.e. the isospin symmetry limit. For the amplitudes of $\mathcal O(p^4)$ similar relations also exist. Furthermore, there are additional terms in both the tree and loop amplitudes that are proportional to the mass differences of the iso-multiplets. For instance, a typical term in the tree amplitudes manifests itself in the term proportional to $L_{7}^r$ as follows:
\begin{equation}
\frac{64 L^r_{7}\left(m_{k^0}^{2}-m_{k^{+}}^{2}\right)^{2}}{3 f_{\pi}^{4}} .
 \end{equation}
In the loop amplitudes, one finds
\begin{equation}
\frac{\left(m_{k^0}^{2}-m_{k^+}^{2}\right)^{2} J_{\eta \pi^{+}}(s)}{27 f_{\pi}^{4}}.
\end{equation}
Such terms would vanish in the limit of isospin symmetry.

In brief, we obtain the partial wave amplitudes with the isospin symmetry breaking properly taken into account. It also confirms the formulation of Ref.~\cite{GomezNicola:2001as} in the isospin symmetry limit. In view of the complication of the full amplitudes, we include all the independent amplitudes in the Appendix for the convenience of readers. In the analytic calculation of loop diagrams, we have used the Mathematica package FeynClac~\cite{Mertig:1990an,Shtabovenko:2016sxi}.

\section{Results and Discussion}

The $P$-wave $\pi^+\pi^-$ scattering phase shift has been measured in experiment with high precision~\cite{Protopopescu:1973sh} where the presence of the $\rho^0$ pole is evident. Unfortunately, the measurements of the phase shifts are only available for the neutral channel. For the mass splitting question that we focus on in this work, the measured mass difference $m_{\rho^0}-m_{\rho^\pm}=-0.7\pm 0.8$ MeV~\cite{Tanabashi:2018oca} is close to zero, which implies that the phase shift for the $P$-wave $\pi^\pm\pi^0$ channel should yield a similar result as the neutral channel. Taking into account that the leading EM contribution is absent in the $\pi^\pm\pi^0$ channel, it shows that the combined isospin breaking effects on the $\rho$ meson mass splitting is small. This can eventually provide a useful strategy for us to fit the chiral parameters and extract the mass difference for the $K^*$. Moreover, note that the isospin breaking and EM effects, as subleading contributions, are small corrections arising at $\mathcal O(p^4)$. Therefore, the low-energy constants $L_i^r$ determined in Ref.~\cite{GomezNicola:2001as} in the isospin symmetry limit should not change dramatically in comparison with those to be fitted here with the isospin breaking considered.

\begin{figure}[h]
  \centering
  \includegraphics[scale=0.25]{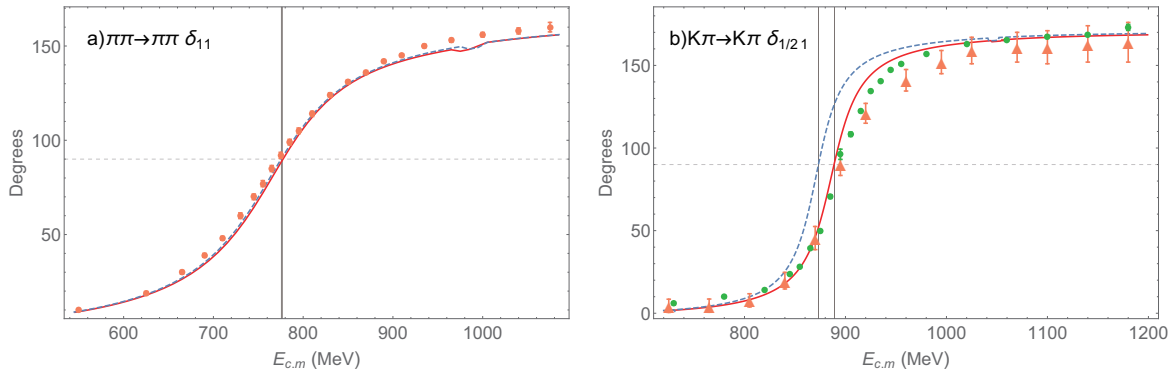}
  \caption{(colored). The curves represent the phase shift using the values of $L_i^r$ given in Ref.~\cite{GomezNicola:2001as}. The red solid line represents the neutral channel and the blue dashed line represents the charged channel in (a) and (b). The experimental data for $\pi\pi$ is extracted from Ref.~\cite{Protopopescu:1973sh} (full dots in (a)) and for $K\pi$ are extracted from Refs.~\cite{Estabrooks:1977xe} (full dots in (b)) and \cite{Mercer:1971kn} (full triangle in (b)).}
  \label{fig:data1}
\end{figure}

As a test of this expectation, we first adopt the values of the low energy constants given in Ref.~\cite{GomezNicola:2001as} to calculate the  $P$-wave $\pi\pi$ and $K\pi$ phase shifts but excluding the EM contributions. The results are plotted in Fig.~\ref{fig:data1} to compare with the experimental data for the $P$-wave $\pi^+\pi^-$ scattering~\cite{Protopopescu:1973sh} and  $K^+\pi^-$ scattering~\cite{Mercer:1971kn,Estabrooks:1977xe}. The neutral channels are illustrated by the red solid lines and the charged channels are shown by the blue dashed lines. It shows that the $P$-wave $\pi^+\pi^-$ scattering data~\cite{Protopopescu:1973sh} can be described well by both curves on the left panel. This comparison is informative and the following points can be learned:
\begin{itemize}
\item Since the red curve on the left panel of Fig.~\ref{fig:data1} do not include the EM contributions, the difference between these two curves are due to the strong isospin breaking.

\item As shown by the measurement of the charged and neutral $\rho$ meson masses, i.e. $m_{\rho^0}-m_{\rho^\pm}=-0.7\pm 0.8$ MeV~\cite{Tanabashi:2018oca}, it suggests that the combined isospin breaking effects from the strong and EM sources should be small. This eventually imposes a constraint on the EM contributions in our treatment. Namely, the inclusion of the EM contributions should not bring in significant deviations from the measured mass difference between the charged and neutral $\rho$ mesons.

\item This above constraint can be implemented into the calculations for the $P$-wave $K\pi$ scatterings with which the combined strong and EM isospin breaking effects can be reliably evaluated.

\item Without the inclusion of the EM, there appears a quite significant mass splitting between the charged and neutral $K\pi$ channels in the phase shifts as shown by the red solid and blue dashed lines on the right panel. Also, note that there are even more significant discrepancies between these two measurements of the $K^+\pi^-$ phase shift from Refs.~\cite{Estabrooks:1977xe} and \cite{Mercer:1971kn} which are denoted by the full dots and full triangles in Fig.~\ref{fig:data1} (b), respectively. This actually makes it difficult to accommodate these two sets of data in the overall fit and also indicates the strong requirement for an update of the $K\pi$ phase shift measurement.

\end{itemize}

Based on the above observations, we re-fit the $\pi^+\pi^-$ and $K^+\pi^-$ phase shifts and take the measured mass difference $m_{\rho^0}-m_{\rho^\pm}=-0.7\pm 0.8$ MeV~\cite{Tanabashi:2018oca} as a constraint on the EM contributions. We then use the fixed low energy constants to predict the phase shifts of the charged $K\pi$ channel from which the mass splitting between the charged and neutral $K^*$ can be extracted.

\begin{table*}[h]
  \begin{center}
  \renewcommand{\arraystretch}{1.3}
  \caption{Low energy constants$\times 10^3$.}
  \begin{threeparttable}
    \begin{tabular}{ccccc}
    \hline \hline
        $L_i^r$ & cutoff fit\cite{Oller:1998hw} & isospin limit\cite{GomezNicola:2001as}& NNLO fit\cite{Bijnens:2014lea}& our fit result  \\
    \hline
        $L_1^r$ & $0.5$   & $0.56\pm 0.10$                  &$0.64\pm0.11$    & $0.54  \pm1.11\times 10^{-2}$       \\
        $L_2^r$ & $1.0$   & $1.21\pm 0.10$                  &$0.59\pm0.04$    & $ 1.19 \pm1.89\times 10^{-2}$       \\
        $L_3^r$ & $-3.25$ & $-2.97\pm0.14$                  &$-2.80\pm0.20$   & $-2.81 \pm1.64\times 10^{-2}$       \\
        $L_4^r$ & $-0.6$  & $-0.36\pm0.17$                  &$0.76\pm0.18$    & $-0.53 \pm9.41\times 10^{-4}$       \\
        $L_5^r$ & $1.7$   & $1.4\pm0.5$                     &$0.50\pm0.07$    & $ 2.00\pm6.26\times 10^{-3}$       \\
        $L_6^r$ & $0.8\tnote{*}$  & $0.07\pm0.08$           &$0.49\pm0.25$    & $ 0.07\pm4.47\times 10^{-2}$        \\
        $L_7^r$ & $0.2$   & $-0.44\pm0.15$                  &$-0.19\pm0.08$   & $-0.43 \pm0.15$       \\
        $L_8^r$ & -       & $0.78\pm0.18$                   &$0.17\pm0.11$    & $ 0.78 \pm0.13$        \\
        $L_9^r$ & -       & -                               &$5.93\pm0.43\tnote{**}$    & $ 3.96 \pm0.45$        \\
     \hline\hline
     \end{tabular}
     \label{tab:fit}
\begin{tablenotes}
\footnotesize
\item[*]This is the value of $2L_6^r+L_8^r$ which is considered as one parameter in Ref.~\cite{Oller:1998hw}.
\item[**] The value of $L_9^r$ is given by Ref.~\cite{Bijnens:2002hp}.
\end{tablenotes}
\end{threeparttable}
\end{center}
\end{table*}

To proceed, let us recall that the leading EM contributions from Figs.~\ref{fig:EM} (a) and (b) contains Coulomb divergence. An empirical treatment is to cut off the forward scattering angle by a cut-off parameter to avoid complicated summation of higher loop contributions. This is understandable since the pion and kaon are not fundamental particles. In the small momentum transfer region, i.e. at the forward scattering region with small scattering angles, the size effects of the hadrons would become non-negligible. Physically, the convolution of the hadron wavefunctions will naturally cut off the Coulomb divergence, which is equivalent to the summation of higher loop contributions in a renormalization scheme. We adopt the minimal value of the scattering angle $\theta$ as the cut-off parameter for the EM contributions. Its value is constrained by requiring that the dynamically generated charged and neutral $\rho$ mesons in the $\pi\pi$ scattering have the difference within the experimental range of $m_{\rho^0}-m_{\rho^\pm}=-0.7\pm 0.8$ MeV~\cite{Tanabashi:2018oca}. This constraint yields $\theta_{min}=30^\circ$.

In Tab.~\ref{tab:fit} we list the fitted low energy constants to compare with those determined in Refs.~\cite{Bijnens:2014lea,Oller:1998hw,GomezNicola:2001as,Bijnens:2002hp}.  It shows that our fitted low energy constants are in good agreement with those fitted by Ref.~\cite{GomezNicola:2001as}. With the isospin symmetry breaking and EM effects considered some of those parameters which are sensitive to certain partial waves in meson-meson scatterings can be well constrained. Several novel features arising from this study can be learned:

\begin{itemize}

\item As shown in Tab.~\ref{tab:fit}, the errors of $L^r_1\sim L^r_5$ are dramatically small in comparison with other studies~\cite{GomezNicola:2001as,Oller:1998hw}. This is reasonable and can be examined explicitly by looking at the corresponding terms proportional to the low energy constants. For instance, by comparing the terms proportional to $L^r_1$ in $\pi\pi \to \pi\pi$ in the charged and neutral channel one can see the difference arising from the strong isospin symmetry breaking, i.e.
    \begin{eqnarray}
    T^{I=1,I_3=0}(s,t,u)_{L^r_1}&=&t^2-u^2+2(m_{\pi}^2+m_{\pi^+}^2)(u-t) \ \ {\text{ for $\rho^0$}},\\
T^{I=1,I_3=1}(s,t,u)_{L^r_1}&=&t^2-u^2+4m_{\pi^+}^2(u-t) \ \ {\text{ for $\rho^+$.}}
\end{eqnarray}
Note that for the same $s$, apart from the pion mass difference the Mandelstam variables $t$ and $u$ are also different in these two channels. It shows that without the consideration of the isospin symmetry breaking effects such discrepancies will be absorbed into the low energy constants and thus result in relatively large errors in the numerical fitting.

\item In contrast, the errors of $L^r_{6,7,8}$ are compatible with those determined in e.g. Ref.~\cite{GomezNicola:2001as} and much larger than those of $L^r_1\sim L^r_5$. We note that for $\pi^+\pi^-\to K^+K^-$, the parameter dependence of the $\mathcal O(p^4)$ amplitude on $L^r_{6,8}$ is in a correlated form, i.e. $2L^r_6+L^r_8$. For $\pi^+\pi^0\to \pi^+ \pi^0$ there is only one term of $\mathcal O(p^4)$ which is proportional to $L_7(m^2_{\pi^0}-m^2_{\pi^+})$. This means that these three constants are process-dependent. In our fitting procedure we have only two processes with experimental data to constrain the parameters. The large errors with $L^r_{6,7,8}$ indicate the need for more data from other channels.

\item Note that $L^r_9$ contributes due to the presence of EM interactions. Its error is correlated with the uncertainty arising from the cut-off angle $\theta$. In another word, since we have fixed $\theta_{min}=30^\circ$ the uncertainties of the EM contributions will be absorbed by $L^r_9$ in the numerical fitting. This problem can be avoid if independent measurements of different isospin channels are available.

\end{itemize}

\begin{figure}[h]
  \centering
   \includegraphics[scale=0.2]{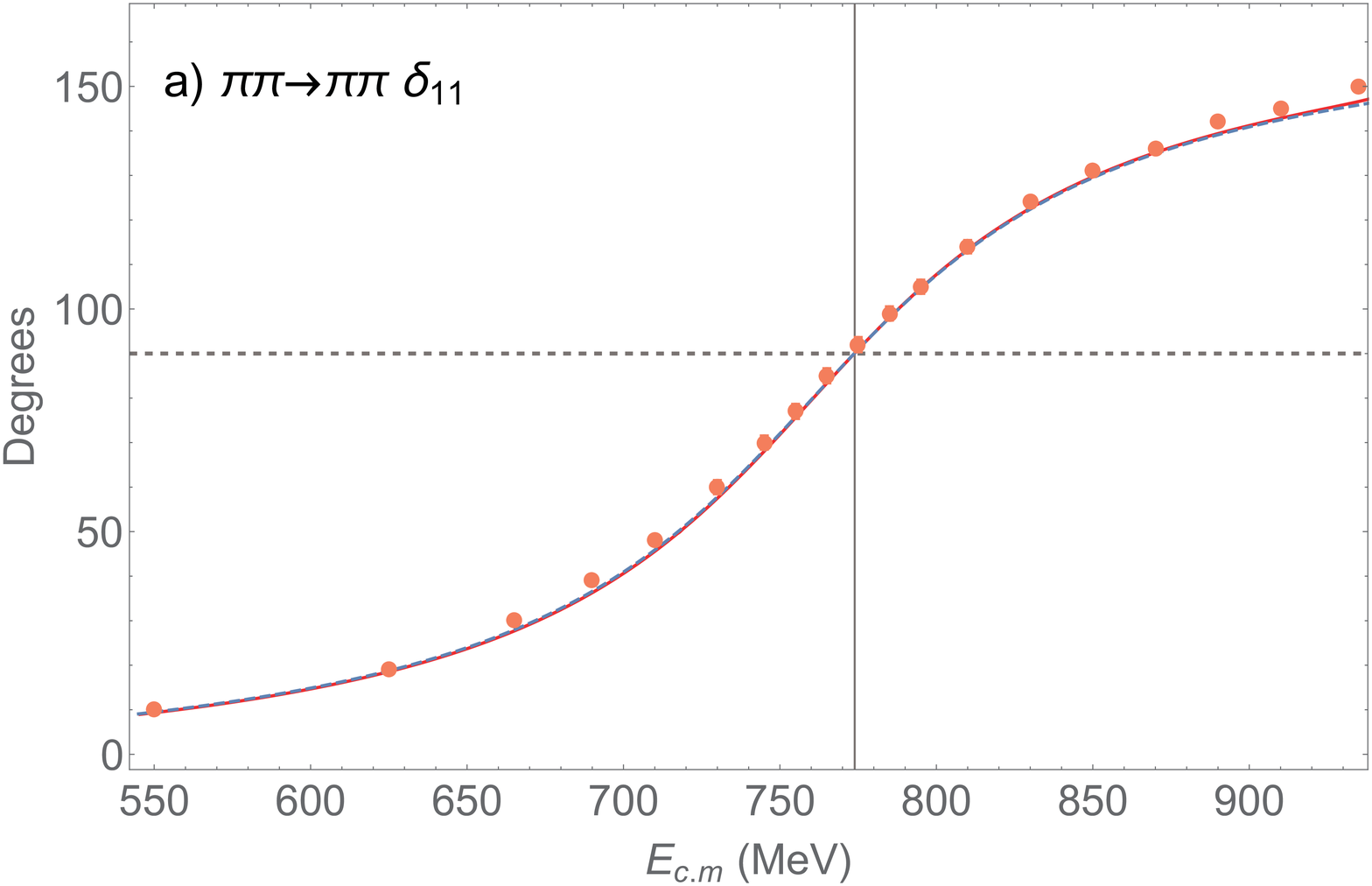}
   \includegraphics[scale=0.2]{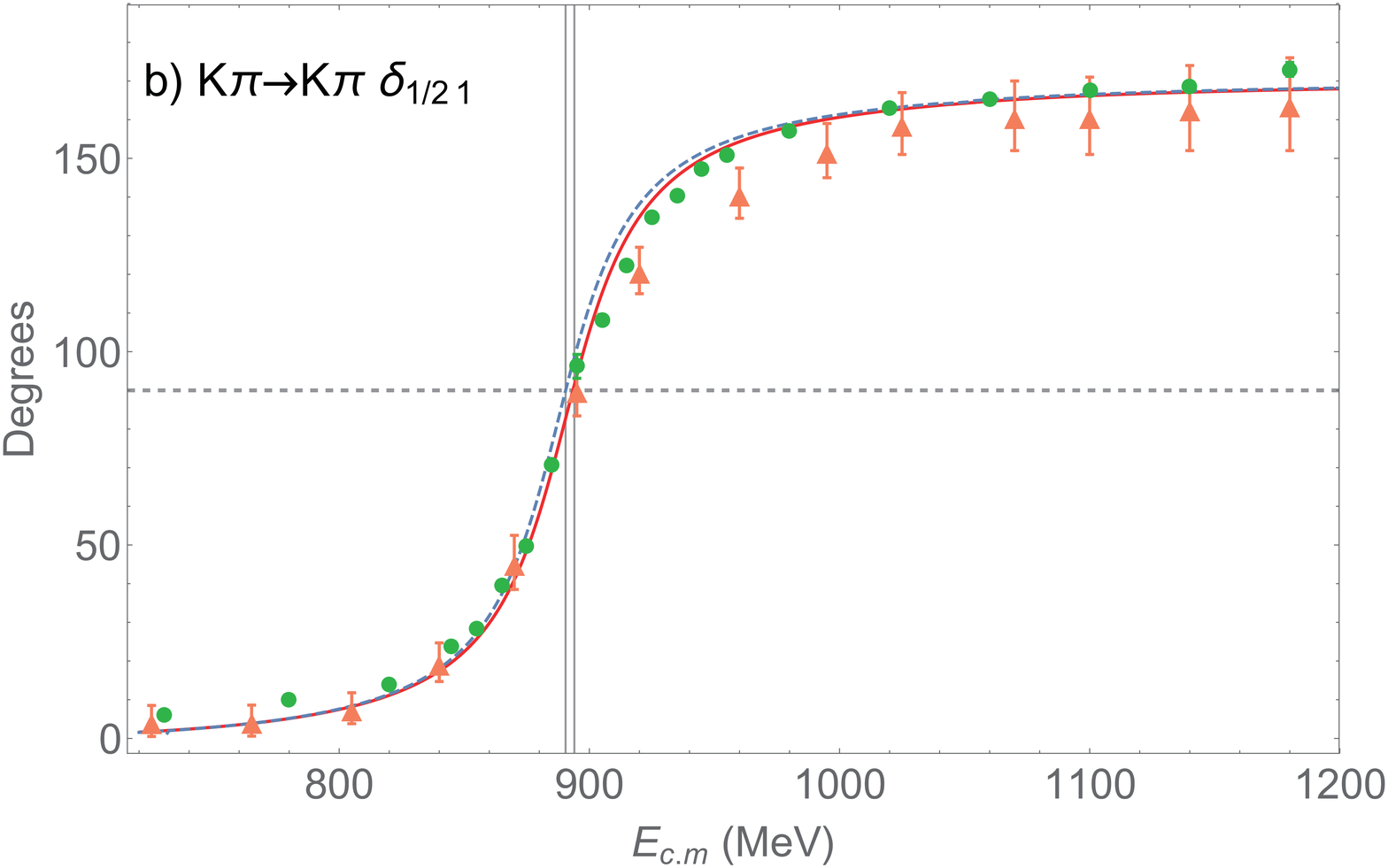}
  \caption{(colored). The phase shifts calculated in our formalism by fitting the experimental data for $\pi^+\pi^-$ from Ref.~\cite{Protopopescu:1973sh} (full dots in (a)) and for $K^+\pi^-$ from Refs.~\cite{Estabrooks:1977xe} (full dots in (b)) and \cite{Mercer:1971kn} (full triangle in (b)). The red solid and blue dashed line represents the neutral and charged channels, respectively. Note that these two curves in (a) are almost overlapping due to small differences. The thin blue dashed line in (b) is the prediction for the $P$-wave $K^+\pi^0$ channel.}
  \label{fig:fit}
\end{figure}

In Fig.~\ref{fig:fit} we show the phase shifts for the $P$-wave $\pi\pi$ (left panel) and $K\pi$ (right panel) scatterings. As described earlier the $\pi^+\pi^-$ and $\pi^+\pi^0$ channels have an additional constraint from the measured mass difference between the neutral and charged $\rho$ meson. Thus, the two curves on the left panel can be regarded as the fitting results for the $\pi\pi$ channel based on our formulation. The red solid line on the right panel denotes the $K^+\pi^-$ channel which is also given by the numerical fitting. With the fitted parameters, we can then make a prediction for the charged $K\pi$ scattering channel as indicated by the blue dashed line on the right panel (Fig.~\ref{fig:fit} (b)). The experimental data for $\pi^+\pi^-$ from Ref.~\cite{Protopopescu:1973sh} (full dots in (a)) and for $K^+\pi^-$ from Refs.~\cite{Estabrooks:1977xe} (full dots in (b)) and \cite{Mercer:1971kn} (full triangle in (b)) are also presented.

Taking a closer look at the results presented in Fig.~\ref{fig:fit}, one finds that the isospin breaking effects between $\rho^0$ and $\rho^\pm$ is rather small and the two phase shifts for the $\pi^+\pi^-$ and $\pi^+\pi^0$ can hardly be distinguished. This is due to the constraint of $m_{\rho^0}-m_{\rho^\pm}=-0.7\pm 0.8$ MeV~\cite{Tanabashi:2018oca}. For the $K\pi$ scattering, the relatively large errors with the data from Ref.~\cite{Mercer:1971kn} and significant discrepancies between the measurements from \cite{Mercer:1971kn} and \cite{Estabrooks:1977xe} result in quite large reduced $\chi^2$, i.e. $\chi^2/d.o.f=20.17$. However, one sees that the red solid curve in Fig.~\ref{fig:fit} can well describe the high-precision data from Ref.~\cite{Estabrooks:1977xe}.

\begin{figure}
  \centering
   \includegraphics[scale=0.25]{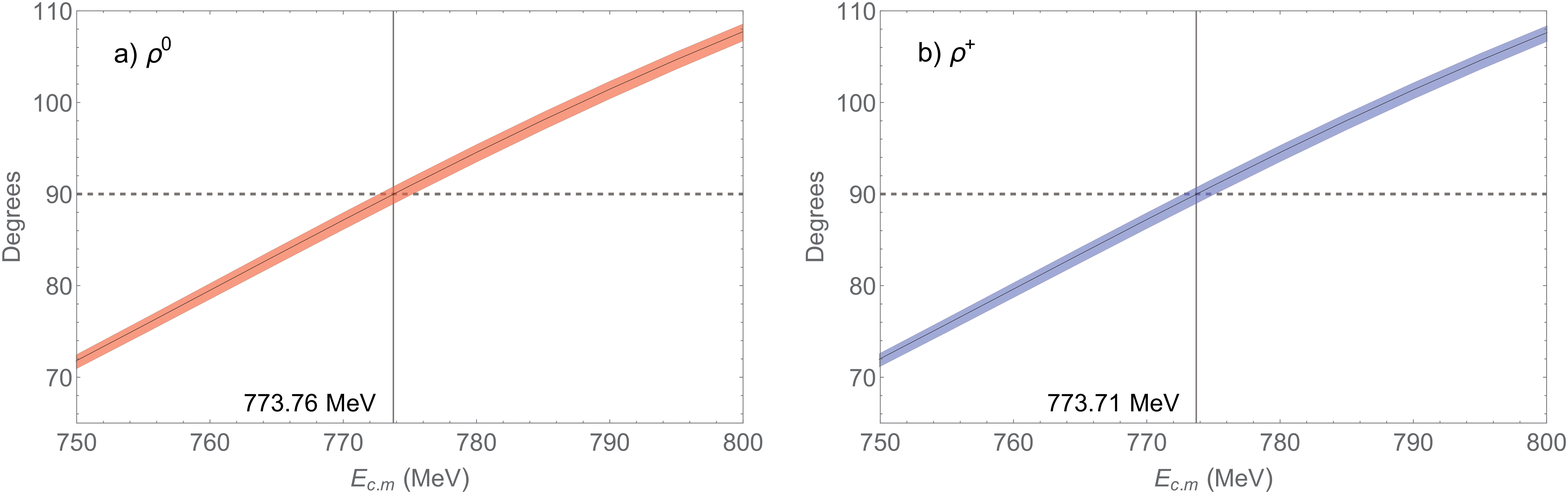}
   \includegraphics[scale=0.25]{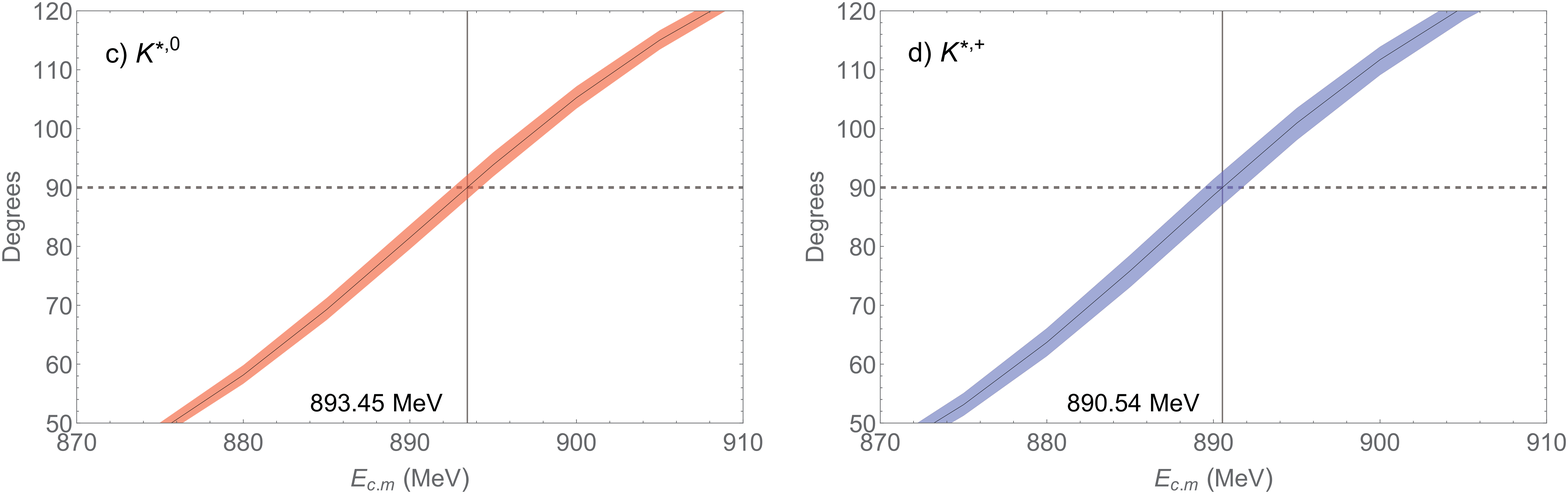}
  \caption{(colored). The phase shifts with uncertainty bands for (a) $\pi^+\pi^-$, (b) $\pi^+\pi^0$, (c) $K^+\pi^-$, and (d) $K^+\pi^0$. The black thin-solid lines represent the center value of the corresponding phase shift.}
  \label{fig:eb}
\end{figure}

With the fitted parameters and phase shifts, we can then extract the pole masses for these iso-multiplets and the results are listed in Table~\ref{tab:mass}. One notices that the central values extracted from the phase shifts are slightly different from those provided by PDG. But they are in agreement with each other within the uncertainties. The uncertainties are given by the errors for the fitted parameters. To illustrate more clearly the uncertainties with the phase shifts, we plot in Fig.~\ref{fig:eb} the phase shifts with uncertainty bands around the pole masses for these four channels. The uncertainty bands are obtained by Monte Carlo sampling. The main procedure contains two steps. First, we use the Monte Carlo sampling to generate one set of sample points in the parameter space which is constituted by the central values of those fitted parameters with errors. With this set of sample points, we can calculate the $\chi^2$. If $\chi^2-\chi_{min}^2<21.67 $ (confidence level 99\%)~\cite{James:1994vla}, we will save this set of sample points. For our calculation $50$ sets of sample points are sufficient. Second, we calculate the phase shift with all sample points at certain energies and save the boundary values. Then the uncertainty bands can be plotted with all boundary values.

With the error transfer formulas (first-order series approximation)
\begin{equation}
x^{+u_x}_{-d_x}-y^{+u_y}_{-d_y}=(x-y)^{+\sqrt{u_x^2+u_y^2}}_{-\sqrt{d_x^2+d_y^2}},
\end{equation}
we finally obtain the mass differences: $m_{K^{*0}}-m_{K^{*+}}= 2.91^{+1.43}_{-1.41}$ MeV, which is close to the result of Ref.~\cite{Bijnens:1997ni}. As a self-consistent check of our formalism, we also extract $m_{\rho^0}-m_{\rho^\pm}=0.050^{+2.04}_{-1.33} $ MeV which is consistent with the experimental value: $m_{\rho^0}-m_{\rho^\pm}=-0.7\pm 0.8$ MeV~\cite{Tanabashi:2018oca}, although the exact sign cannot be determined here.

\begin{table}
  \begin{center}
  \caption{The pole masses of $\rho^0$, $\rho^+$, $K^{*,0}$ and $K^{*,+}$ with uncertainties extracted from the phase shifts. The experimental values from PDG~\cite{Tanabashi:2018oca} are also listed as a comparison.}
    \begin{tabular}{cllll}
    \hline \hline
       Mesons     &$\rho^0$                 &$\rho^+$                 &$K^{*,0}$   &$K^{*,+}$   \\
    \hline
       mass (MeV)&$773.76_{-1.41}^{+1.01}$ &$773.71_{-1.48}^{+0.87}$&$893.45_{-0.81}^{+0.89}$& $890.54_{-1.18}^{+1.10}$ \\
    \hline
       Exp. (MeV)&$775.26\pm 0.25$ &$775.11\pm 0.34$ & $895.81\pm 0.19$ & $895.47\pm 0.20\pm 0.74$ (via $\tau$ decay)\\
       &  &  &  & $891.66\pm 0.26$ (hadroproduced) \\
     \hline\hline
     \end{tabular}
     \label{tab:mass}
    \end{center}
\end{table}

\section{Summary}

In the framework of chiral perturbation theory and the coupled channel inverse amplitude method, we do a full calculation of the $P$-wave $\pi \pi$ and $K\pi$ scattering amplitudes up to $\mathcal O(p^4)$ including the strong isospin breaking effects and EM contributions. With the experimental data for the $\pi^+\pi^-$ and $K^+\pi^-$ phase shifts and constraint on the charged and neutral $\rho$ meson masses, we succeed in re-fitting the low energy constants of the ChPT,  which allows us to extract the mass difference between the neutral and charged $K^*$ meson, i.e. $m_{K^{*0}}-m_{K^{*+}}= 2.91^{+1.43}_{-1.41}$ MeV. This result favors a relatively large mass splitting between the neutral and charged $K^*$ and should be useful for clarifying this long-standing puzzle. Meanwhile, our study also shed some lights on the determination of these low energy constants. We show that the isospin symmetry breaking can impose more stringent constraints on some of these low energy constants. This calls for experimental measurements of the phase shifts in charged channels which seem to have been overlooked in recent experimental efforts. In particular, we would suggest that the large data sample for $\tau$ at BESIII should be able to provide a high-precision measurement of the $P$-wave $K^-\pi^0$ phase shift.

\section*{ACKOWLEDGMENTS}

We are grateful to Jose Pel\'aez for very helpful discussions. We also thank Feng-kun Guo and Zhi Yang for help on treating the error bands in the numerical calculation. This work is supported, in part, by the National Natural Science Foundation of China (NSFC) under Grant Nos. 11425525 and 11521505, by DFG and NSFC through funds provided to the Sino-German CRC 110 ``Symmetries and the Emergence of Structure in QCD'' (NSFC Grant No. 11621131001), and by the National Key Basic Research Program of China under Contract No.~2015CB856700.

\newpage

\begin{appendix}

\section{Loop integrals}

We list below the necessary loop integrals as the supplemental details for the calculations:
\begin{eqnarray}
I_a&=&i\mu^{4-d}\int\frac{d^{d}l}{(2\pi)^{d}}\frac{1}{l^2-m_a^2+i\epsilon}\nonumber\\
&=&\frac{m_a^2}{16\pi^2}\left[-L+\ln{(\frac{m_a^2}{\mu^2})}\right] \ ;
\end{eqnarray}

\begin{eqnarray}
I_{aa}(p^2)
& = & i\mu^{4-d}\int\frac{d^{d}l}{(2\pi)^{d}}\frac{1}{(l^{2}-m_a^{2}+i\epsilon)[(p-l)^{2}-m_a^{2}+i\epsilon]}\nonumber\\
&=&\frac{1}{16\pi^2}\left[-L+1+\ln{(\frac{m_a^2}{\mu^2})}+J_{aa}(p^2)\right] \ ,
\end{eqnarray}
where
\begin{eqnarray}
J_{aa}(p^2)=
\begin{cases}
&-2+2\sqrt{\frac{4m_a^2}{p^2}-1}~\mathrm{arccot}\left(\sqrt{\frac{4m_a^2}{p^2}-1}\right),
~~~~~~~~~~~~~~~~0\leq p^2<4m_a^2 \ , \\
&\\
&-2-\sigma\ln{\left|\frac{1-\sigma}{1+\sigma}\right|}
-i\pi\sigma\theta(p^2-4m_a^2),
~~~~~~~~~~~~~~~~~~~~~~~\text{otherwise} \ ,
\end{cases}
\end{eqnarray}
and
\begin{eqnarray}
L&=&\frac{1}{\epsilon}-\gamma_E+\ln{4\pi}+1,\\
\sigma &=&\sqrt{1-\frac{4m_a^2}{p^2}},~~~~~p^2\notin [0,4m_a^2] \ ;
\end{eqnarray}

\begin{eqnarray}
I_{ab}(p^2)
& = & i\mu^{4-d}\int\frac{d^{d}l}{(2\pi)^{d}}\frac{1}{(l^{2}-m_a^{2}+i\epsilon)[(p-l)^{2}-m_b^{2}+i\epsilon]}\nonumber\\
&=&\frac{1}{16\pi^2}\left[-L+\frac{m_a^2}{m_a^2-m_b^2}\ln{(\frac{m_a^2}{\mu^2})}
-\frac{m_b^2}{m_a^2-m_b^2}\ln{(\frac{m_b^2}{\mu^2})}+J_{ab}(p^2)\right] \ ,
\end{eqnarray}
where
\begin{eqnarray}
J_{ab}(p^2)
&=&-1-\left[\frac{m_a^2+m_b^2}{2(m_a^2-m_b^2)}-\frac{m_a^2-m_b^2}{2p^2}\right]\ln{(\frac{m_a^2}{m_b^2})}\nonumber\\
&&+
\begin{cases}
&
-\frac{\Lambda}{p^2}\arctan{\left(\frac{\Lambda}{p^2-m_a^2-m_b^2}\right)} \ ,
~~~~~~~~~~~~~~~~~~~~~(m_a-m_b)^2<p^2<(m_a+m_b)^2 \ ,\\
&\\
&
\frac{\Lambda}{p^2}\ln{\left|\frac{p^2-m_a^2-m_b^2+\Lambda}{2m_am_b}\right|}
-\frac{i\pi\Lambda}{p^2}\theta(p^2-(m_a+m_b)^2) \ ,
~~~~~~~~~~~\text{otherwise} ,
\end{cases}
\end{eqnarray}
and
\begin{eqnarray}
\Lambda = \sqrt{\left|[p^2-(m_a+m_b)^2][p^2-(m_a-m_b)^2]\right|} \ .
\end{eqnarray}

\section{Amplitudes}
In this Appendix we collect the analytic expressions of the related meson-meson scattering amplitudes. Since we consider the explicit isospin symmetry breaking effects and the electromagnetic contributions, we need to calculate 13 independent amplitudes. However, with the SU(3) chiral symmetry, the numbers of independent amplitudes will reduce to 8. In the following expressions, we use $m_{\pi, k, \eta}$ to represent the masses of the corresponding neutral mesons, while $m_{\pi^+, k^+}$ are the masses of the charged ones.
\begin{eqnarray}
T_{\pi^{+}\pi^{-}\rightarrow\pi^{+}\pi^{-}}^{EM}(s,t,u) & = &
e^2 \left(\frac{u-s}{t}+\frac{u-t}{s}\right)+\frac{4 e^2 L_9 \left(3 u-4 m_{\pi ^+}^2\right)}{f_{\pi }^2}~\nonumber \\ &&
-\frac{4 e^2 \left(s^2+t^2+u^2-4 m_{\pi ^+}^2 u\right)}{f_{\pi }^2 s t}\left[4L_4 \left(m_{\pi }^2+m_{\pi ^+}^2+3 m_k^2+m_{k^+}^2\right) \right.~\nonumber \\ &&
\left.+4L_5 \left(m_{\pi }^2+m_{\pi ^+}^2\right)-2\mu _{\pi } -\mu _k\right] \ ;
\end{eqnarray}
\begin{eqnarray}
T_{\pi^{+}\pi^{-}\rightarrow\pi^{+}\pi^{-}}^{H}(s,t,u) & =&
-\frac{ u-2 m_{\pi ^+}^2}{ f_{\pi }^2}+\frac{2 s t+u \left(u-4 m_{\pi ^+}^2\right)-\left(4 m_{\pi ^+}^2-3 u\right) \left(m_k^2+m_{k^+}^2+4
m_{\pi ^+}^2\right)}{192 \pi ^2 f_{\pi }^4}~\nonumber \\ &&
+\frac{8 L_1 \left(-8 m_{\pi ^+}^4+4 u m_{\pi ^+}^2+s^2+t^2\right)}{f_{\pi }^4}+\frac{4 L_2 \left(s^2+t^2+2 u^2-4 u m_{\pi ^+}^2\right)}{f_{\pi
}^4}~\nonumber \\ &&
+\frac{4 L_3 \left(-8 m_{\pi ^+}^4+4 u m_{\pi ^+}^2+s^2+t^2\right)}{f_{\pi }^4}-\frac{8 L_5 \left(2 m_{\pi ^+}^4-2 m_{\pi }^2m_{\pi ^+}^2+u
m_{\pi }^2 \right)}{f_{\pi }^4}~\nonumber \\ &&
-\frac{8 L_4 \left[2 m_{\pi ^+}^4+u m_{\pi ^+}^2+\left(m_{\pi }^2+m_k^2-m_{k^+}^2\right)\left(u-2 m_{\pi ^+}^2\right)\right]}{f_{\pi }^4}+\frac{64
L_6 m_{\pi ^+}^4}{f_{\pi }^4}~\nonumber \\ &&
+\frac{32 L_8 m_{\pi ^+}^4}{f_{\pi }^4}+\frac{2 \left(-8 m_{\pi }^2-22 m_{\pi ^+}^2+15 u\right) \mu _{\pi }}{15 f_{\pi }^4}+\frac{4\text{  }m_{\pi
^+}^2\mu _{\pi ^+}}{3 f_{\pi }^4}~\nonumber \\ &&
+\frac{\left(9 u-16 m_{\pi ^+}^2\right) \mu _k}{6 f_{\pi }^4}-\frac{\left(3 u-8 m_{\pi ^+}^2\right) \mu _{k^+}}{6 f_{\pi }^4}-\frac{\left(u-2
m_{\pi ^+}^2\right){}^2 J_{\pi ^+ \pi ^+}(u)}{2 f_{\pi }^4}~\nonumber \\ &&
+\left\{\frac{\left[2 (t-u) m_k^2+s \left(-2 m_{\pi ^+}^2-s+u\right)\right] J_{k k}(s)}{24 f_{\pi }^4}-\frac{\left(2 m_{\pi }^2+m_{\pi ^+}^2-3
s\right){}^2 J_{\pi  \pi }(s)}{18 f_{\pi }^4}\right.~\nonumber \\ &&
-\frac{\left(m_k^2-m_{k^+}^2\right){}^2 J_{\pi  \eta }(s)}{27 f_{\pi }^4}+\frac{\left[2 (t-u) m_{k^+}^2+s \left(-2 m_{\pi ^+}^2-s+u\right)\right]
J_{k^+ k^+}(s)}{24 f_{\pi }^4}~\nonumber \\ &&
\left.-\frac{\left[8 m_{\pi ^+}^4-4 t m_{\pi ^+}^2+s (s-u)\right]J_{\pi ^+ \pi ^+}(s)}{6 f_{\pi }^4}-\frac{ m_{\pi ^+}^4J_{\eta  \eta }(s)}{18
f_{\pi }^4}+[s\longleftrightarrow t]\right\} \ ;
\end{eqnarray}

\begin{eqnarray}
T_{\pi^{+}\pi^{-}\rightarrow K^{+}K^{-}}^{EM}(s,t,u) & = &
\frac{e^2 (u-t)}{s}+\frac{4 e^2 L_9 (u-t)}{f_{\pi }^2}+\frac{ e^2\text{  }(u-t)}{f_{\pi }^2 s}\left[-21\mu _{\pi }-3\mu _k-3\mu _{\eta
}\right.~\nonumber \\ &&
\left.+16 L_4 \left(m_{k^+}^2+3 m_k^2+m_{\pi }^2+m_{\pi ^+}^2\right)+8L_5 \left(m_{k^+}^2-m_k^2+3 m_{\pi }^2+m_{\pi ^+}^2\right) \right] \ ;
\end{eqnarray}
\begin{eqnarray}
T_{\pi^{+}\pi^{-}\rightarrow K^{+}K^{-}}^{H}(s,t,u)&=&
\frac{m_{k^+}^2+m_{\pi ^+}^2-u}{2 f_{\pi }^2} -\frac{1}{768 \pi ^2 t f_{\pi }^4}\left\{2 m_{k^+}^6+m_{\pi }^2 m_{k^+}^4+3 m_{\eta
}^2 m_{k^+}^4-2 m_{\pi ^+}^2 m_{k^+}^4\right.~\nonumber \\ &&
-2 t m_{k^+}^4-2 m_{\pi ^+}^4 m_{k^+}^2+8 t^2 m_{k^+}^2-2 m_{\pi }^2 m_{\pi ^+}^2 m_{k^+}^2-6 m_{\eta }^2 m_{\pi ^+}^2 m_{k^+}^2~\nonumber \\ &&
+8 t m_{\pi ^+}^2 m_{k^+}^2+2 s t m_{k^+}^2-6 t u m_{k^+}^2+2 m_{\pi ^+}^6+m_{\pi }^2 m_{\pi ^+}^4+3 m_{\eta }^2 m_{\pi ^+}^4~\nonumber \\ &&
+2 t m_{\pi ^+}^4-4 s t^2-2 t u^2+s t m_{\pi }^2-t u m_{\pi }^2+3 s t m_{\eta }^2-3 t u m_{\eta }^2~\nonumber \\ &&
\left.+6 t^2 m_{\pi ^+}^2-8 t u m_{\pi ^+}^2+4 m_k^2 \left[m_{k^+}^4-2 m_{\pi ^+}^2 m_{k^+}^2+m_{\pi ^+}^4+(s-t) t\right]\right\}~\nonumber \\ &&
+\frac{8 L_1 \left(s-2 m_{k^+}^2\right) \left(s-2 m_{\pi ^+}^2\right)}{f_{\pi }^4}+\frac{32 L_6 m_{k^+}^2 m_{\pi ^+}^2}{f_{\pi }^4}~\nonumber \\ &&
+\frac{4 L_2 \left[\left(m_{k^+}^2+m_{\pi ^+}^2-t\right){}^2+\left(m_{k^+}^2+m_{\pi ^+}^2-u\right){}^2\right]}{f_{\pi }^4}~\nonumber \\ &&
+\frac{2 L_3 \left[\left(m_{k^+}^2+m_{\pi ^+}^2-t\right){}^2+\left(s-2 m_{k^+}^2\right) \left(s-2 m_{\pi ^+}^2\right)\right]}{f_{\pi }^4}~\nonumber \\ &&
+\frac{4 L_5 \left[\left(2 m_{\pi }^2-m_k^2\right)\left(m_{k^+}^2+m_{\pi ^+}^2-u\right)-2 m_{k^+}^2 m_{\pi ^+}^2 \right]}{f_{\pi }^4}~\nonumber \\ &&
+\frac{4 L_4 }{f_{\pi }^4}\left[-m_{k^+}^4-10 m_{\pi ^+}^2 m_{k^+}^2+2 s m_{k^+}^2+u m_{k^+}^2-m_{\pi ^+}^4+2 s m_{\pi ^+}^2+u m_{\pi ^+}^2\right.~\nonumber \\ &&
\left.+\left(m_k^2+m_{\pi }^2 \right)\left(m_{k^+}^2+m_{\pi ^+}^2-u\right)\right]+\frac{16 L_8 m_{k^+}^2 m_{\pi ^+}^2}{f_{\pi }^4}~\nonumber \\ &&
+\frac{1}{360 t^2 f_{\pi }^4}\left\{-\left[30 m_{k^+}^4+30 \left(t-2 m_{\pi ^+}^2\right) m_{k^+}^2+30 m_{\pi ^+}^4+37 t^2-15 t u\right] m_k^2\right.~\nonumber \\ &&
-3 m_{\pi }^2 \left[-10 m_{k^+}^4+10 \left(2 m_{\pi ^+}^2+t\right) m_{k^+}^2-10 m_{\pi ^+}^4+29 t^2-20 t m_{\pi ^+}^2+5 t u\right]+~\nonumber \\ &&
\left.t \left[-90 m_{k^+}^4+\left(60 m_{\pi ^+}^2-818 t\right) m_{k^+}^2+3 \left(10 m_{\pi ^+}^4-336 t m_{\pi ^+}^2+5 t (t+65 u)\right)\right]\right\}
\mu _{\pi }~\nonumber \\ &&
+\frac{1}{12 t^2 f_{\pi }^4}\left\{-2 m_{k^+}^6+6 m_{\pi ^+}^2 m_{k^+}^4+\left[-6 m_{\pi ^+}^4+2 t m_{\pi ^+}^2+t (2 t-s)\right] m_{k^+}^2\right.~\nonumber \\ &&
\left.+2 m_{\pi ^+}^6-2 t m_{\pi ^+}^4+3 s t^2+s t m_{\pi ^+}^2\right\} \mu _{\pi ^+}~\nonumber \\ &&
-\frac{1}{720 t^2 f_{\pi }^4}\left\{180 m_{\eta }^2 m_{k^+}^4-136 t^2 m_{k^+}^2+90 t m_{\eta }^2 m_{k^+}^2-360 m_{\eta }^2 m_{\pi ^+}^2 m_{k^+}^2\right.~\nonumber \\ &&
+180 m_{\eta }^2 m_{\pi ^+}^4-72 t^3+168 s t^2+15 t^2 m_{\eta }^2+45 s t m_{\eta }^2-45 t u m_{\eta }^2~\nonumber \\ &&
-136 t^2 m_{\pi ^+}^2-90 t m_{\eta }^2 m_{\pi ^+}^2+48 t^2 u+30 m_{\pi }^2 \left[2 m_{k^+}^4-2 \left(2 m_{\pi ^+}^2+t\right) m_{k^+}^2\right.~\nonumber \\ &&
\left.+2 m_{\pi ^+}^4+t^2+4 t m_{\pi ^+}^2-t u\right]-40 m_k^2 \left[6 m_{k^+}^4-\left(12 m_{\pi ^+}^2+t\right) m_{k^+}^2\right.~\nonumber \\ &&
\left.\left.+6 m_{\pi ^+}^4-t m_{\pi ^+}^2+2 s t-t u\right]\right\} \mu _k~\nonumber \\ &&
+\frac{1}{12 t^2 f_{\pi }^4}\left[2 m_{k^+}^6-2 \left(3 m_{\pi ^+}^2+t\right) m_{k^+}^4+\left(6 m_{\pi ^+}^4+2 t m_{\pi ^+}^2+s t\right) m_{k^+}^2\right.~\nonumber \\ &&\left.
-2m_{\pi ^+}^6+3 s t^2+t (2 t-s) m_{\pi ^+}^2\right] \mu _{k^+}~\nonumber \\ &&
+\frac{1}{360 t^2 f_{\pi }^4}\left\{5 \left[-18 m_{k^+}^4-6 \left(t-6 m_{\pi ^+}^2\right) m_{k^+}^2-18 m_{\pi ^+}^4-12 t m_{\pi ^+}^2+t (t+9
u)\right] m_k^2\right.~\nonumber \\ &&
+m_{\eta }^2 \left[90 m_{k^+}^4+90 \left(t-2 m_{\pi ^+}^2\right) m_{k^+}^2+90 m_{\pi ^+}^4-3 t (11 t+15 u)\right]~\nonumber \\ &&
\left.+t \left[-30 m_{k^+}^4+18 \left(10 m_{\pi ^+}^2+13 t\right) m_{k^+}^2-150 m_{\pi ^+}^4+424 t m_{\pi ^+}^2+45 t (t-7 u)\right]\right\}
\mu _{\eta }~\nonumber \\ &&
-\frac{1}{144 t^2 f_{\pi }^4}\left\{\left[6 m_{k^+}^4+4 \left(t-3 m_{\pi ^+}^2\right) m_{k^+}^2+6 m_{\pi ^+}^4-2 t m_{\pi ^+}^2+s t-2 t u\right]
m_k^4\right.~\nonumber \\ &&
-2 \left\{3 \left[2 m_{k^+}^4-2 \left(2 m_{\pi ^+}^2+t\right) m_{k^+}^2+2 m_{\pi ^+}^4+s t\right]m_{\pi }^2\right.~\nonumber \\ &&
\left.+t \left[-6 m_{k^+}^4+14 t m_{k^+}^2+3 \left(2 m_{\pi ^+}^4-4 t m_{\pi ^+}^2+t^2-t u\right)\right]\right\} m_k^2~\nonumber \\ &&
+t^2 \left[14 m_{k^+}^4-6 \left(7 t-2 m_{\pi ^+}^2\right) m_{k^+}^2+6 m_{\pi ^+}^4+39 t^2-24 t m_{\pi ^+}^2-3 t u\right]~\nonumber \\ &&
-6 t m_{\pi }^2 \left[3 m_{k^+}^4-2 \left(m_{\pi ^+}^2+3 t\right) m_{k^+}^2-m_{\pi ^+}^4+4 t^2+4 t m_{\pi ^+}^2-t u\right]~\nonumber \\ &&
\left.+3 m_{\pi }^4 \left[2 m_{k^+}^4-2 \left(2 m_{\pi ^+}^2+t\right) m_{k^+}^2+2 m_{\pi ^+}^4+t^2+4 t m_{\pi ^+}^2-t u\right]\right\}J_{k \pi
}(t)~\nonumber \\ &&
-\frac{1}{432 t^2 f_{\pi }^4}\left\{\left[54 m_{k^+}^4-4 \left(27 m_{\pi ^+}^2+2 t\right) m_{k^+}^2+54 m_{\pi ^+}^4+10 t m_{\pi ^+}^2+13 s t-14
t u\right] m_k^4\right.~\nonumber \\ &&
+2 \left\{t \left[-18 m_{k^+}^4+14 t m_{k^+}^2+18 m_{\pi ^+}^4-64 t m_{\pi ^+}^2+9 t (t+3 u)\right]\right.~\nonumber \\ &&
\left.-3 m_{\eta }^2 \left[18 m_{k^+}^4+2 \left(t-18 m_{\pi ^+}^2\right) m_{k^+}^2+18 m_{\pi ^+}^4-4 t m_{\pi ^+}^2+5 s t-4 t u\right]\right\}
m_k^2~\nonumber \\ &&
+t^2 \left[14 m_{k^+}^4+2 \left(9 t-10 m_{\pi ^+}^2\right) m_{k^+}^2+38 m_{\pi ^+}^4-36 t m_{\pi ^+}^2+27 t (t-u)\right]~\nonumber \\ &&
+9 m_{\eta }^4 \left[6 m_{k^+}^4+4 \left(t-3 m_{\pi ^+}^2\right) m_{k^+}^2+6 m_{\pi ^+}^4-2 t m_{\pi ^+}^2+t (s-2 u)\right]~\nonumber \\ &&
\left.+6 t m_{\eta }^2\left[-3 m_{k^+}^4+\left(18 m_{\pi ^+}^2-22 t\right) m_{k^+}^2-15 m_{\pi ^+}^4+8 t m_{\pi ^+}^2+9 t u\right]\right\}J_{k \eta
}(t)~\nonumber \\ &&
-\frac{1}{72 f_{\pi }^4}\left[4 m_{\pi }^4-2 \left(2 m_{k^+}^2+m_{\pi ^+}^2+6 s\right) m_{\pi }^2-2 m_{\pi ^+}^4+9 s^2+6 s m_{k^+}^2\right.~\nonumber \\ &&
\left.-2 m_{k^+}^2 m_{\pi ^+}^2+3 s m_{\pi ^+}^2+m_k^2 \left(4 m_{\pi }^2+2 m_{\pi ^+}^2-6 s\right)\right] J_{\pi \pi }(s)~\nonumber \\ &&
-\frac{\left(m_k^2-m_{k^+}^2\right) \left(2 m_k^2+3 m_{\pi }^2+3 m_{\eta }^2+6 m_{k^+}^2-2 m_{\pi ^+}^2-9 s\right) J_{\pi  \eta }(s)}{108 f_{\pi
}^4}~\nonumber \\ &&
-\frac{\left[2 m_{k^+}^4+\left(2 m_{\pi ^+}^2-3 t+u\right) m_{k^+}^2+s \left(m_{\pi ^+}^2+s-u\right)\right] J_{k^+ k^+}(s)}{12 f_{\pi }^4}~\nonumber \\ &&
-\frac{1}{24 t^2 f_{\pi }^4}\left\{2 m_{k^+}^8-2 \left(4 m_{\pi ^+}^2+t\right) m_{k^+}^6+\left(12 m_{\pi ^+}^4+2 t m_{\pi ^+}^2+s t\right) m_{k^+}^4\right.~\nonumber \\ &&
-2 \left[4 m_{\pi ^+}^6-t m_{\pi ^+}^4+(s-3 t) t m_{\pi ^+}^2+s t^2\right] m_{k^+}^2~\nonumber \\ &&
\left.+\left(t-m_{\pi ^+}^2\right) \left[-2 m_{\pi ^+}^6+t (2 t-s) m_{\pi ^+}^2+t^2 (t-u)\right]\right\}J_{k^+ \pi ^+}(t)~\nonumber \\ &&
-\frac{\left(m_{k^+}^2+m_{\pi ^+}^2-u\right){}^2 J_{k^+ \pi ^+}(u)}{4 f_{\pi }^4}-\frac{\left(2 (t-u) m_k^2+s \left(m_{k^+}^2+m_{\pi ^+}^2+s-t\right)\right)
J_{k k}(s)}{24 f_{\pi }^4}~\nonumber \\ &&
+\frac{m_{\pi ^+}^2\left(-2 m_k^2+6 m_{\eta }^2+2 m_{k^+}^2+2 m_{\pi ^+}^2-9 s\right) J_{\eta  \eta }(s) }{72 f_{\pi }^4}~\nonumber \\ &&
-\frac{\left[2 m_{\pi ^+}^4+(u-3 t) m_{\pi ^+}^2+s (s-u)+m_{k^+}^2 \left(2 m_{\pi ^+}^2+s\right)\right] J_{\pi ^+ \pi ^+}(s)}{12 f_{\pi }^4}
\end{eqnarray}

\begin{eqnarray}
T_{K^{+}K^{-}\rightarrow K^{+}K^{-}}^{EM}(s,t,u)&=&
e^2 \left(\frac{u-s}{t}+\frac{u-t}{s}\right)+\frac{4 e^2 L_9 \left(3 u-4 m_{k^+}^2\right)}{f_{\pi }^2}~\nonumber \\ &&
-\frac{e^2\text{  }\left(s^2+t^2+u^2-4 u m_{k^+}^2\right)}{f_{\pi }^2 s t}\left[16 L_4 \left(m_{k^+}^2+3 m_k^2+m_{\pi }^2+m_{\pi ^+}^2\right)\right.~\nonumber \\ &&
\left.-16L_5 \left(-m_{k^+}^2+m_k^2-2 m_{\pi }^2\right)-13\mu _{\pi }-2\mu _k+3\mu _{\eta }\right] \ ;
\end{eqnarray}
\begin{eqnarray}
T_{K^{+}K^{-}\rightarrow K^{+}K^{-}}^{H}(s,t,u)&=&
-\frac{u-2 m_{k^+}^2}{f_{\pi }^2}-\frac{-2 s t-u \left(u-4 m_{k^+}^2\right)+\left(4 m_{k^+}^2-3 u\right) \left(m_k^2+4 m_{k^+}^2+m_{\pi
^+}^2\right)}{192 \pi ^2 f_{\pi }^4}~\nonumber \\ &&
+\frac{8 L_1 \left(-8 m_{k^+}^4+4 u m_{k^+}^2+s^2+t^2\right)}{f_{\pi }^4}+\frac{4 L_2 \left(s^2+t^2+2 u^2-4 u m_{k^+}^2\right)}{f_{\pi }^4}~\nonumber \\ &&
+\frac{4 L_3 \left(-8 m_{k^+}^4+4 u m_{k^+}^2+s^2+t^2\right)}{f_{\pi }^4}+\frac{64 L_6 m_{k^+}^4}{f_{\pi }^4}+\frac{32 L_8 m_{k^+}^4}{f_{\pi
}^4}~\nonumber \\ &&
-\frac{8 L_4 \left[2 m_{k^+}^4+u m_{k^+}^2+\left(m_{\pi }^2-m_{\pi ^+}^2+m_k^2\right) \left(u-2 m_{k^+}^2\right)\right]}{f_{\pi }^4}~\nonumber \\ &&
+\frac{8 L_5 \left(-2 m_{k^+}^4+2 m_k^2 \left(u-2 m_{k^+}^2\right)-3 m_{\pi }^2 \left(u-2 m_{k^+}^2\right)\right)}{f_{\pi }^4}~\nonumber \\ &&
-\frac{\left(28 m_k^2+33 m_{\pi }^2+15 m_{\eta }^2+3012 m_{k^+}^2-28 m_{\pi ^+}^2-1530 u\right) \mu _{\pi }}{180 f_{\pi }^4}~\nonumber \\ &&
+\frac{\left(8 m_{k^+}^2-3 u\right) \mu _{\pi ^+}}{6 f_{\pi }^4}+\frac{4 \mu _{k^+} m_{k^+}^2}{3 f_{\pi }^4}+\frac{\left(8 m_{k^+}^2-3 u\right)
\mu _k}{6 f_{\pi }^4}~\nonumber \\ &&
-\frac{\left(-28 m_k^2+15 m_{\pi }^2+129 m_{\eta }^2-1764 m_{k^+}^2+28 m_{\pi ^+}^2+810 u\right) \mu _{\eta }}{180 f_{\pi }^4}~\nonumber \\ &&
+\left\{-\frac{\left(-2 m_k^2-2 m_{\pi }^2+2 m_{k^+}^2+2 m_{\pi ^+}^2+3 s\right){}^2 J_{\pi  \pi }(s)}{288 f_{\pi }^4}\right.~\nonumber \\ &&
+\frac{\left[-2 s m_{k^+}^2+2 (t-u) m_{\pi ^+}^2+s (u-s)\right] J_{\pi ^+ \pi ^+}(s)}{24 f_{\pi }^4}~\nonumber \\ &&
+\frac{\left[2 (t-u) m_k^2+s \left(-2 m_{k^+}^2-s+u\right)\right] J_{k k}(s)}{24 f_{\pi }^4}~\nonumber \\ &&
+\frac{\left[-8 m_{k^+}^4+4 t m_{k^+}^2+s (u-s)\right]J_{k^+ k^+}(s)}{6 f_{\pi }^4}~\nonumber \\ &&
-\frac{\left(2 m_k^2+3 m_{\pi }^2+3 m_{\eta }^2+6 m_{k^+}^2-2 m_{\pi ^+}^2-9 s\right){}^2 J_{\pi  \eta }(s)}{432 f_{\pi }^4}~\nonumber \\ &&
\left.-\frac{\left(-2 m_k^2+6 m_{\eta }^2+2 m_{k^+}^2+2 m_{\pi ^+}^2-9 s\right){}^2 J_{\eta  \eta }(s)}{288 f_{\pi }^4}+[s\longleftrightarrow
t]\right\}~\nonumber \\ &&
-\frac{\left(u-2 m_{k^+}^2\right){}^2 J_{k^+ k^+}(u)}{2 f_{\pi }^4} \ ;
\end{eqnarray}

\begin{eqnarray}
T_{\pi^{+}\pi^{0}\rightarrow \pi^{+}\pi^{0}}^{H}(s,t,u)&=&
\frac{3 t-2 m_{\pi }^2-m_{\pi ^+}^2}{3\text{  }f_{\pi }^2}+\frac{1}{192 \pi ^2 s u f_{\pi }^4}\left\{-2 s m_{\pi }^6-2 u m_{\pi }^6-s
m_{k^+}^2 m_{\pi }^4-u m_{k^+}^2 m_{\pi }^4\right.~\nonumber \\&&
+2 s u m_{\pi }^4+2 s u^2 m_{\pi }^2+2 s^2 u m_{\pi }^2-4 s t u m_{\pi }^2+s u^2 m_{k^+}^2+s^2 u m_{k^+}^2~\nonumber \\&&
-2 s t u m_{k^+}^2+s u [s (t-2 u)+t u]-2 m_{\pi ^+}^6 \left(2 m_{\pi }^2+2 m_{\pi ^+}^2-t\right)~\nonumber \\&&
+m_{\pi ^+}^4 \left[2 s u+\left(2 m_{\pi }^2-m_{k^+}^2\right) \left(2 m_{\pi }^2+2 m_{\pi ^+}^2-t\right)\right]~\nonumber \\&&
+m_k^2 \left[s u \left(2 m_{\pi }^2+2 m_{\pi ^+}^2-3 t\right)-\left(m_{\pi }^2-m_{\pi ^+}^2\right){}^2 \left(2 m_{\pi }^2+2 m_{\pi ^+}^2-t\right)\right]~\nonumber \\&&
\left.+2 m_{\pi ^+}^2 \left\{\left[\left(m_{\pi }^2+m_{k^+}^2\right) \left(2 m_{\pi }^2+2 m_{\pi ^+}^2-t\right)-2 s u\right] m_{\pi }^2+s u
\left(2 m_{\pi }^2+2 m_{\pi ^+}^2-3 t\right)\right\}\right\}~\nonumber \\&&
+\frac{8 L_1 \left(t-2 m_{\pi }^2\right) \left(t-2 m_{\pi ^+}^2\right)}{f_{\pi }^4}+\frac{4 L_2 \left[\left(m_{\pi }^2+m_{\pi ^+}^2-s\right){}^2+\left(m_{\pi
}^2+m_{\pi ^+}^2-u\right){}^2\right]}{f_{\pi }^4}~\nonumber \\&&
+\frac{4 L_3 \left(t-2 m_{\pi }^2\right) \left(t-2 m_{\pi ^+}^2\right)}{f_{\pi }^4}-\frac{8 L_5 \left[2 m_{\pi }^4+\left(2 m_{\pi ^+}^2-3 t\right)
m_{\pi }^2+2 m_{\pi ^+}^4\right]}{3 f_{\pi }^4}+~\nonumber \\&&
\frac{8 L_4}{3 f_{\pi }^4} \left[-2 m_{\pi }^4+\left(2 m_{k^+}^2-5 m_{\pi ^+}^2+3 t\right) m_{\pi }^2-5 m_{\pi ^+}^4-3 t m_{k^+}^2+m_{k^+}^2
m_{\pi ^+}^2+3 t m_{\pi ^+}^2\right.~\nonumber \\&&
\left.+m_k^2 \left(-2 m_{\pi }^2-m_{\pi ^+}^2+3 t\right)\right]+\frac{32 L_6 m_{\pi ^+}^4}{f_{\pi }^4}+\frac{64 L_7 \left(m_k^2-m_{k^+}^2\right){}^2}{3
f_{\pi }^4}~\nonumber \\&&
+\frac{L_8 \left(48 m_{\pi ^+}^4+32 \left(m_k^2-m_{k^+}^2\right){}^2\right)}{3 f_{\pi }^4}+\frac{1}{45 s^2 u^2 f_{\pi }^4}\left\{30 (s-u)^2
m_{\pi }^6\right.~\nonumber \\&&
-30 \left[\left(3 s^2+2 u s+3 u^2\right) m_{\pi ^+}^2-s t u\right] m_{\pi }^4+\left[30 \left(3 s^2+2 u s+3 u^2\right) m_{\pi ^+}^4\right.~\nonumber \\&&
\left.+s u \left(-15 s^2+146 u s-15 u^2\right)\right] m_{\pi }^2-30 (s-u)^2 m_{\pi ^+}^6-30 s t u m_{\pi ^+}^4~\nonumber \\&&
\left.-180 s^2 t u^2+s u \left(15 s^2+34 u s+15 u^2\right) m_{\pi ^+}^2\right\} \mu _{\pi }~\nonumber \\&&
+\frac{1}{15 s^2 u^2 f_{\pi }^4}\left\{-10 \left(s^2+u^2\right) m_{\pi }^6+10 \left(3 s^2+4 u s+3 u^2\right) m_{\pi ^+}^2 m_{\pi }^4\right.~\nonumber \\&&
+\left[-30 \left(s^2+u^2\right) m_{\pi ^+}^4-20 s t u m_{\pi ^+}^2+s u \left(5 s^2-22 u s+5 u^2\right)\right] m_{\pi }^2~\nonumber \\&&
\left.+10 \left(s^2-4 u s+u^2\right) m_{\pi ^+}^6+20 s t u m_{\pi ^+}^4+30 s^2 t u^2+s u \left(-5 s^2+2 u s-5 u^2\right) m_{\pi ^+}^2\right\}\mu
_{\pi ^+}~\nonumber \\&&
-\frac{1}{180 s^2 u^2 f_{\pi }^4}\left\{60 s u m_{\pi }^6+30 \left[2 \left(s^2-u s+u^2\right) m_{k^+}^2-s u \left(2 m_{\pi ^+}^2+t\right)\right]
m_{\pi }^4\right.~\nonumber \\&&
+2 \left\{2 s u \left(-30 m_{\pi ^+}^4+15 t m_{\pi ^+}^2+19 s u\right)-15 m_{k^+}^2 \left[2 \left(2 s^2+u s+2 u^2\right) m_{\pi ^+}^2-s t u\right]\right\}m_{\pi
}^2~\nonumber \\&&
+30 s u^2 m_{\pi ^+}^4+60 s^2 m_{k^+}^2 m_{\pi ^+}^4+60 u^2 m_{k^+}^2 m_{\pi ^+}^4+30 s^2 u m_{\pi ^+}^4-132 s^2 u^3~\nonumber \\&&
-132 s^3 u^2+138 s^2 t u^2-15 s u^3 m_{k^+}^2+2 s^2 u^2 m_{k^+}^2+15 s t u^2 m_{k^+}^2~\nonumber \\&&
-15 s^3 u m_{k^+}^2+15 s^2 t u m_{k^+}^2+152 s^2 u^2 m_{\pi ^+}^2+30 s u^2 m_{k^+}^2 m_{\pi ^+}^2+30 s^2 u m_{k^+}^2 m_{\pi ^+}^2~\nonumber \\&&
-m_k^2 \left\{60 \left(s^2-u s+u^2\right) m_{\pi }^4-30 \left[4 \left(s^2+u^2\right) m_{\pi ^+}^2-s t u\right] m_{\pi }^2+60 \left(s^2+u s+u^2\right)
m_{\pi ^+}^4\right.~\nonumber \\&&
\left.\left.-30 s t u m_{\pi ^+}^2+s u \left(-15 s^2+(15 t+2 u) s+15 (t-u) u\right)\right\}\right\} \mu _k~\nonumber \\&&
+\frac{1}{180 s^2 u^2 f_{\pi }^4}\left\{-60 s u m_{\pi }^6+30 \left[2 \left(s^2-u s+u^2\right) m_{k^+}^2+s u \left(2 m_{\pi ^+}^2+t\right)\right]
m_{\pi }^4\right.~\nonumber \\&&
-2 \left\{15 \left[2 \left(2 s^2+u s+2 u^2\right) m_{\pi ^+}^2-s t u\right] m_{k^+}^2+2 s u \left(-30 m_{\pi ^+}^4+15 t m_{\pi ^+}^2+19 s u\right)\right\}
m_{\pi }^2~\nonumber \\&&
-30 s u^2 m_{\pi ^+}^4+60 s^2 m_{k^+}^2 m_{\pi ^+}^4+60 u^2 m_{k^+}^2 m_{\pi ^+}^4-30 s^2 u m_{\pi ^+}^4+12 s^2 u^3~\nonumber \\&&
+12 s^3 u^2+102 s^2 t u^2-15 s u^3 m_{k^+}^2+2 s^2 u^2 m_{k^+}^2+15 s t u^2 m_{k^+}^2~\nonumber \\&&
-15 s^3 u m_{k^+}^2+15 s^2 t u m_{k^+}^2-32 s^2 u^2 m_{\pi ^+}^2+30 s u^2 m_{k^+}^2 m_{\pi ^+}^2+30 s^2 u m_{k^+}^2 m_{\pi ^+}^2~\nonumber \\&&
+m_k^2 \left\{-60 \left(s^2-u s+u^2\right) m_{\pi }^4+30 \left[4 \left(s^2+u^2\right) m_{\pi ^+}^2-s t u\right]m_{\pi }^2-60 \left(s^2+u s+u^2\right)
m_{\pi ^+}^4\right.~\nonumber \\&&
\left.\left.+30 s t u m_{\pi ^+}^2+s u \left[15 s^2-(15 t+2 u) s+15 u (u-t)\right]\right\}\right\}\mu _{k^+}~\nonumber \\&&
+\frac{m_{\pi ^+}^2\left(2 m_{\pi }^2+m_{\pi ^+}^2-3 t\right) J_{\pi  \pi }(t) }{6 f_{\pi }^4}+\frac{t \left(2 m_{\pi }^2+m_{\pi ^+}^2-3 t\right)
J_{\pi ^+ \pi ^+}(t)}{6 f_{\pi }^4}~\nonumber \\&&
-\frac{t \left(2 m_k^2-2 m_{\pi }^2-2 m_{k^+}^2+2 m_{\pi ^+}^2+3 t\right) J_{k k}(t)}{48 f_{\pi }^4}-\frac{\left(m_k^2-m_{k^+}^2\right){}^2
J_{\pi  \eta }(t)}{9 f_{\pi }^4}~\nonumber \\&&
-\frac{t \left(-2 m_k^2-2 m_{\pi }^2+2 m_{k^+}^2+2 m_{\pi ^+}^2+3 t\right) J_{k^+ k^+}(t)}{48 f_{\pi }^4}-\frac{m_{\pi ^+}^4J_{\eta  \eta }(t)
}{18 f_{\pi }^4}~\nonumber \\&&
-\left\{\frac{1}{72 s^2 f_{\pi }^4}\left\{2 \left[6 m_{\pi }^4-2 \left(6 m_{\pi ^+}^2+s\right) m_{\pi }^2+6 m_{\pi ^+}^4+4 s m_{\pi ^+}^2+s
(t-2 u)\right] m_k^4\right.\right.~\nonumber \\&&
-2 \left\{3 \left(4 m_{k^+}^2+s\right) m_{\pi }^4-2 \left[2 \left(6 m_{\pi ^+}^2+s\right) m_{k^+}^2+3 s m_{\pi ^+}^2\right] m_{\pi }^2\right.~\nonumber \\&&
\left.+3 s \left(m_{\pi ^+}^4+s t-s u\right)+2 m_{k^+}^2 \left(6 m_{\pi ^+}^4+4 s m_{\pi ^+}^2+s t-2 s u\right)\right\} m_k^2+s^2 m_{k^+}^4~\nonumber \\&&
+3 s t m_{k^+}^4-3 s u m_{k^+}^4+12 m_{k^+}^4 m_{\pi ^+}^4+3 s^2 m_{\pi ^+}^4-6 s m_{k^+}^2 m_{\pi ^+}^4-6 s^2 t m_{k^+}^2~\nonumber \\&&
+6 s^2 u m_{k^+}^2+6 s m_{k^+}^4 m_{\pi ^+}^2+3 s^3 t-3 s^3 u+3 m_{\pi }^4 \left(4 m_{k^+}^4-2 s m_{k^+}^2+s^2\right)~\nonumber \\&&
\left.-6 m_{\pi }^2 \left[\left(4 m_{\pi ^+}^2+s\right) m_{k^+}^4-2 s m_{\pi ^+}^2 m_{k^+}^2+s^2 m_{\pi ^+}^2\right]\right\} J_{k k^+}(s)~\nonumber \\&&
-\frac{1}{18 s^2 f_{\pi }^4}\left\{-6 m_{\pi }^8+6 \left(4 m_{\pi ^+}^2+s\right) m_{\pi }^6+\left[s (3 u-5 s)-36 m_{\pi ^+}^4\right] m_{\pi
}^4\right.~\nonumber \\&&
-2 \left[-12 m_{\pi ^+}^6+9 s m_{\pi ^+}^4+s (3 u-11 s) m_{\pi ^+}^2+3 s^2 u\right] m_{\pi }^2-6 m_{\pi ^+}^8~\nonumber \\&&
\left.+12 s m_{\pi ^+}^6+s (3 u-11 s) m_{\pi ^+}^4+6 s^2 (s-u) m_{\pi ^+}^2+3 s^3 (u-s)\right\}J_{\pi  \pi ^+}(s)~\nonumber \\&&
\left.+\frac{\left(m_k^2-m_{k^+}^2\right){}^2 J_{\eta  \pi ^+}(s)}{27 f_{\pi }^4}+[s\longleftrightarrow u]\right\} \ ;
\end{eqnarray}

\begin{eqnarray}
T_{\pi^{+}\pi^{0}\rightarrow K^{+}\bar{K}^{0}}^{H}(s,t,u)&=&
\frac{m_{k^+}^2-m_k^2-3 t+3 u}{6 \sqrt{2} f_{\pi }^2}+\frac{1}{768 \sqrt{2} \pi ^2 s t u f_{\pi }^4}\left\{2 \left[t (s+2 u) m_{\pi
}^2-s (t-2 u) m_{k^+}^2\right.\right.~\nonumber \\&&
\left.-2 u m_{\pi ^+}^2 \left(m_k^2+m_{\pi }^2+m_{k^+}^2+m_{\pi ^+}^2-u\right)\right] m_k^4+\left\{t (s+8 u) m_{\pi }^4\right.~\nonumber \\&&
+3 s \left[t m_{\eta }^2+(t-u) m_{k^+}^2+u \left(m_{\pi ^+}^2-2 t\right)\right] m_{\pi }^2-2 (s+4 t) u m_{\pi ^+}^4~\nonumber \\&&
+3 s u \left(2 t-m_{\eta }^2\right) m_{\pi ^+}^2+s \left[2 (u-2 t) m_{k^+}^4+3 (u-t) m_{\eta }^2 m_{k^+}^2\right.~\nonumber \\&&
\left.\left.+2 t u \left(m_k^2+m_{\pi }^2+m_{k^+}^2+m_{\pi ^+}^2+2 t-5 u\right)\right]\right\} m_k^2~\nonumber \\&&
+2 \left[s (u-t) m_{\pi }^2+t (s+4 u) m_{k^+}^2\right] m_{\pi ^+}^4+m_{\pi ^+}^2 \left[4 t \left(m_k^2+m_{\pi }^2+m_{k^+}^2+m_{\pi ^+}^2-t\right)
m_{k^+}^4\right.~\nonumber \\&&
\left.-3 s t \left(m_{\pi }^2-m_{\eta }^2+2 u\right) m_{k^+}^2+s (t-u) \left(-m_{\pi }^4-3 m_{\eta }^2 m_{\pi }^2+10 t u\right)\right]~\nonumber \\&&
+u \left\{-2 (s+2 t) m_{\pi }^2 m_{k^+}^4-\left[(s+8 t) m_{\pi }^4+3 s \left(m_{\eta }^2-2 t\right) m_{\pi }^2+2 s t \left(m_k^2+m_{\pi }^2\right.\right.\right.~\nonumber \\&&
\left.\left.\left.\left.+m_{k^+}^2+m_{\pi ^+}^2-5 t+2 u\right)\right]m_{k^+}^2+3 s t (u-t) \left(-3 m_{\pi }^2-m_{\eta }^2+2 s\right)\right\}\right\}~\nonumber \\&&
+\frac{\sqrt{2} L_3 \left[\left(m_{\pi }^2+m_{k^+}^2-u\right) \left(m_k^2+m_{\pi ^+}^2-u\right)-\left(m_k^2+m_{\pi }^2-t\right) \left(m_{k^+}^2+m_{\pi
^+}^2-t\right)\right]}{f_{\pi }^4}~\nonumber \\&&
-\frac{2 \sqrt{2} L_4 \left(m_k^2-m_{k^+}^2+3 t-3 u\right) \left(m_k^2+m_{\pi }^2-m_{k^+}^2-m_{\pi ^+}^2\right)}{3 f_{\pi }^4}~\nonumber \\&&
+\frac{\sqrt{2} L_5 }{3 f_{\pi }^4}\left[m_k^4+\left(-8 m_{\pi }^2-5 m_{k^+}^2+2 m_{\pi ^+}^2+9 t-3 u\right) m_k^2\right.~\nonumber \\&&
\left.+m_{\pi }^2 \left(5 m_{k^+}^2-12 t+12 u\right)+m_{k^+}^2 \left(-2 m_{k^+}^2-5 m_{\pi ^+}^2+3 s\right)\right]~\nonumber \\&&
-\frac{16 \sqrt{2} L_7 \left(m_k^2-m_{k^+}^2\right) \left(m_k^2+m_{k^+}^2+m_{\pi ^+}^2\right)}{3 f_{\pi }^4}-\frac{8 \sqrt{2} L_8 \left(m_k^2-m_{k^+}^2\right)
\left(m_k^2+m_{k^+}^2+m_{\pi ^+}^2\right)}{3 f_{\pi }^4}~\nonumber \\&&
+\frac{1}{240 \sqrt{2} s^2 t^2 u^2 f_{\pi }^4}\left\{5 s^2 u \left[4 u m_{k^+}^2-4 u m_{\pi ^+}^2+t \left(m_k^2+m_{\pi }^2+m_{k^+}^2+m_{\pi
^+}^2-6 t\right)\right] m_k^4\right.~\nonumber \\&&
+\left\{-20 t^2 \left(s^2+8 u^2\right) m_{\pi }^4+5 \left\{8 s^2 \left(t^2-u^2\right) m_{k^+}^2+u \left[8 \left(s^2+8 t^2\right) u m_{\pi ^+}^2\right.\right.\right.~\nonumber \\&&
\left.\left.+s t \left(s^2+7 t s+3 u s+40 t u\right)\right]\right\} m_{\pi }^2+t \left\{-20 s^2 t m_{k^+}^4+10 s^2 (t-u) u m_{k^+}^2\right.~\nonumber \\&&
+u\left[-160 t u m_{\pi ^+}^4+5 s \left(s^2+7 t s+11 u s+8 t u\right) m_{\pi ^+}^2\right.~\nonumber \\&&
\left.\left.\left.+s^2 \left(-5 s^2+5 t^2-5 u^2+136 t u\right)\right]\right\}\right\} m_k^2+10 m_{\pi }^4 \left\{\left[2 \left(t^2-u^2\right)
m_{\pi ^+}^2\right.\right.~\nonumber \\&&
\left.+t (t-u) u] s^2+2 \left(s^2+8 t^2\right) u^2 m_{k^+}^2\right\}-5 t m_{\pi }^2 \left\{\left[8 t \left(s^2+8 u^2\right) m_{\pi ^+}^2\right.\right.~\nonumber \\&&
\left.\left.+s u \left(s^2+3 t s+7 u s+40 t u\right)\right] m_{k^+}^2+2 s (t-u) u \left(-s^2+m_{\pi ^+}^2 s+4 t u\right)\right\}~\nonumber \\&&
+t \left\{-5 s^2 \left[u \left(m_k^2+m_{\pi }^2+m_{k^+}^2+m_{\pi ^+}^2-6 u\right)-4 t m_{\pi ^+}^2\right] m_{k^+}^4\right.~\nonumber \\&&
-u \left[-160 t u m_{\pi ^+}^4+5 s \left(s^2+11 t s+7 u s+8 t u\right) m_{\pi ^+}^2\right.~\nonumber \\&&
\left.+s^2 \left(-5 s^2-5 t^2+5 u^2+136 t u\right)\right] m_{k^+}^2~\nonumber \\&&
\left.\left.+20 s (t-u) u \left(-s m_{\pi ^+}^4+2 t u m_{\pi ^+}^2+39 s t u\right)\right\}\right\} \mu _{\pi }~\nonumber \\&&
+\frac{1}{180 \sqrt{2} s^2 t^2 u^2 f_{\pi }^4}\left\{-2 s^2 t^2 \left(-15 m_{\pi }^2+15 m_{k^+}^2+7 u\right) m_k^4\right.~\nonumber \\&&
+\left\{120 t^2 u^2 m_{\pi }^4+2 \left[15 s^2 (t-u) m_{\pi ^+}^4-30 t^2 \left(s^2+4 u^2\right) m_{\pi ^+}^2+s t^2 (8 s-15 u) u\right] m_{\pi
}^2\right.~\nonumber \\&&
+30 s^2 u^2 m_{k^+}^4+s^2 m_{k^+}^2 \left[60 \left(t^2-u^2\right) m_{\pi ^+}^2+t u (t+30 u)\right]~\nonumber \\&&
+u \left[30 \left(s^2+4 t^2\right) u m_{\pi ^+}^4+s t (31 s t-150 u t-15 s u) m_{\pi ^+}^2\right.~\nonumber \\&&
\left.\left.-s^2 t^2 (31 s+16 t)\right]\right\} m_k^2-30 m_{\pi }^4 \left[s^2 (u-t) m_{\pi ^+}^4+4 t^2 u^2 m_{k^+}^2\right]~\nonumber \\&&
+t \left\{45 s^2 u^2 m_{k^+}^4-\left[30 t \left(s^2+4 u^2\right) m_{\pi ^+}^4-15 s t u (s+10 u) m_{\pi ^+}^2\right.\right.~\nonumber \\&&
\left.+s^2 u^2 (31 t+15 u)\right]m_{k^+}^2+15 s (t-u) u \left[s m_{\pi ^+}^4+\left(s^2-2 t u\right) m_{\pi ^+}^2\right.~\nonumber \\&&
-7 s t u]\}-15 m_{\pi }^2 \left\{2 s^2 u^2 m_{k^+}^4-\left[2 s^2 (t-u) m_{\pi ^+}^4+4 \left(s^2+4 t^2\right) u^2 m_{\pi ^+}^2\right.\right.~\nonumber \\&&
\left.\left.\left.+s t (s+2 t) u^2\right] m_{k^+}^2+2 s (t-u) \left(-s m_{\pi ^+}^6+s^2 m_{\pi ^+}^4-t^2 u^2\right)\right\}\right\} \mu _{\pi
^+}~\nonumber \\&&
+\frac{1}{120 \sqrt{2} s^2 t^2 u^2 f_{\pi }^4}\left\{10 s t^2 u m_k^6-10 \left\{t^2 \left(2 s^2-u s+4 u^2\right) m_{\pi }^2\right.\right.~\nonumber \\&&
\left.-s \left[u t^2+2 s \left(t^2-2 u^2\right)\right] m_{k^+}^2-u \left[\left(4 u s^2+t^2 s+4 t^2 u\right) m_{\pi ^+}^2-s t^3\right]\right\}m_k^4~\nonumber \\&&
+\left\{5 \left\{2 \left(5 s^2+8 t^2\right) u^2 m_{k^+}^2+s \left[2 s \left(4 t^2-5 u^2\right) m_{\pi ^+}^2+t u (2 s t+4 u t-s u)\right]\right\}
m_{\pi }^2\right.~\nonumber \\&&
+2\left\{5 s^2 u^2 \left(3 m_{k^+}^2-3 m_{\pi ^+}^2+t\right) m_{\eta }^2+t\left\{-10 \left\{2 t \left(s^2+2 u^2\right) m_{\pi ^+}^2\right.\right.\right.~\nonumber \\&&
+s u [s (t-2 u)+t u]\} m_{k^+}^2-s u \left\{5 (u-2 t) s^2\right.~\nonumber \\&&
+\left(-5 t^2+u t-5 u^2\right) s+5 [2 t u+s (t+4 u)] m_{\pi ^+}^2~\nonumber \\&&
+5 t (t-u) u\}\}\}\} m_k^2+10 s^2 u^2 m_{\pi }^4 \left(-m_{k^+}^2+m_{\pi ^+}^2+t\right)~\nonumber \\&&
+m_{\pi }^2 \left\{t \left\{-40 t u^2 m_{k^+}^4-5 s (s+4 t) u^2 m_{k^+}^2+s^2 \left[-20 t m_{\pi ^+}^4-10 (t-u) u m_{\pi ^+}^2\right.\right.\right.~\nonumber \\&&
\left.\left.\left.+(19 t-5 u) u^2\right]\right\}-5 s^2 u^2 m_{\eta }^2 \left(6 m_{k^+}^2-6 m_{\pi ^+}^2+t\right)\right\}~\nonumber \\&&
+t\left\{5 s^2 u^2 m_{\eta }^2 \left(2 m_{k^+}^2-m_{\pi ^+}^2+s-2 u\right)-2\left\{5 u^2 \left[s (3 s-t)-4 t m_{\pi ^+}^2\right] m_{k^+}^4\right.\right.~\nonumber \\&&
-s \left[10 s t m_{\pi ^+}^4+10 u (s t+u t+2 s u) m_{\pi ^+}^2+u^2 \left(5 s^2-4 t s\right.\right.~\nonumber \\&&
\left.\left.+5 u s+5 t^2-5 t u\right)\right]m_{k^+}^2+s^2 u \left\{5 u m_{\pi ^+}^4\right.~\nonumber \\&&
\left.\left.\left.\left.+\left[5 t^2-3 u t+5 u^2+5 s (2 t+u)\right]m_{\pi ^+}^2+15 t u (u-t)\right\}\right\}\right\}\right\} \mu _k~\nonumber \\&&
-\frac{1}{360 \sqrt{2} s^2 t^2 u^2 f_{\pi }^4}\left\{30 t^2 u \left[-4 u m_{\pi }^2+4 u m_{\pi ^+}^2+s (u-3 s)\right] m_k^4\right.~\nonumber \\&&
+\left\{-30 s^2 t^2 m_{\pi }^4-15 t^2 \left[6 s^2 m_{\eta }^2-2 \left(5 s^2+8 u^2\right) m_{k^+}^2+s u (s+4 u)\right] m_{\pi }^2\right.~\nonumber \\&&
+2 \left\{15 s \left(-4 s t^2+u^2 t+2 s u^2\right) m_{k^+}^4-30 u \left[2 \left(s^2+2 t^2\right) u m_{\pi ^+}^2\right.\right.~\nonumber \\&&
+s t (-2 s t+u t+s u)] m_{k^+}^2+15 s^2 t^2 m_{\eta }^2 \left(3 m_{k^+}^2+u\right)~\nonumber \\&&
+s u \left\{30 s u m_{\pi ^+}^4+30 t [t u+s (2 t+u)]m_{\pi ^+}^2\right.~\nonumber \\&&
\left.\left.\left.+t^2 \left[15 s^2+(15 t+28 u) s+15 u (u-t)\right]\right\}\right\}\right\} m_k^2~\nonumber \\&&
+30 s^2 t^2 m_{\pi }^4 \left(m_{\pi ^+}^2+u\right)-3 m_{\pi }^2 \left\{10 \left(2 s^2-t s+4 t^2\right) u^2 m_{k^+}^4\right.~\nonumber \\&&
+5 s \left[2 s \left(5 t^2-4 u^2\right) m_{\pi ^+}^2+t u (s t-4 u t-2 s u)\right] m_{k^+}^2~\nonumber \\&&
\left.+5 s^2 t^2 m_{\eta }^2 \left(u-6 m_{\pi ^+}^2\right)+s^2 u \left[20 u m_{\pi ^+}^4-10 t (t-u) m_{\pi ^+}^2+t^2 (5 t-19 u)\right]\right\}~\nonumber \\&&
+t \left\{15 s^2 t m_{\eta }^2 \left[2 \left(u-3 m_{\pi ^+}^2\right) m_{k^+}^2+u \left(-m_{\pi ^+}^2+s-2 t\right)\right]\right.~\nonumber \\&&
-2 \left\{-15 s u^2 m_{k^+}^6-15 \left[\left(4 t s^2+u^2 s+4 t u^2\right) m_{\pi ^+}^2-s u^3\right]m_{k^+}^4\right.~\nonumber \\&&
+s u \left\{15 (t-2 u) s^2+\left(-15 t^2+43 u t-15 u^2\right) s\right.~\nonumber \\&&
\left.+15 [2 t u+s (4 t+u)] m_{\pi ^+}^2+15 t u (u-t)\right\} m_{k^+}^2~\nonumber \\&&
\left.\left.\left.+3 s^2 u \left\{5 t m_{\pi ^+}^4+\left[5 t^2-3 u t+5 u^2+5 s (t+2 u)\right] m_{\pi ^+}^2+25 t u (u-t)\right\}\right\}\right\}\right\}\mu
_{k^+}~\nonumber \\&&
-\frac{1}{240 \sqrt{2} t^2 u^2 f_{\pi }^4}\left\{5 t u m_k^6-5 u \left\{2 \left[(t-6 u) m_{\pi ^+}^2+(2 s-3 t) t\right]\right.\right.~\nonumber \\&&
\left.-(t-12 u) m_{k^+}^2\right\} m_k^4+\left\{-5 t u m_{\pi }^4-10 \left\{-6 t^2 m_{\eta }^2+6 \left(t^2-u^2\right) m_{k^+}^2\right.\right.~\nonumber \\&&
\left.+u \left[2 (t+3 u) m_{\pi ^+}^2+t (s+2 t)\right]\right\} m_{\pi }^2+5 t (12 t-u) m_{k^+}^4~\nonumber \\&&
+10 \left[t (t-u) u-6 \left(t^2-u^2\right) m_{\eta }^2\right] m_{k^+}^2+u \left\{-15 t m_{\pi ^+}^4\right.~\nonumber \\&&
\left.\left.-10 \left[6 u m_{\eta }^2+t (t-5 u)\right] m_{\pi ^+}^2+3 t \left(5 s^2-5 t^2+5 u^2+16 t u\right)\right\}\right\} m_k^2~\nonumber \\&&
-30 m_{\eta }^2 \left\{\left[2 u^2 m_{k^+}^2+2 \left(t^2-u^2\right) m_{\pi ^+}^2+t u (u-t)\right] m_{\pi }^2+t\left\{\left[u (u-t)-2 t m_{k^+}^2\right]m_{\pi
^+}^2\right.\right.~\nonumber \\&&
+s (t-u) u\}\}+t \left\{-5 u m_{k^+}^6+10 \left[(u-6 t) m_{\pi ^+}^2+(2 s-3 u) u\right] m_{k^+}^4\right.~\nonumber \\&&
+\left\{5 u m_{\pi }^4+10 \left[2 (3 t+u) m_{\pi ^+}^2+u (s+2 u)\right] m_{\pi }^2+u \left[15 m_{\pi ^+}^4+10 (u-5 t) m_{\pi ^+}^2\right.\right.~\nonumber \\&&
\left.\left.-3 \left(5 s^2+5 t^2-5 u^2+16 t u\right)\right]\right\}m_{k^+}^2~\nonumber \\&&
\left.\left.+20 (t-u) u \left(m_{\pi ^+}^4+4 m_{\pi }^2 m_{\pi ^+}^2+12 t u\right)\right\}\right\} \mu _{\eta }~\nonumber \\&&
-\frac{1}{144 \sqrt{2} t^2 f_{\pi }^4}\left\{-3 \left(2 m_{k^+}^2-2 m_{\pi ^+}^2+t\right) m_k^6+\left[9 \left(2 m_{\pi }^2+t\right) m_{k^+}^2\right.\right.~\nonumber \\&&
\left.-6 \left(3 m_{\pi }^2+4 t\right) m_{\pi ^+}^2+t (23 t+3 u)\right] m_k^4+\left\{9 \left(-2 m_{k^+}^2+2 m_{\pi ^+}^2+t\right) m_{\pi }^4\right.~\nonumber \\&&
+2 t \left(3 m_{k^+}^2+12 m_{\pi ^+}^2-7 t-3 u\right) m_{\pi }^2+t \left[-6 m_{k^+}^4+4 \left(3 m_{\pi ^+}^2+5 t\right) m_{k^+}^2\right.~\nonumber \\&&
\left.\left.-6 m_{\pi ^+}^4-51 t^2+26 t m_{\pi ^+}^2-6 t u\right]\right\} m_k^2+3 t m_{\pi }^4 \left(-5 m_{k^+}^2+5 t+u\right)~\nonumber \\&&
-6 m_{\pi }^6 \left(-m_{k^+}^2+m_{\pi ^+}^2+t\right)+2 t m_{\pi }^2 \left\{3 m_{k^+}^4+\left(t-6 m_{\pi ^+}^2\right) m_{k^+}^2+3 \left[m_{\pi
^+}^4+(s-2 t) t\right]\right\}~\nonumber \\&&
\left.+t^2\left\{-10 m_{k^+}^4+\left(4 m_{\pi ^+}^2+9 t\right) m_{k^+}^2+3\left[2 m_{\pi ^+}^4-8 t m_{\pi ^+}^2+t (5 t+u)\right]\right\}\right\}J_{k \pi
}(t)~\nonumber \\&&
+\frac{1}{864 \sqrt{2} t^2 f_{\pi }^4}\left\{18 \left(6 m_{k^+}^2-6 m_{\pi ^+}^2+t\right) m_k^6+6 \left[6 \left(-3 m_{k^+}^2+3 m_{\pi ^+}^2+t\right)
m_{\pi }^2\right.\right.~\nonumber \\&&
\left.-12 m_{\eta }^2 \left(3 m_{k^+}^2-3 m_{\pi ^+}^2+t\right)+t \left(-15 m_{k^+}^2+12 m_{\pi ^+}^2+t-9 u\right)\right] m_k^4~\nonumber \\&&
+\left\{36 \left(3 m_{k^+}^2-3 m_{\pi ^+}^2+t\right) m_{\eta }^4-3 t \left(24 m_{k^+}^2-21 m_{\pi ^+}^2+5 t-36 u\right) m_{\eta }^2\right.~\nonumber \\&&
+2 t \left[18 m_{k^+}^4+4 \left(7 t-9 m_{\pi ^+}^2\right) m_{k^+}^2+18 m_{\pi ^+}^4-58 t m_{\pi ^+}^2+9 t (t+6 u)\right]~\nonumber \\&&
\left.-12 m_{\pi }^2 \left[3 \left(-6 m_{k^+}^2+6 m_{\pi ^+}^2+t\right) m_{\eta }^2+t \left(-3 m_{k^+}^2-3 m_{\pi ^+}^2+11 t\right)\right]\right\}m_k^2~\nonumber \\&&
+2 t^2 \left[-2 m_{k^+}^4+3 \left(-4 m_{\pi }^2-4 m_{\pi ^+}^2+9 t\right) m_{k^+}^2+14 m_{\pi ^+}^4+\left(24 m_{\pi }^2-36 t\right) m_{\pi ^+}^2\right.~\nonumber \\&&
+27 t (t-u)]-3 t m_{\eta }^2 \left[12 m_{k^+}^4+33 \left(t-m_{\pi ^+}^2\right) m_{k^+}^2+3 m_{\pi ^+}^4+15 t^2\right.~\nonumber \\&&
+9 s m_{\pi ^+}^2-26 t m_{\pi ^+}^2+9 u m_{\pi ^+}^2+15 s t-21 t u~\nonumber \\&&
\left.-3 m_{\pi }^2 \left(8 m_{k^+}^2-13 m_{\pi ^+}^2+9 t\right)\right]~\nonumber \\&&
\left.+18 m_{\eta }^4 \left[t \left(2 m_{k^+}^2-m_{\pi ^+}^2+s-2 u\right)-m_{\pi }^2 \left(6 m_{k^+}^2-6 m_{\pi ^+}^2+t\right)\right]\right\}J_{k \eta
}(t)~\nonumber \\&&
+\frac{1}{24 \sqrt{2} s^2 f_{\pi }^4}\left\{4 \left(m_{\pi }^2-m_{\pi ^+}^2\right) m_k^6+\left[-\left(12 m_{k^+}^2+s\right) m_{\pi }^2+3 s m_{\pi
^+}^2-2 s u\right.\right.~\nonumber \\&&
\left.+4 m_{k^+}^2 \left(3 m_{\pi ^+}^2+s\right)\right] m_k^4+\left[-4 \left(3 m_{\pi ^+}^2+s\right) m_{k^+}^4+2 s (u-t) m_{k^+}^2\right.~\nonumber \\&&
\left.-m_{\pi }^2 \left(s^2-12 m_{k^+}^4\right)+s^2 \left(3 m_{\pi ^+}^2-3 t+u\right)\right] m_k^2+s^3 (t-u)~\nonumber \\&&
+s m_{k^+}^4 \left(m_{\pi }^2-3 m_{\pi ^+}^2+2 t\right)+s^2 m_{k^+}^2 \left(m_{\pi }^2-3 m_{\pi ^+}^2-t+3 u\right)~\nonumber \\&&
\left.-4 m_{k^+}^6 \left(m_{\pi }^2-m_{\pi ^+}^2\right)\right\} J_{k k^+}(s)~\nonumber \\&&
+\frac{1}{24 \sqrt{2} u^2 f_{\pi }^4}\left\{-2 \left(-m_{\pi }^2+m_{k^+}^2+u\right) m_k^6+\left[-2 \left(3 m_{\pi ^+}^2+u\right) m_{\pi }^2\right.\right.~\nonumber \\&&
\left.+u \left(3 m_{\pi ^+}^2-s+5 u\right)+m_{k^+}^2 \left(6 m_{\pi ^+}^2+u\right)\right]m_k^4+\left[6 m_{\pi }^2 m_{\pi ^+}^4\right.~\nonumber \\&&
\left.+u \left(2 m_{\pi }^2+2 s-7 u\right) m_{\pi ^+}^2+u^2 (u-2 t)+2 m_{k^+}^2 \left(u^2-3 m_{\pi ^+}^4\right)\right] m_k^2~\nonumber \\&&
+\left(u-m_{\pi ^+}^2\right) \left[\left(u-2 m_{k^+}^2\right) m_{\pi ^+}^4+u \left(-m_{k^+}^2+s-3 u\right) m_{\pi ^+}^2\right.~\nonumber \\&&
\left.\left.+(t-4 u) u^2+m_{\pi }^2 \left(2 m_{\pi ^+}^4+2 u m_{\pi ^+}^2+u^2\right)\right]\right\} J_{k \pi ^+}(u)~\nonumber \\&&
+\frac{1}{288 \sqrt{2} u^2 f_{\pi }^4}\left\{-6 \left(2 m_{\pi ^+}^2+u\right) m_{\pi }^6+6 \left[\left(6 m_{\pi ^+}^2+4 u\right) m_{k^+}^2\right.\right.~\nonumber \\&&
\left.+u \left(m_{\pi ^+}^2-s+6 u\right)\right]m_{\pi }^4+\left\{3 \left(u-12 m_{\pi ^+}^2\right) m_{k^+}^4\right.~\nonumber \\&&
+u \left(51 m_{\pi ^+}^2-12 t-23 u\right) m_{k^+}^2+3 u \left[4 m_{\pi ^+}^4+3 u m_{\pi ^+}^2\right.~\nonumber \\&&
-4 u (t+2 u)]\} m_{\pi }^2-3 u m_{k^+}^6+18 u^4+36 u^2 m_{k^+}^4~\nonumber \\&&
-3 s u m_{k^+}^4+3 t u m_{k^+}^4+9 u^2 m_{\pi ^+}^4-9 u m_{k^+}^2 m_{\pi ^+}^4-12 s u^3~\nonumber \\&&
-6 t u^3-83 u^3 m_{k^+}^2+7 s u^2 m_{k^+}^2-5 t u^2 m_{k^+}^2+12 m_{k^+}^6 m_{\pi ^+}^2~\nonumber \\&&
-42 u m_{k^+}^4 m_{\pi ^+}^2-33 u^3 m_{\pi ^+}^2+3 s u^2 m_{\pi ^+}^2+3 t u^2 m_{\pi ^+}^2~\nonumber \\&&
+39 u^2 m_{k^+}^2 m_{\pi ^+}^2-3 s u m_{k^+}^2 m_{\pi ^+}^2-3 t u m_{k^+}^2 m_{\pi ^+}^2~\nonumber \\&&
-4 u m_k^4 \left(-3 m_{\pi }^2+3 m_{k^+}^2+5 u\right)+m_k^2 \left[12 m_{\pi }^6-12 \left(3 m_{k^+}^2+2 u\right) m_{\pi }^4\right.~\nonumber \\&&
+4 \left(9 m_{k^+}^4+3 u m_{k^+}^2+4 u^2-6 u m_{\pi ^+}^2\right) m_{\pi }^2-12 m_{k^+}^6+21 u m_{k^+}^4~\nonumber \\&&
\left.\left.+5 u^2 \left(m_{\pi ^+}^2+6 u\right)+3 u m_{k^+}^2 \left(9 m_{\pi ^+}^2+11 u\right)\right]\right\} J_{\pi  k^+}(u)~\nonumber \\&&
+\frac{1}{36 \sqrt{2} s^2 f_{\pi }^4}\left\{-12 m_{k^+}^2 m_{\pi }^6+3 \left[\left(12 m_{\pi ^+}^2+5 s\right) m_{k^+}^2+s (t-u)\right]m_{\pi
}^4\right.~\nonumber \\&&
-2 \left[\left(18 m_{\pi ^+}^4+7 s^2\right) m_{k^+}^2+3 s (t-u) \left(m_{\pi ^+}^2+s\right)\right] m_{\pi }^2+12 m_{k^+}^2 m_{\pi ^+}^6~\nonumber \\&&
-15 s m_{k^+}^2 m_{\pi ^+}^4+3 s t m_{\pi ^+}^4-3 s u m_{\pi ^+}^4+3 s^3 m_{k^+}^2+8 s^2 m_{k^+}^2 m_{\pi ^+}^2~\nonumber \\&&
-6 s^2 t m_{\pi ^+}^2+6 s^2 u m_{\pi ^+}^2+3 s^3 t-3 s^3 u~\nonumber \\&&
+m_k^2 \left[12 m_{\pi }^6-3 \left(12 m_{\pi ^+}^2+5 s\right) m_{\pi }^4+2 \left(18 m_{\pi ^+}^4+7 s^2\right) m_{\pi }^2\right.~\nonumber \\&&
\left.\left.-12 m_{\pi ^+}^6+15 s m_{\pi ^+}^4-3 s^3-8 s^2 m_{\pi ^+}^2\right]\right\} J_{\pi  \pi ^+}(s)~\nonumber \\&&
-\frac{1}{864 \sqrt{2} u^2 f_{\pi }^4}\left\{-108 m_{\pi ^+}^2 m_{k^+}^6-9 u m_{k^+}^6+84 u^2 m_{k^+}^4+9 u m_{\pi }^2 m_{k^+}^4\right.~\nonumber \\&&
+108 m_{\pi }^2 m_{\pi ^+}^2 m_{k^+}^4+18 u m_{\pi ^+}^2 m_{k^+}^4+27 s u m_{k^+}^4-27 t u m_{k^+}^4~\nonumber \\&&
+9 u m_{\pi ^+}^4 m_{k^+}^2-33 u^3 m_{k^+}^2-51 s u^2 m_{k^+}^2+57 t u^2 m_{k^+}^2~\nonumber \\&&
-81 u^2 m_{\pi }^2 m_{k^+}^2-59 u^2 m_{\pi ^+}^2 m_{k^+}^2+9 u m_{\pi }^2 m_{\pi ^+}^2 m_{k^+}^2~\nonumber \\&&
+27 s u m_{\pi ^+}^2 m_{k^+}^2+27 t u m_{\pi ^+}^2 m_{k^+}^2+54 u^4+7 u^2 m_{\pi ^+}^4~\nonumber \\&&
-54 t u^3-51 u^3 m_{\pi ^+}^2+21 s u^2 m_{\pi ^+}^2+21 t u^2 m_{\pi ^+}^2+27 u^2 m_{\pi }^2 m_{\pi ^+}^2~\nonumber \\&&
+u m_k^4 \left(-36 m_{\eta }^2+63 m_{k^+}^2-7 u\right)+18 m_{\eta }^4 \left[-\left(u-6 m_{\pi ^+}^2\right) m_{\pi }^2\right.~\nonumber \\&&
\left.+2 m_{k^+}^2 \left(u-3 m_{\pi ^+}^2\right)+u \left(-m_{\pi ^+}^2+s-2 t\right)\right]~\nonumber \\&&
+m_k^2 \left\{108 m_{k^+}^6-90 u m_{k^+}^4+77 u^2 m_{k^+}^2-72 u m_{\pi ^+}^2 m_{k^+}^2\right.~\nonumber \\&&
-27 s u m_{k^+}^2-27 t u m_{k^+}^2+57 u^3+3 s u^2+3 t u^2~\nonumber \\&&
-48 u^2 m_{\pi ^+}^2+36 m_{\eta }^4 \left(3 m_{k^+}^2+u\right)-9 m_{\pi }^2 \left[12 m_{\eta }^4\right.~\nonumber \\&&
\left.-8 \left(3 m_{k^+}^2+u\right) m_{\eta }^2+12 m_{k^+}^4+3 u^2-7 u m_{k^+}^2\right]~\nonumber \\&&
\left.-72 m_{\eta }^2 \left(3 m_{k^+}^4+u m_{k^+}^2+2 u^2-u m_{\pi ^+}^2\right)\right\}~\nonumber \\&&
+12 m_{\eta }^2 \left\{-6 \left(u-3 m_{\pi ^+}^2\right) m_{k^+}^4+u \left(3 m_{\pi ^+}^2+9 t-5 u\right) m_{k^+}^2\right.~\nonumber \\&&
+u \left(-3 m_{\pi ^+}^4+5 u m_{\pi ^+}^2+9 t u\right)~\nonumber \\&&
\left.\left.+3 m_{\pi }^2 \left[u \left(u-4 m_{\pi ^+}^2\right)-m_{k^+}^2 \left(6 m_{\pi ^+}^2+u\right)\right]\right\}\right\}J_{\eta  k^+}(u)~\nonumber \\&&
-\frac{\left(m_k^2-m_{k^+}^2\right) \left(4 m_k^2+3 m_{\eta }^2+4 m_{k^+}^2+m_{\pi ^+}^2-9 s\right) J_{\eta \pi ^+}(s)}{54 \sqrt{2} f_{\pi
}^4}~\nonumber \\&&
+\frac{1}{24 \sqrt{2} t^2 f_{\pi }^4}\left\{2 \left(m_k^2+t\right) m_{k^+}^6+\left[t \left(-3 m_{\pi ^+}^2+s-5 t\right)\right.\right.~\nonumber \\&&
\left.-m_k^2 \left(6 m_{\pi ^+}^2+t\right)\right] m_{k^+}^4-\left\{2 \left(t^2-3 m_{\pi ^+}^4\right) m_k^2\right.~\nonumber \\&&
\left.+t \left[(2 s-7 t) m_{\pi ^+}^2+t (t-2 u)\right]\right\} m_{k^+}^2~\nonumber \\&&
+\left(t-m_{\pi ^+}^2\right) \left[-\left(t-2 m_k^2\right) m_{\pi ^+}^4+t \left(m_k^2-s+3 t\right) m_{\pi ^+}^2\right.~\nonumber \\&&
\left.+t^2 (4 t-u)\right]-m_{\pi }^2 \left[2 m_{k^+}^6-2 \left(3 m_{\pi ^+}^2+t\right) m_{k^+}^4\right.~\nonumber \\&&
\left.\left.+2 m_{\pi ^+}^2 \left(3 m_{\pi ^+}^2+t\right) m_{k^+}^2-2 m_{\pi ^+}^6+t^3+t^2 m_{\pi ^+}^2\right]\right\} J_{k^+ \pi ^+}(t) \ ;
\end{eqnarray}

\begin{eqnarray}
T_{K^{0}\pi^{+}\rightarrow K^{0}\pi^{+}}^{H}(s,t,u)&=&
\frac{m_k^2+m_{\pi ^+}^2-u}{2 f_{\pi }^2}-\frac{1}{768 \pi ^2 s f_{\pi }^4}\left\{2 m_k^6+\left(m_{\pi }^2+3 m_{\eta }^2+4 m_{k^+}^2-2
m_{\pi ^+}^2+2 s\right) m_k^4\right.~\nonumber \\&&
-2 \left[m_{\pi ^+}^4+\left(m_{\pi }^2+3 m_{\eta }^2+4 m_{k^+}^2-6 s\right) m_{\pi ^+}^2+s (4 u-3 s)\right] m_k^2+2 m_{\pi ^+}^6~\nonumber \\&&
+3 m_{\eta }^2 m_{\pi ^+}^4+4 m_{k^+}^2 m_{\pi ^+}^4+2 s m_{\pi ^+}^4-2 s u^2+3 s t m_{\eta }^2-3 s u m_{\eta }^2~\nonumber \\&&
-4 s^2 m_{k^+}^2+4 s t m_{k^+}^2+6 s^2 m_{\pi ^+}^2-8 s u m_{\pi ^+}^2-4 s^2 t~\nonumber \\&&
\left.+m_{\pi }^2 \left[m_{\pi ^+}^4+s (t-u)\right]\right\}+\frac{8 L_1 \left(t-2 m_k^2\right) \left(t-2 m_{\pi ^+}^2\right)}{f_{\pi }^4}~\nonumber \\&&
+\frac{4 L_2 \left(\left(m_k^2+m_{\pi ^+}^2-s\right){}^2+\left(m_k^2+m_{\pi ^+}^2-u\right){}^2\right)}{f_{\pi }^4}+\frac{32 L_6 m_k^2 m_{\pi
^+}^2}{f_{\pi }^4}~\nonumber \\&&
+\frac{2 L_3 \left(-3 m_k^4+2 \left(u-m_{\pi ^+}^2\right) m_k^2-3 m_{\pi ^+}^4+s^2+t^2+2 u m_{\pi ^+}^2\right)}{f_{\pi }^4}~\nonumber \\&&
+\frac{4 L_5 \left(-m_k^4+\left(2 m_{\pi }^2-3 m_{\pi ^+}^2+u\right) m_k^2+2 m_{\pi }^2 \left(m_{\pi ^+}^2-u\right)\right)}{f_{\pi }^4}+\frac{16
L_8 m_k^2 m_{\pi ^+}^2}{f_{\pi }^4}~\nonumber \\&&
+\frac{4 L_4}{f_{\pi }^4} \left[-m_k^4+\left(m_{\pi }^2-m_{k^+}^2-10 m_{\pi ^+}^2+s+3 t\right) m_k^2-m_{\pi ^+}^4\right.~\nonumber \\&&
\left.+u m_{k^+}^2-m_{k^+}^2 m_{\pi ^+}^2+2 t m_{\pi ^+}^2+u m_{\pi ^+}^2+m_{\pi }^2 \left(m_{\pi ^+}^2-u\right)\right]~\nonumber \\&&
-\frac{1}{720 s^2 f_{\pi }^4}\{60 \left(-m_{\pi }^2+m_{k^+}^2+3 s\right) m_k^4+2 \left[12 \left(5 m_{\pi ^+}^2+17 s\right) m_{\pi }^2\right.~\nonumber \\&&
\left.+s \left(92 s-15 m_{\pi ^+}^2\right)+15 m_{k^+}^2 \left(s-4 m_{\pi ^+}^2\right)\right] m_k^2+60 m_{k^+}^2 m_{\pi ^+}^4~\nonumber \\&&
+30 s m_{\pi ^+}^4+696 s^3+89 s^2 m_{k^+}^2+15 s t m_{k^+}^2-15 s u m_{k^+}^2~\nonumber \\&&
+519 s^2 m_{\pi ^+}^2-30 s m_{k^+}^2 m_{\pi ^+}^2-45 s t m_{\pi ^+}^2-45 s u m_{\pi ^+}^2~\nonumber \\&&
\left.+726 s^2 t-1224 s^2 u-6 m_{\pi }^2 \left(10 m_{\pi ^+}^4-38 s m_{\pi ^+}^2+29 s t+24 s u\right)\right\} \mu _{\pi }~\nonumber \\&&
+\frac{1}{12 s^2 f_{\pi }^4}\left\{-2 m_k^6+6 m_{\pi ^+}^2 m_k^4+\left[-6 m_{\pi ^+}^4+2 s m_{\pi ^+}^2+s (2 s-t)\right] m_k^2\right.~\nonumber \\&&
\left.+2 m_{\pi ^+}^6-2 s m_{\pi ^+}^4+s t m_{\pi ^+}^2+3 s^2 t\right\} \mu _{\pi ^+}~\nonumber \\&&
+\frac{1}{12 s^2 f_{\pi }^4}\left\{2 m_k^6-2 \left(3 m_{\pi ^+}^2+s\right) m_k^4+\left[6 m_{\pi ^+}^4+2 s m_{\pi ^+}^2+s (t-8 s)\right] m_k^2\right.~\nonumber \\&&
\left.-2 m_{\pi ^+}^6-s (6 s+t) m_{\pi ^+}^2+s^2 (3 t+8 u)\right\} \mu _k~\nonumber \\&&
-\frac{1}{720 s^2 f_{\pi }^4}\left\{60 \left(m_{\pi }^2+3 m_{\eta }^2-4 m_{k^+}^2\right) m_k^4-2 \left[30 \left(2 m_{\pi ^+}^2+s\right) m_{\pi
}^2\right.\right.~\nonumber \\&&
\left.-45 m_{\eta }^2 \left(s-4 m_{\pi ^+}^2\right)+4 \left(47 s^2-60 m_{k^+}^2 m_{\pi ^+}^2\right)\right] m_k^2~\nonumber \\&&
+180 m_{\eta }^2 m_{\pi ^+}^4-240 m_{k^+}^2 m_{\pi ^+}^4-192 s^3+15 s^2 m_{\eta }^2+45 s t m_{\eta }^2~\nonumber \\&&
-45 s u m_{\eta }^2+20 s^2 m_{k^+}^2-60 s t m_{k^+}^2+60 s u m_{k^+}^2-376 s^2 m_{\pi ^+}^2~\nonumber \\&&
\left.-90 s m_{\eta }^2 m_{\pi ^+}^2+48 s^2 t+408 s^2 u+30 m_{\pi }^2 \left[2 m_{\pi ^+}^4+4 s m_{\pi ^+}^2+s (s-u)\right]\right\} \mu _{k^+}~\nonumber \\&&
-\frac{1}{720 s^2 f_{\pi }^4}\{60 \left(-3 m_{\eta }^2+3 m_{k^+}^2+s\right) m_k^4-6 \left[15 \left(s-4 m_{\pi ^+}^2\right) m_{\eta }^2\right.~\nonumber \\&&
\left.+5 m_{k^+}^2 \left(12 m_{\pi ^+}^2+s\right)+3 s \left(25 m_{\pi ^+}^2+12 s\right)\right] m_k^2+180 m_{k^+}^2 m_{\pi ^+}^4~\nonumber \\&&
+210 s m_{\pi ^+}^4-216 s^3+35 s^2 m_{k^+}^2+45 s t m_{k^+}^2-45 s u m_{k^+}^2-551 s^2 m_{\pi ^+}^2~\nonumber \\&&
+30 s m_{k^+}^2 m_{\pi ^+}^2+45 s t m_{\pi ^+}^2+45 s u m_{\pi ^+}^2-126 s^2 t+504 s^2 u~\nonumber \\&&
\left.+3 m_{\eta }^2\left\{-60 m_{\pi ^+}^4+30 s m_{\pi ^+}^2+s [7 s+15 (u-t)]\right\}\right\}\mu _{\eta }~\nonumber \\&&
+\frac{\left(2 m_{\pi }^2+m_{\pi ^+}^2-3 t\right) \left(2 m_k^2-2 m_{\pi }^2-2 m_{k^+}^2+2 m_{\pi ^+}^2+3 t\right) J_{\pi  \pi }(t)}{72 f_{\pi
}^4}~\nonumber \\&&
+\frac{\left[-t m_k^2+t (u-t)+m_{\pi ^+}^2 \left(4 m_k^2+4 m_{\pi ^+}^2-3 t-4 u\right)\right]J_{\pi ^+ \pi ^+}(t)}{12 f_{\pi }^4}~\nonumber \\&&
-\frac{\left[-4 m_k^4+\left(-4 m_{\pi ^+}^2+3 t+4 u\right) m_k^2+t \left(m_{\pi ^+}^2+t-u\right)\right] J_{k k}(t)}{12 f_{\pi }^4}~\nonumber \\&&
-\frac{\left[t m_k^2+2 (s-u) m_{k^+}^2+t \left(m_{\pi ^+}^2-s+t\right)\right] J_{k^+ k^+}(t)}{24 f_{\pi }^4}~\nonumber \\&&
+\frac{m_{\pi ^+}^2\left(2 m_k^2+6 m_{\eta }^2-2 m_{k^+}^2+2 m_{\pi ^+}^2-9 t\right) J_{\eta  \eta }(t) }{72 f_{\pi }^4}~\nonumber \\&&
+\frac{\left(m_k^2-m_{k^+}^2\right) \left(6 m_k^2+3 m_{\pi }^2+3 m_{\eta }^2+2 m_{k^+}^2-2 m_{\pi ^+}^2-9 t\right) J_{\pi  \eta }(t)}{108 f_{\pi
}^4}~\nonumber \\&&
-\frac{1}{24 s^2 f_{\pi }^4}\left\{2 m_k^8-2 \left(4 m_{\pi ^+}^2+s\right) m_k^6+\left(12 m_{\pi ^+}^4+2 s m_{\pi ^+}^2+s t\right) m_k^4\right.~\nonumber \\&&
-2\left[4 m_{\pi ^+}^6-s m_{\pi ^+}^4+s (t-3 s) m_{\pi ^+}^2+s^2 t\right] m_k^2+\left(s-m_{\pi ^+}^2\right) \left[-2 m_{\pi ^+}^6\right.~\nonumber \\&&
\left.\left.+s (2 s-t) m_{\pi ^+}^2+s^2 (s-u)\right]\right\} J_{k \pi ^+}(s)-\frac{\left(m_k^2+m_{\pi ^+}^2-u\right){}^2 J_{k \pi ^+}(u)}{4
f_{\pi }^4}~\nonumber \\&&
-\frac{1}{288 s^2 f_{\pi }^4}\left\{54 s^4-15 m_{k^+}^2 s^3+9 m_{\pi ^+}^2 s^3-24 t s^3-30 u s^3+m_{k^+}^4 s^2\right.~\nonumber \\&&
-6 m_{\pi ^+}^4 s^2-3 t m_{k^+}^2 s^2+9 u m_{k^+}^2 s^2+45 m_{k^+}^2 m_{\pi ^+}^2 s^2+9 t m_{\pi ^+}^2 s^2~\nonumber \\&&
+9 u m_{\pi ^+}^2 s^2+3 t m_{k^+}^4 s-3 u m_{k^+}^4 s-6 m_{k^+}^2 m_{\pi ^+}^4 s-6 m_{k^+}^4 m_{\pi ^+}^2 s~\nonumber \\&&
-9 t m_{k^+}^2 m_{\pi ^+}^2 s-9 u m_{k^+}^2 m_{\pi ^+}^2 s+12 m_{k^+}^4 m_{\pi ^+}^4+4 m_k^4 \left[3 m_{\pi }^4\right.~\nonumber \\&&
\left.-3 \left(2 m_{k^+}^2+3 s\right) m_{\pi }^2+3 m_{k^+}^4+7 s^2+6 s \right]+6 m_{\pi }^4 \left[2 m_{\pi ^+}^4\right.~\nonumber \\&&
\left.+4 s m_{\pi ^+}^2+s (s-u)\right]+12 m_{\pi }^2 \left[s \left(m_{\pi ^+}^4-12 s m_{\pi ^+}^2+4 s t+5 s u\right)\right.~\nonumber \\&&
\left.-m_{k^+}^2 \left(2 m_{\pi ^+}^4+s t\right)\right]-2 m_k^2 \left\{6 \left(2 m_{\pi ^+}^2+s\right) m_{\pi }^4+12 \left[s \left(s-m_{\pi
^+}^2\right)\right.\right.~\nonumber \\&&
\left.-m_{k^+}^2 \left(2 m_{\pi ^+}^2+s\right)\right] m_{\pi }^2+s m_{k^+}^2 \left(25 s-9 m_{\pi ^+}^2\right)~\nonumber \\&&
\left.\left.-3 m_{k^+}^4 \left(s-4 m_{\pi ^+}^2\right)+3 s^2 \left(6 s-m_{\pi ^+}^2\right)\right\}\right\} J_{\pi  k^+}(s)~\nonumber \\&&
-\frac{1}{864 s^2 f_{\pi }^4}\left\{54 s^4-15 m_{k^+}^2 s^3-51 m_{\pi ^+}^2 s^3-54 u s^3+m_{k^+}^4 s^2\right.~\nonumber \\&&
+34 m_{\pi ^+}^4 s^2-51 t m_{k^+}^2 s^2+57 u m_{k^+}^2 s^2-127 m_{k^+}^2 m_{\pi ^+}^2 s^2~\nonumber \\&&
+21 t m_{\pi ^+}^2 s^2+21 u m_{\pi ^+}^2 s^2+27 t m_{k^+}^4 s-27 u m_{k^+}^4 s~\nonumber \\&&
+18 m_{k^+}^2 m_{\pi ^+}^4 s+18 m_{k^+}^4 m_{\pi ^+}^2 s+27 t m_{k^+}^2 m_{\pi ^+}^2 s~\nonumber \\&&
+27 u m_{k^+}^2 m_{\pi ^+}^2 s+108 m_{k^+}^4 m_{\pi ^+}^4+4 m_k^4 \left[27 m_{\eta }^4\right.~\nonumber \\&&
\left.-9 \left(6 m_{k^+}^2+s\right) m_{\eta }^2+27 m_{k^+}^4+7 s^2-18 s m_{k^+}^2\right]~\nonumber \\&&
+18 m_{\eta }^4 \left[6 m_{\pi ^+}^4-2 s m_{\pi ^+}^2+s (t-2 u)\right]~\nonumber \\&&
+12 m_{\eta }^2 \left[\left(-18 m_{\pi ^+}^4+4 s m_{\pi ^+}^2-5 s t+4 s u\right) m_{k^+}^2\right.~\nonumber \\&&
\left.+s \left(-15 m_{\pi ^+}^4+8 s m_{\pi ^+}^2+9 s u\right)\right]~\nonumber \\&&
+2 m_k^2 \left\{36 \left(s-3 m_{\pi ^+}^2\right) m_{\eta }^4-12\left[\left(s-18 m_{\pi ^+}^2\right) m_{k^+}^2\right.\right.~\nonumber \\&&
\left.+s \left(11 s-9 m_{\pi ^+}^2\right)\right]m_{\eta }^2+s^2 \left(18 s-41 m_{\pi ^+}^2\right)~\nonumber \\&&
\left.\left.+s m_{k^+}^2 \left(79 s-27 m_{\pi ^+}^2\right)-9 m_{k^+}^4 \left(12 m_{\pi ^+}^2+s\right)\right\}\right\} J_{\eta  k^+}(s) \ ;
\end{eqnarray}

\begin{eqnarray}
T_{K^{+}\pi^{0}\rightarrow K^{+}\pi^{0}}^{H}(s,t,u)&=&
\frac{-2 m_k^2-2 m_{\pi }^2+2 m_{k^+}^2+2 m_{\pi ^+}^2+3 t}{12 f_{\pi }^2}+\frac{1}{1536 \pi ^2 s u f_{\pi }^4}\left\{-\left(2 m_{\pi
}^2+2 m_{k^+}^2-t\right) m_{\pi }^6\right.~\nonumber \\&&
+\left[4 s u+\left(-2 m_{\pi }^2-2 m_{k^+}^2+t\right) \left(3 m_{\eta }^2+2 m_{k^+}^2+2m_{\pi ^+}^2\right)\right] m_{\pi }^4~\nonumber \\&&
+\left\{\left[\left(2 m_{\pi }^2+2 m_{k^+}^2-t\right) \left(6 m_{\eta }^2+7 m_{k^+}^2+4 m_{\pi ^+}^2\right)-8 s u\right]m_{k^+}^2\right.~\nonumber \\&&
\left.+s u \left(2 m_{\pi }^2+2 m_{k^+}^2-3 t\right)\right\} m_{\pi }^2-4 s m_{k^+}^6-4 u m_{k^+}^6~\nonumber \\&&
-3 s m_{\eta }^2 m_{k^+}^4-3 u m_{\eta }^2 m_{k^+}^4+4 s u m_{k^+}^4+3 s u^2 m_{\eta }^2+3 s^2 u m_{\eta }^2~\nonumber \\&&
-6 s t u m_{\eta }^2+4 s u^2 m_{k^+}^2+4 s^2 u m_{k^+}^2-8 s t u m_{k^+}^2~\nonumber \\&&
+2 \left[s u \left(2 m_{\pi }^2+2 m_{k^+}^2-3 t\right)-m_{k^+}^4 \left(2 m_{\pi }^2+2 m_{k^+}^2-t\right)\right] m_{\pi ^+}^2~\nonumber \\&&
+2 s u [s (t-2 u)+t u]+2 m_k^2 \left[s u \left(2 m_{\pi }^2+2 m_{k^+}^2-3 t\right)\right.~\nonumber \\&&
\left.\left.-\left(m_{\pi }^2-m_{k^+}^2\right){}^2 \left(2 m_{\pi }^2+2 m_{k^+}^2-t\right)\right]\right\}+\frac{32 L_6 m_{k^+}^2 m_{\pi ^+}^2}{f_{\pi
}^4}~\nonumber \\&&
+\frac{8 L_1 \left(t-2 m_{\pi }^2\right) \left(t-2 m_{k^+}^2\right)}{f_{\pi }^4}+\frac{4 L_2 \left(\left(m_{\pi }^2+m_{k^+}^2-s\right){}^2+\left(m_{\pi
}^2+m_{k^+}^2-u\right){}^2\right)}{f_{\pi }^4}~\nonumber \\&&
+\frac{L_3 \left(-2 m_{\pi }^4-2 \left(t-2 m_{k^+}^2\right) m_{\pi }^2-2 m_{k^+}^4+s^2+2 t^2+u^2-2 t m_{k^+}^2\right)}{f_{\pi }^4}~\nonumber \\&&
+\frac{1}{3 f_{\pi }^4}2 L_4\left[-2 m_k^4+\left(-4 m_{\pi }^2+4 m_{k^+}^2+4 m_{\pi ^+}^2+3 t\right) m_k^2-2 m_{\pi }^4-2 m_{k^+}^4\right.~\nonumber \\&&
-2 m_{\pi ^+}^4+9 t m_{k^+}^2-28 m_{k^+}^2 m_{\pi ^+}^2+9 t m_{\pi ^+}^2~\nonumber \\&&
\left.+m_{\pi }^2 \left(-20 m_{k^+}^2+4 m_{\pi ^+}^2+3 t\right)\right]+\frac{32 L_7 \left(m_k^2-m_{k^+}^2\right) \left(m_k^2-3 m_{k^+}^2-m_{\pi
^+}^2\right)}{3 f_{\pi }^4}~\nonumber \\&&
+\frac{16 L_8 \left(m_k^4-\left(4 m_{k^+}^2+m_{\pi ^+}^2\right) m_k^2+3 m_{k^+}^4+4 m_{k^+}^2 m_{\pi ^+}^2\right)}{3 f_{\pi }^4}~\nonumber \\&&
-\frac{2 L_5 }{3 f_{\pi }^4}\left[4 m_{\pi }^4+\left(2 m_{k^+}^2-4 m_{\pi ^+}^2-6 t\right) m_{\pi }^2\right.~\nonumber \\&&
\left.+6 m_k^2 \left(t-m_{k^+}^2\right)+m_{k^+}^2 \left(6 m_{k^+}^2+10 m_{\pi ^+}^2-3 t\right)\right]~\nonumber \\&&
+\frac{1}{720 s^2 u^2 f_{\pi }^4}\left\{30 (s-u)^2 m_{\pi }^6-30 \left[\left(3 s^2+8 u s+3 u^2\right) m_{k^+}^2\right.\right.~\nonumber \\&&
\left.-s u \left(t-4 m_{\pi ^+}^2\right)\right] m_{\pi }^4+\left[90 (s-u)^2 m_{k^+}^4\right.~\nonumber \\&&
+30 s u \left(3 t-4 m_{\pi ^+}^2\right) m_{k^+}^2+s u \left(-15 s^2+2 u s\right.~\nonumber \\&&
\left.\left.-15 u^2+60 t m_{\pi ^+}^2\right)\right] m_{\pi }^2-30 s^2 m_{k^+}^6-30 u^2 m_{k^+}^6~\nonumber \\&&
+120 s u^2 m_{k^+}^4+120 s^2 u m_{k^+}^4+402 s^2 u^3+402 s^3 u^2~\nonumber \\&&
-768 s^2 t u^2+15 s u^3 m_{k^+}^2-1670 s^2 u^2 m_{k^+}^2+15 s^3 u m_{k^+}^2~\nonumber \\&&
-896 s^2 u^2 m_{\pi ^+}^2+60 s u^2 m_{k^+}^2 m_{\pi ^+}^2+60 s^2 u m_{k^+}^2 m_{\pi ^+}^2~\nonumber \\&&
\left.+4 s u m_k^2 \left(30 m_{\pi }^4-15 t m_{\pi }^2-30 m_{k^+}^4+15 t m_{k^+}^2+194 s u\right)\right\}\mu _{\pi }~\nonumber \\&&
+\frac{1}{720 s^2 u^2 f_{\pi }^4}\left\{-60 s u m_{\pi }^6+30 \left[-2 s u m_{k^+}^2+2 \left(s^2+3 u s+u^2\right) m_{\pi ^+}^2\right.\right.~\nonumber \\&&
+s u (3 s+t+3 u)]m_{\pi }^4-2 \left\{s u \left(45 t m_{\pi ^+}^2+152 s u\right)\right.~\nonumber \\&&
\left.-15 m_{k^+}^2 \left[\left(-4 s^2+6 u s-4 u^2\right) m_{\pi ^+}^2+3 s u (s+u)\right]\right\}m_{\pi }^2~\nonumber \\&&
+30 s u^2 m_{k^+}^4+30 s^2 u m_{k^+}^4+48 s^2 u^3+48 s^3 u^2+408 s^2 t u^2~\nonumber \\&&
-45 s u^3 m_{k^+}^2+702 s^2 u^2 m_{k^+}^2-45 s t u^2 m_{k^+}^2-45 s^3 u m_{k^+}^2~\nonumber \\&&
-45 s^2 t u m_{k^+}^2+60 s^2 m_{k^+}^4 m_{\pi ^+}^2+60 u^2 m_{k^+}^4 m_{\pi ^+}^2-15 s u^3 m_{\pi ^+}^2~\nonumber \\&&
+58 s^2 u^2 m_{\pi ^+}^2+15 s t u^2 m_{\pi ^+}^2-90 s u^2 m_{k^+}^2 m_{\pi ^+}^2-90 s^2 u m_{k^+}^2 m_{\pi ^+}^2~\nonumber \\&&
-15 s^3 u m_{\pi ^+}^2+15 s^2 t u m_{\pi ^+}^2+3 m_k^2 \left\{-20 \left(s^2-5 u s+u^2\right) m_{\pi }^4\right.~\nonumber \\&&
+10 \left[4 \left(s^2+u^2\right) m_{k^+}^2-5 s t u\right] m_{\pi }^2-20 \left(s^2+5 u s+u^2\right) m_{k^+}^4~\nonumber \\&&
\left.\left.+50 s t u m_{k^+}^2+s u \left(5 s^2-5 t s-126 u s+5 u^2-5 t u\right)\right\}\right\} \mu _{\pi ^+}~\nonumber \\&&
+\frac{1}{720 s^2 u^2 f_{\pi }^4}\left\{-60 s u m_{\pi }^6-30 \left[-6 s u m_{k^+}^2+2 \left(s^2+3 u s+u^2\right) m_{\pi ^+}^2\right.\right.~\nonumber \\&&
+s u (3 s-t+3 u)] m_{\pi }^4+2 \left\{120 s u m_{k^+}^4\right.~\nonumber \\&&
-15 \left[\left(-4 s^2+6 u s-4 u^2\right) m_{\pi ^+}^2+s u (3 s+4 t+3 u)\right] m_{k^+}^2~\nonumber \\&&
\left.+s u \left(45 t m_{\pi ^+}^2+124 s u\right)\right\} m_{\pi }^2-90 s u^2 m_{k^+}^4-90 s^2 u m_{k^+}^4~\nonumber \\&&
+144 s^2 u^3+144 s^3 u^2-36 s^2 t u^2+45 s u^3 m_{k^+}^2-118 s^2 u^2 m_{k^+}^2~\nonumber \\&&
+45 s t u^2 m_{k^+}^2+45 s^3 u m_{k^+}^2+45 s^2 t u m_{k^+}^2-60 s^2 m_{k^+}^4 m_{\pi ^+}^2~\nonumber \\&&
-60 u^2 m_{k^+}^4 m_{\pi ^+}^2+15 s u^3 m_{\pi ^+}^2-106 s^2 u^2 m_{\pi ^+}^2-15 s t u^2 m_{\pi ^+}^2~\nonumber \\&&
+90 s u^2 m_{k^+}^2 m_{\pi ^+}^2+90 s^2 u m_{k^+}^2 m_{\pi ^+}^2+15 s^3 u m_{\pi ^+}^2-15 s^2 t u m_{\pi ^+}^2~\nonumber \\&&
-15 m_k^2 \left\{-4 \left(s^2-5 u s+u^2\right) m_{\pi }^4+2 \left[4 \left(s^2+u^2\right) m_{k^+}^2-5 s t u\right] m_{\pi }^2\right.~\nonumber \\&&
-4 \left(s^2+5 u s+u^2\right) m_{k^+}^4+10 s t u m_{k^+}^2~\nonumber \\&&
\left.\left.+s u \left(s^2-t s+10 u s+u^2-t u\right)\right\}\right\}\mu _k~\nonumber \\&&
-\frac{1}{480 s^2 u^2 f_{\pi }^4}\left\{20 \left(s^2+3 u s+u^2\right) m_{\pi }^6+10 \left[6 \left(s^2-u s+u^2\right) m_{\eta }^2\right.\right.~\nonumber \\&&
\left.-4 \left(3 s^2+8 u s+3 u^2\right) m_{k^+}^2-s u \left(16 m_{\pi ^+}^2+3 t\right)\right] m_{\pi }^4~\nonumber \\&&
-\left\{-20 \left(9 s^2-19 u s+9 u^2\right) m_{k^+}^4+10 s u \left(16 m_{\pi ^+}^2-19 t\right) m_{k^+}^2\right.~\nonumber \\&&
+30 m_{\eta }^2 \left[2 \left(2 s^2+u s+2 u^2\right) m_{k^+}^2-s t u\right]~\nonumber \\&&
\left.+s u \left(25 s^2+15 t s-118 u s+25 u^2-80 t m_{\pi ^+}^2+15 t u\right)\right\} m_{\pi }^2~\nonumber \\&&
-80 s^2 m_{k^+}^6-80 u^2 m_{k^+}^6+160 s u^2 m_{k^+}^4+60 s^2 m_{\eta }^2 m_{k^+}^4+60 u^2 m_{\eta }^2 m_{k^+}^4~\nonumber \\&&
+160 s^2 u m_{k^+}^4-16 s^2 u^3-16 s^3 u^2-136 s^2 t u^2-15 s u^3 m_{\eta }^2~\nonumber \\&&
+10 s^2 u^2 m_{\eta }^2+15 s t u^2 m_{\eta }^2-15 s^3 u m_{\eta }^2+15 s^2 t u m_{\eta }^2+40 s u^3 m_{k^+}^2~\nonumber \\&&
-64 s^2 u^2 m_{k^+}^2+30 s u^2 m_{\eta }^2 m_{k^+}^2+30 s^2 u m_{\eta }^2 m_{k^+}^2+40 s^3 u m_{k^+}^2~\nonumber \\&&
-16 s^2 u^2 m_{\pi ^+}^2+80 s u^2 m_{k^+}^2 m_{\pi ^+}^2+80 s^2 u m_{k^+}^2 m_{\pi ^+}^2~\nonumber \\&&
\left.+16 s u m_k^2 \left(10 m_{\pi }^4-5 t m_{\pi }^2-10 m_{k^+}^4+5 t m_{k^+}^2+s u\right)\right\} \mu _{k^+}~\nonumber \\&&
-\frac{1}{1440 s^2 u^2 f_{\pi }^4}\left\{180 s u m_{\pi }^6+30 \left\{-6 \left(s^2-u s+u^2\right) m_{\eta }^2\right.\right.~\nonumber \\&&
\left.+6 \left(s^2+u s+u^2\right) m_{k^+}^2+s u \left[8 m_{\pi ^+}^2+3 (s-t+u)\right]\right\}m_{\pi }^4~\nonumber \\&&
-6 \left\{4 \left[15 \left(s^2+u^2\right) m_{k^+}^4-10 s u m_{\pi ^+}^2 m_{k^+}^2+s u \left(5 t m_{\pi ^+}^2+6 s u\right)\right]\right.~\nonumber \\&&
\left.-15 m_{\eta }^2 \left[2 \left(2 s^2+u s+2 u^2\right) m_{k^+}^2-s t u\right]\right\}m_{\pi }^2+180 s^2 m_{k^+}^6~\nonumber \\&&
+180 u^2 m_{k^+}^6-180 s^2 m_{\eta }^2 m_{k^+}^4-180 u^2 m_{\eta }^2 m_{k^+}^4+108 s^2 u^3~\nonumber \\&&
+108 s^3 u^2-432 s^2 t u^2+45 s u^3 m_{\eta }^2+42 s^2 u^2 m_{\eta }^2~\nonumber \\&&
-45 s t u^2 m_{\eta }^2+45 s^3 u m_{\eta }^2-45 s^2 t u m_{\eta }^2-90 s u^3 m_{k^+}^2~\nonumber \\&&
-540 s^2 u^2 m_{k^+}^2-90 s u^2 m_{\eta }^2 m_{k^+}^2-90 s^2 u m_{\eta }^2 m_{k^+}^2~\nonumber \\&&
-90 s^3 u m_{k^+}^2-688 s^2 u^2 m_{\pi ^+}^2-120 s u^2 m_{k^+}^2 m_{\pi ^+}^2-120 s^2 u m_{k^+}^2 m_{\pi ^+}^2~\nonumber \\&&
\left.+8 s u m_k^2 \left(-30 m_{\pi }^4+15 t m_{\pi }^2+30 m_{k^+}^4-15 t m_{k^+}^2+56 s u\right)\right\}\mu _{\eta }~\nonumber \\&&
-\frac{ m_{\pi ^+}^2\left(-2 m_k^2-2 m_{\pi }^2+2 m_{k^+}^2+2 m_{\pi ^+}^2+3 t\right) J_{\pi  \pi }(t)}{24 f_{\pi }^4}~\nonumber \\&&
+\frac{t \left(2 m_{\pi }^2+m_{\pi ^+}^2-3 t\right) J_{\pi ^+ \pi ^+}(t)}{12 f_{\pi }^4}-\frac{t \left(2 m_k^2-2 m_{\pi }^2-2 m_{k^+}^2+2 m_{\pi
^+}^2+3 t\right) J_{k k}(t)}{48 f_{\pi }^4}~\nonumber \\&&
-\frac{t \left(-2 m_k^2-2 m_{\pi }^2+2 m_{k^+}^2+2 m_{\pi ^+}^2+3 t\right) J_{k^+ k^+}(t)}{24 f_{\pi }^4}~\nonumber \\&&
+\frac{ m_{\pi ^+}^2\left(-2 m_k^2+6 m_{\eta }^2+2 m_{k^+}^2+2 m_{\pi ^+}^2-9 t\right) J_{\eta  \eta }(t)}{72 f_{\pi }^4}~\nonumber \\&&
-\frac{\left(m_k^2-m_{k^+}^2\right) \left(2 m_k^2+3 m_{\pi }^2+3 m_{\eta }^2+6 m_{k^+}^2-2 m_{\pi ^+}^2-9 t\right) J_{\pi  \eta }(t)}{36 f_{\pi
}^4}~\nonumber \\&&
-\frac{1}{288 s^2 f_{\pi }^4}\left\{\left\{54 s^4+39 m_{k^+}^2 s^3-45 m_{\pi ^+}^2 s^3-24 t s^3-30 u s^3\right.\right.~\nonumber \\&&
-2 m_{k^+}^4 s^2+9 m_{\pi ^+}^4 s^2+3 t m_{k^+}^2 s^2+3 u m_{k^+}^2 s^2~\nonumber \\&&
+33 m_{k^+}^2 m_{\pi ^+}^2 s^2+3 t m_{\pi ^+}^2 s^2+15 u m_{\pi ^+}^2 s^2-18 m_{k^+}^2 m_{\pi ^+}^4 s~\nonumber \\&&
+3 t m_{\pi ^+}^4 s-3 u m_{\pi ^+}^4 s+6 m_{k^+}^4 m_{\pi ^+}^2 s-9 t m_{k^+}^2 m_{\pi ^+}^2 s~\nonumber \\&&
-9 u m_{k^+}^2 m_{\pi ^+}^2 s+12 m_{k^+}^4 m_{\pi ^+}^4+m_k^4 \left[12 m_{\pi }^4+4 \left(5 s-6 m_{k^+}^2\right) m_{\pi }^2\right.~\nonumber \\&&
\left.+12 m_{k^+}^4+80 s m_{k^+}^2-22 s t-28 s u\right]+12 m_{\pi }^4 \left(m_{\pi ^+}^4+s m_{\pi ^+}^2+s^2\right)~\nonumber \\&&
-6 m_{\pi }^2 \left[\left(4 m_{\pi ^+}^4-3 s m_{\pi ^+}^2+s^2\right) m_{k^+}^2+3 s m_{\pi ^+}^2 \left(3 s-m_{\pi ^+}^2\right)\right]~\nonumber \\&&
+4 m_k^2 \left\{-6 \left(m_{\pi ^+}^2+s\right) m_{\pi }^4+6 \left[\left(2 m_{\pi ^+}^2+s\right) m_{k^+}^2+s \left(m_{\pi ^+}^2-4 s\right)\right]
m_{\pi }^2\right.~\nonumber \\&&
\left.\left.-56 s^2 m_{k^+}^2-6 m_{k^+}^4 m_{\pi ^+}^2+3 s \left[(2 s-t) m_{\pi ^+}^2+s (7 t+8 u)\right]\right\}\right\}J_{k \pi ^+}(s)~\nonumber \\&&
+\left\{6 m_{\pi }^8-6 \left(4 m_{k^+}^2+s\right) m_{\pi }^6+\left[36 m_{k^+}^4-18 s m_{k^+}^2+s \left(-12 m_{\pi ^+}^2+5 s-3 u\right)\right]
m_{\pi }^4\right.~\nonumber \\&&
+2 \left[-12 m_{k^+}^6+27 s m_{k^+}^4+s \left(12 m_{\pi ^+}^2-7 s+3 u\right) m_{k^+}^2+s^2 \left(2 m_{\pi ^+}^2+3 u\right)\right] m_{\pi }^2~\nonumber \\&&
+6 m_{k^+}^8-30 s m_{k^+}^6+3 s^4+8 s^2 m_k^4+49 s^2 m_{k^+}^4-3 s u m_{k^+}^4+8 s^2 m_{\pi ^+}^4~\nonumber \\&&
-22 s^3 m_{k^+}^2+2 s^2 t m_{k^+}^2+8 s^2 u m_{k^+}^2-12 s m_{k^+}^4 m_{\pi ^+}^2-10 s^3 m_{\pi ^+}^2~\nonumber \\&&
+36 s^2 m_{k^+}^2 m_{\pi ^+}^2+2 s^2 t m_{\pi ^+}^2+2 s^2 u m_{\pi ^+}^2-3 s^3 u~\nonumber \\&&
\left.+4 s m_k^2 \left[3 m_{\pi }^4-2 \left(3 m_{k^+}^2+s\right) m_{\pi }^2+3 m_{k^+}^4+3 s^2-10 s m_{k^+}^2-4 s m_{\pi ^+}^2\right]\right\}
J_{\pi  k^+}(s)~\nonumber \\&&
+\frac{1}{6}\left\{108 m_{k^+}^8-216 m_{\eta }^2 m_{k^+}^6-108 s m_{k^+}^6+108 m_{\eta }^4 m_{k^+}^4+162 s^2 m_{k^+}^4\right.~\nonumber \\&&
-54 s m_{\eta }^2 m_{k^+}^4-72 s m_{\pi ^+}^2 m_{k^+}^4-54 s u m_{k^+}^4+54 s m_{\eta }^4 m_{k^+}^2-36 s^3 m_{k^+}^2~\nonumber \\&&
-117 s^2 m_{\eta }^2 m_{k^+}^2-27 s t m_{\eta }^2 m_{k^+}^2+81 s u m_{\eta }^2 m_{k^+}^2+24 s^2 m_{\pi ^+}^2 m_{k^+}^2~\nonumber \\&&
+72 s m_{\eta }^2 m_{\pi ^+}^2 m_{k^+}^2-36 s^2 t m_{k^+}^2+72 s^2 u m_{k^+}^2+54 s^4+16 s^2 m_k^4~\nonumber \\&&
+9 s^2 m_{\eta }^4+27 s t m_{\eta }^4-27 s u m_{\eta }^4+16 s^2 m_{\pi ^+}^4-45 s^3 m_{\eta }^2-45 s^2 t m_{\eta }^2~\nonumber \\&&
+63 s^2 u m_{\eta }^2-60 s^3 m_{\pi ^+}^2+24 s^2 m_{\eta }^2 m_{\pi ^+}^2+12 s^2 t m_{\pi ^+}^2+12 s^2 u m_{\pi ^+}^2~\nonumber \\&&
-54 s^3 u+36 m_{\pi }^4 \left[3 m_{\eta }^4-3 \left(2 m_{k^+}^2+s\right) m_{\eta }^2+3 m_{k^+}^4+s^2\right]~\nonumber \\&&
+8 s m_k^2 \left[9 m_{k^+}^4-6 s m_{k^+}^2+9 s^2-4 s m_{\pi ^+}^2-3 m_{\eta }^2 \left(3 m_{k^+}^2+s\right)\right.~\nonumber \\&&
\left.-3 m_{\pi }^2 \left(-3 m_{\eta }^2+3 m_{k^+}^2+s\right)\right]+18 m_{\pi }^2 \left\{-3 \left(4 m_{k^+}^2+s\right) m_{\eta }^4\right.~\nonumber \\&&
+\left[24 m_{k^+}^4+3 s m_{k^+}^2+s \left(9 s-4 m_{\pi ^+}^2\right)\right] m_{\eta }^2~\nonumber \\&&
\left.\left.\left.+4 m_{k^+}^2 \left[-3 m_{k^+}^4+3 s m_{k^+}^2+s \left(m_{\pi ^+}^2-3 s\right)\right]\right\}\right\}J_{\eta  k^+}(s)+[s\longleftrightarrow
u]\right\} \ ;
\end{eqnarray}

\begin{eqnarray}
T_{K^{0}\pi^{0}\rightarrow K^{0}\pi^{0}}^{H}(s,t,u)&=&
\frac{2 m_k^2-2 m_{\pi }^2-2 m_{k^+}^2+2 m_{\pi ^+}^2+3 t}{12 f_{\pi }^2}+\frac{1}{1536 \pi ^2 s u f_{\pi }^4}\left\{-4 \left(2 m_k^2+2
m_{\pi }^2-t\right) m_k^6\right.~\nonumber \\&&
+\left[4 s u+\left(2 m_k^2+2 m_{\pi }^2-t\right) \left(7 m_{\pi }^2-3 m_{\eta }^2-2 m_{k^+}^2-2m_{\pi ^+}^2\right)\right] m_k^4~\nonumber \\&&
+2 \left\{\left[\left(-2 m_k^2-2 m_{\pi }^2+t\right) \left(m_{\pi }^2-3 m_{\eta }^2-2m_{k^+}^2-2m_{\pi ^+}^2\right)-4 s u\right]m_{\pi }^2\right.~\nonumber \\&&
\left.+2 s u \left(2 m_k^2+2 m_{\pi }^2-3 t\right)\right\} m_k^2+s u m_{\pi }^2 \left(2 m_k^2+2 m_{\pi }^2-3 t\right)~\nonumber \\&&
-m_{\pi }^6 \left(2 m_k^2+2 m_{\pi }^2-t\right)+m_{\pi }^4 \left[4 s u+\left(-2 m_k^2-2 m_{\pi }^2+t\right) \left(3 m_{\eta }^2+2 m_{k^+}^2+2m_{\pi
^+}^2\right)\right]~\nonumber \\&&
\left.+s u \left[2 (s t-2 s u+t u)+\left(2 m_k^2+2 m_{\pi }^2-3 t\right) \left(3 m_{\eta }^2+2 m_{k^+}^2+2m_{\pi ^+}^2\right)\right]\right\}~\nonumber \\&&
+\frac{8 L_1 \left(t-2 m_k^2\right) \left(t-2 m_{\pi }^2\right)}{f_{\pi }^4}+\frac{4 L_2 \left[\left(m_k^2+m_{\pi }^2-s\right){}^2+\left(m_k^2+m_{\pi
}^2-u\right){}^2\right]}{f_{\pi }^4}~\nonumber \\&&
+\frac{L_3 \left[-2 m_k^4-2 \left(t-2 m_{\pi }^2\right) m_k^2-2 m_{\pi }^4+s^2+2 t^2+u^2-2 t m_{\pi }^2\right]}{f_{\pi }^4}~\nonumber \\&&
+\frac{2 L_4 }{3 f_{\pi }^4}\left[2 m_k^4+\left(-24 m_{\pi }^2-4 m_{k^+}^2-24 m_{\pi ^+}^2+15 t\right) m_k^2-2 m_{\pi }^4\right.~\nonumber \\&&
\left.+2 m_{k^+}^4-2 m_{\pi ^+}^4-3 t m_{k^+}^2+9 t m_{\pi ^+}^2+m_{\pi }^2 \left(4 m_{\pi ^+}^2+3 t\right)\right]~\nonumber \\&&
-\frac{2 L_5 }{3 f_{\pi }^4}\left[8 m_k^4-2 \left(5 m_{k^+}^2-6 m_{\pi ^+}^2\right) m_k^2+4 m_{\pi }^4\right.~\nonumber \\&&
\left.+m_{\pi }^2 \left(2 m_{k^+}^2-4 m_{\pi ^+}^2-6 t\right)+m_{k^+}^2 \left(2 m_{k^+}^2-2 m_{\pi ^+}^2+3 t\right)\right]~\nonumber \\&&
+\frac{32 L_6 m_k^2 m_{\pi ^+}^2}{f_{\pi }^4}+\frac{32 L_7 \left(m_k^2-m_{k^+}^2\right) \left(3 m_k^2-m_{k^+}^2+m_{\pi ^+}^2\right)}{3 f_{\pi
}^4}~\nonumber \\&&
+\frac{16 L_8 \left[3 m_k^4+4 \left(m_{\pi ^+}^2-m_{k^+}^2\right) m_k^2+m_{k^+}^4-m_{k^+}^2 m_{\pi ^+}^2\right]}{3 f_{\pi }^4}~\nonumber \\&&
+\frac{1}{720 s^2 u^2 f_{\pi }^4}\left\{-30 \left(s^2-8 u s+u^2\right) m_k^6+30 \left[\left(3 s^2+2 u s+3 u^2\right) m_{\pi }^2\right.\right.~\nonumber \\&&
\left.-4 s u \left(m_{k^+}^2-m_{\pi ^+}^2+t\right)\right] m_k^4+\left[-30 \left(3 s^2+8 u s+3 u^2\right) m_{\pi }^4\right.~\nonumber \\&&
\left.+90 s t u m_{\pi }^2+s u \left(15 s^2-866 u s+15 u^2+60 t m_{k^+}^2-60 t m_{\pi ^+}^2\right)\right] m_k^2~\nonumber \\&&
+30 (s-u)^2 m_{\pi }^6+30 s u m_{\pi }^4 \left(4 m_{k^+}^2-4 m_{\pi ^+}^2+t\right)~\nonumber \\&&
-2 s^2 u^2 \left(-388 m_{k^+}^2+448 m_{\pi ^+}^2+585 t\right)~\nonumber \\&&
\left.+s u m_{\pi }^2 \left(-15 s^2+806 u s-15 u^2-60 t m_{k^+}^2+60 t m_{\pi ^+}^2\right)\right\} \mu _{\pi }~\nonumber \\&&
+\frac{1}{720 s^2 u^2 f_{\pi }^4}\left\{60 s u m_k^6-30 \left[-2 m_{\pi ^+}^2 s^2+3 u s^2+3 u^2 s-2 u m_{\pi }^2 s\right.\right.~\nonumber \\&&
\left.+6 u m_{\pi ^+}^2 s+t u s+2 \left(s^2+5 u s+u^2\right) m_{k^+}^2-2 u^2 m_{\pi ^+}^2\right]m_k^4~\nonumber \\&&
+6 \left\{-10 s u m_{\pi }^4-5 \left[-4 \left(s^2+u^2\right) m_{k^+}^2+4 \left(s^2+u^2\right) m_{\pi ^+}^2+3 s u (s+u)\right] m_{\pi }^2\right.~\nonumber \\&&
\left.+s u \left(25 t m_{k^+}^2+15 t m_{\pi ^+}^2+148 s u\right)\right\} m_k^2-60 s u m_{\pi }^6~\nonumber \\&&
+s u m_{\pi }^2 \left(45 s^2+45 t s-118 u s+45 u^2-150 t m_{k^+}^2-90 t m_{\pi ^+}^2+45 t u\right)~\nonumber \\&&
-30 m_{\pi }^4 \left[2 \left(s^2-5 u s+u^2\right) m_{k^+}^2-2 \left(s^2+3 u s+u^2\right) m_{\pi ^+}^2-s t u\right]~\nonumber \\&&
+s u \left\{3\left[5 s^2-(5 t+126 u) s+5 u (u-t)\right] m_{k^+}^2\right.~\nonumber \\&&
\left.\left.+\left[-15 s^2+(15 t+58 u) s+15 (t-u) u\right] m_{\pi ^+}^2+360 s t u\right\}\right\} \mu _{\pi ^+}~\nonumber \\&&
+\frac{1}{1440 s^2 u^2 f_{\pi }^4}\left\{240 \left(s^2-4 u s+u^2\right) m_k^6-60 \left[\left(9 s^2-3 u s+9 u^2\right) m_{\pi }^2\right.\right.~\nonumber \\&&
\left.+3 \left(s^2+u s+u^2\right) m_{\eta }^2-8 s u \left(m_{k^+}^2-m_{\pi ^+}^2+t\right)\right] m_k^4~\nonumber \\&&
+\left\{120 \left(3 s^2+8 u s+3 u^2\right) m_{\pi }^4+30 \left[12 \left(s^2+u^2\right) m_{\eta }^2-s u \left(16 m_{\pi ^+}^2+19 t\right)\right]
m_{\pi }^2\right.~\nonumber \\&&
\left.-2 s u \left[4 \left(15 s^2+4 u s+15 u^2+30 t m_{k^+}^2-30 t m_{\pi ^+}^2\right)-45 t m_{\eta }^2\right]\right\} m_k^2~\nonumber \\&&
-60 \left(s^2+3 u s+u^2\right) m_{\pi }^6-30 m_{\pi }^4 \left[6 \left(s^2-u s+u^2\right) m_{\eta }^2+s u \left(16 m_{k^+}^2-3 t\right)\right]~\nonumber \\&&
+s u m_{\pi }^2 \left(75 s^2+240 m_{\pi ^+}^2 s+45 t s+62 u s+75 u^2-90 t m_{\eta }^2+240 t m_{k^+}^2\right.~\nonumber \\&&
\left.+240 u m_{\pi ^+}^2+45 t u\right)+s u \left\{15 \left[3 s^2-(3 t+2 u) s+3 u (u-t)\right] m_{\eta }^2\right.~\nonumber \\&&
\left.\left.-8 s u \left(-34 m_{k^+}^2+34 m_{\pi ^+}^2+15 t\right)\right\}\right\} \mu _k~\nonumber \\&&
-\frac{1}{720 s^2 u^2 f_{\pi }^4}\left\{180 s u m_k^6-30 \left[-2 m_{\pi ^+}^2 s^2+3 u s^2+3 u^2 s+2 u m_{\pi }^2 s\right.\right.~\nonumber \\&&
\left.+6 u m_{\pi ^+}^2 s+3 t u s+2 \left(s^2+5 u s+u^2\right) m_{k^+}^2-2 u^2 m_{\pi ^+}^2\right] m_k^4~\nonumber \\&&
-30 \left\{6 s u m_{\pi }^4+\left[-4 \left(s^2+u^2\right) m_{k^+}^2+4 \left(s^2+u^2\right) m_{\pi ^+}^2+s u (3 s-4 t+3 u)\right] m_{\pi }^2\right.~\nonumber \\&&
\left.+s u \left(-5 t m_{k^+}^2-3 t m_{\pi ^+}^2+8 s u\right)\right\} m_k^2+60 s u m_{\pi }^6+s u m_{\pi }^2 \left(45 s^2\right.~\nonumber \\&&
\left.+45 t s-286 u s+45 u^2-150 t m_{k^+}^2-90 t m_{\pi ^+}^2+45 t u\right)~\nonumber \\&&
-30 m_{\pi }^4 \left[2 \left(s^2-5 u s+u^2\right) m_{k^+}^2-2 \left(s^2+3 u s+u^2\right) m_{\pi ^+}^2+s t u\right]~\nonumber \\&&
+s u \left\{5 \left[3 s^2+(62 u-3 t) s+3 u (u-t)\right]m_{k^+}^2\right.~\nonumber \\&&
\left.\left.-3 \left[\left(5 s^2-5 t s+18 u s+5 u^2-5 t u\right) m_{\pi ^+}^2+20 s t u\right]\right\}\right\}\mu _{k^+}~\nonumber \\&&
+\frac{1}{1440 s^2 u^2 f_{\pi }^4}\left\{-180 \left(s^2+u^2\right) m_k^6+60 \left[3 \left(2 s^2+u s+2 u^2\right) m_{\pi }^2\right.\right.~\nonumber \\&&
\left.+3 \left(s^2+u s+u^2\right) m_{\eta }^2-4 s u m_{k^+}^2\right] m_k^4+6 \left\{-30 \left(s^2+u^2\right) m_{\pi }^4\right.~\nonumber \\&&
-15 \left[4 \left(s^2+u^2\right) m_{\eta }^2+s t u\right]m_{\pi }^2+s u \left(15 s^2+20 m_{\pi ^+}^2 s\right.~\nonumber \\&&
\left.\left.+54 u s+15 u^2-15 t m_{\eta }^2+20 t m_{k^+}^2+20 u m_{\pi ^+}^2\right)\right\}m_k^2~\nonumber \\&&
-180 s u m_{\pi }^6+30 m_{\pi }^4 \left[6 \left(s^2-u s+u^2\right) m_{\eta }^2+s u \left(8 m_{k^+}^2+3 t\right)\right]~\nonumber \\&&
-3 s u m_{\pi }^2 \left(15 s^2+40 m_{\pi ^+}^2 s+15 t s+54 u s+15 u^2-30 t m_{\eta }^2\right.~\nonumber \\&&
\left.+40 t m_{k^+}^2+40 u m_{\pi ^+}^2+15 t u\right)+s u \left\{\left[-45 s^2+(45 t-42 u) s\right.\right.~\nonumber \\&&
\left.\left.+45 (t-u) u] m_{\eta }^2+4 s u \left(-112 m_{k^+}^2+172 m_{\pi ^+}^2+135 t\right)\right\}\right\} \mu _{\eta }~\nonumber \\&&
+\frac{t \left(2 m_{\pi }^2+m_{\pi ^+}^2-3 t\right) J_{\pi ^+ \pi ^+}(t)}{12 f_{\pi }^4}-\frac{m_{\pi ^+}^2\left(2 m_k^2-2 m_{\pi }^2-2 m_{k^+}^2+2
m_{\pi ^+}^2+3 t\right) J_{\pi  \pi }(t) }{24 f_{\pi }^4}~\nonumber \\&&
-\frac{t \left(2 m_k^2-2 m_{\pi }^2-2 m_{k^+}^2+2 m_{\pi ^+}^2+3 t\right) J_{k k}(t)}{24 f_{\pi }^4}~\nonumber \\&&
-\frac{t \left(-2 m_k^2-2 m_{\pi }^2+2 m_{k^+}^2+2 m_{\pi ^+}^2+3 t\right) J_{k^+ k^+}(t)}{48 f_{\pi }^4}~\nonumber \\&&
+\frac{\left(2 m_k^2+6 m_{\eta }^2-2 m_{k^+}^2+2 m_{\pi ^+}^2-9 t\right) J_{\eta  \eta }(t) m_{\pi ^+}^2}{72 f_{\pi }^4}~\nonumber \\&&
+\frac{\left(m_k^2-m_{k^+}^2\right) \left(6 m_k^2+3 m_{\pi }^2+3 m_{\eta }^2+2 m_{k^+}^2-2 m_{\pi ^+}^2-9 t\right) J_{\pi  \eta }(t)}{36 f_{\pi
}^4}~\nonumber \\&&
-\frac{1}{144 s^2 f_{\pi }^4}\left\{\frac{1}{2}\left\{6 m_k^8-6 \left(4 m_{\pi }^2+5 s\right) m_k^6\right.\right.~\nonumber \\&&
+\left[36 m_{\pi }^4+54 s m_{\pi }^2+s \left(12 m_{k^+}^2-12 m_{\pi ^+}^2+5 s-3 u\right)\right] m_k^4~\nonumber \\&&
+2 \left[-12 m_{\pi }^6-9 s m_{\pi }^4+s \left(-12 m_{k^+}^2+12 m_{\pi ^+}^2-29 s+3 u\right) m_{\pi }^2\right.~\nonumber \\&&
\left.+s^2 \left(-20 m_{k^+}^2+20 m_{\pi ^+}^2+12 t+15 u\right)\right] m_k^2+6 m_{\pi }^8~\nonumber \\&&
-6 s m_{\pi }^6+s m_{\pi }^4 \left(12 m_{k^+}^2-12 m_{\pi ^+}^2+5 s-3 u\right)~\nonumber \\&&
+2 s^2 m_{\pi }^2 \left(-4 m_{k^+}^2+4 m_{\pi ^+}^2+3 u\right)+s^2 \left[8 m_{k^+}^4+4 \left(3 s-4 m_{\pi ^+}^2\right) m_{k^+}^2\right.~\nonumber \\&&
\left.\left.+8 m_{\pi ^+}^4-12 s m_{\pi ^+}^2+3 s (s-u)\right]\right\} J_{k \pi }(s)~\nonumber \\&&
+\frac{1}{6}\left\{54 m_k^8-54 \left(2 m_{\pi }^2+2 m_{\eta }^2+s\right) m_k^6+9 \left[6 m_{\pi }^4+12 \left(2 m_{\eta }^2+s\right) m_{\pi }^2\right.\right.~\nonumber \\&&
\left.+6 m_{\eta }^4-26 s m_{\eta }^2+s \left(4 m_{k^+}^2-4 m_{\pi ^+}^2+5 s-3 u\right)\right] m_k^4~\nonumber \\&&
+6 \left\{-18 m_{\eta }^2 m_{\pi }^4-6 \left[3 m_{\eta }^4+5 s m_{\eta }^2+s \left(m_{k^+}^2-m_{\pi ^+}^2+4 s\right)\right]m_{\pi }^2\right.~\nonumber \\&&
+s \left[6 m_{\eta }^4+3 \left(-2 m_{k^+}^2+2 m_{\pi ^+}^2+5 t+8 u\right) m_{\eta }^2\right.~\nonumber \\&&
\left.\left.+s \left(-4 m_{k^+}^2+4 m_{\pi ^+}^2+9 u\right)\right]\right\}m_k^2+18 m_{\pi }^4 \left(3 m_{\eta }^4-3 s m_{\eta }^2+s^2\right)~\nonumber \\&&
+6 s m_{\pi }^2 \left[-3 m_{\eta }^4+6 \left(m_{k^+}^2-m_{\pi ^+}^2+s\right) m_{\eta }^2+2 s \left(m_{\pi ^+}^2-m_{k^+}^2\right)\right]~\nonumber \\&&
+s \left\{9 (t-2 u) m_{\eta }^4+6 s \left(-2 m_{k^+}^2+2 m_{\pi ^+}^2+9 u\right) m_{\eta }^2\right.~\nonumber \\&&
+s \left[8 m_{k^+}^4+4 \left(9 s-4 m_{\pi ^+}^2\right) m_{k^+}^2+8 m_{\pi ^+}^4\right.~\nonumber \\&&
\left.\left.\left.-36 s m_{\pi ^+}^2+27 s (s-u)\right]\right\}\right\}J_{k \eta }(s)~\nonumber \\&&
+\left\{2 \left(3 m_{k^+}^4-6 m_{\pi ^+}^2 m_{k^+}^2+3 m_{\pi ^+}^4+s^2-3 s m_{\pi ^+}^2\right) m_k^4\right.~\nonumber \\&&
+2 \left\{\left(20 s-6 m_{\pi }^2\right) m_{k^+}^4+\left[6 m_{\pi }^2 \left(2 m_{\pi ^+}^2+s\right)-56 s^2\right] m_{k^+}^2\right.~\nonumber \\&&
\left.-3 \left(2 m_{\pi }^2+s\right) m_{\pi ^+}^4+36 s^3+12 s^2 m_{\pi ^+}^2\right\}m_k^2~\nonumber \\&&
+6 m_{\pi }^4 \left[m_{k^+}^4-2 \left(m_{\pi ^+}^2+s\right) m_{k^+}^2+m_{\pi ^+}^4+s^2+s m_{\pi ^+}^2\right]~\nonumber \\&&
+2 s m_{\pi }^2 \left[5 m_{k^+}^4+6 \left(m_{\pi ^+}^2-4 s\right) m_{k^+}^2+3 \left(2 m_{\pi ^+}^4-4 s m_{\pi ^+}^2+9 s^2\right)\right]~\nonumber \\&&
-s \left\{(11 t+14 u) m_{k^+}^4-6 \left[(2 s-t) m_{\pi ^+}^2+s (7 t+8 u)\right] m_{k^+}^2\right.~\nonumber \\&&
+3 \left[(u-s) m_{\pi ^+}^4+2 s (4 s-u) m_{\pi ^+}^2\right.~\nonumber \\&&
\left.\left.\left.\left.+s^2 (13 t+14 u)\right]\right\}\right\} J_{k^+ \pi ^+}(s)+[s\longleftrightarrow u]\right\} \ ;
\end{eqnarray}

\begin{eqnarray}
T_{K^{+}K^{-}\rightarrow K^{0}\bar{K}^{0}}^{H}(s,t,u)&=&
\frac{ m_k^2+m_{k^+}^2-u}{2 f_{\pi }^2}-\frac{1}{384 \pi ^2 t f_{\pi }^4}\left\{m_k^6+\left(2 m_{\pi }^2-m_{k^+}^2+2 m_{\pi ^+}^2+t\right)
m_k^4\right.~\nonumber \\&&
-\left[m_{k^+}^4+4 \left(m_{\pi }^2+m_{\pi ^+}^2-t\right) m_{k^+}^2+t (4 u-3 t)\right] m_k^2+m_{k^+}^6~\nonumber \\&&
+2 m_{\pi }^2 m_{k^+}^4-t m_{k^+}^4-2 s t^2-t u^2+2 s t m_{\pi }^2-2 t u m_{\pi }^2+4 t^2 m_{k^+}^2~\nonumber \\&&
\left.+s t m_{k^+}^2-3 t u m_{k^+}^2+2 m_{k^+}^4 m_{\pi ^+}^2-2 t^2 m_{\pi ^+}^2+2 s t m_{\pi ^+}^2\right\}~\nonumber \\&&
+\frac{8 L_1 \left(s-2 m_k^2\right) \left(s-2 m_{k^+}^2\right)}{f_{\pi }^4}+\frac{4 L_2\left[\left(m_k^2+m_{k^+}^2-t\right){}^2+\left(m_k^2+m_{k^+}^2-u\right){}^2\right]}{f_{\pi
}^4}~\nonumber \\&&
+\frac{2 L_3 \left(\left(m_k^2+m_{k^+}^2-t\right){}^2+\left(s-2 m_k^2\right) \left(s-2 m_{k^+}^2\right)\right)}{f_{\pi }^4}+\frac{32 L_6 m_k^2
m_{k^+}^2}{f_{\pi }^4}~\nonumber \\&&
+\frac{4 L_5 \left(-2 m_k^4+\left(3 m_{\pi }^2-4 m_{k^+}^2+2 u\right) m_k^2+3 m_{\pi }^2 \left(m_{k^+}^2-u\right)\right)}{f_{\pi }^4}~\nonumber \\&&
+\frac{4 L_4}{f_{\pi }^4} \left[-m_k^4+\left(m_{\pi }^2-10 m_{k^+}^2-m_{\pi ^+}^2+3 s+t\right) m_k^2-m_{k^+}^4+2 s m_{k^+}^2\right.~\nonumber \\&&
\left.+u m_{k^+}^2-m_{k^+}^2 m_{\pi ^+}^2+u m_{\pi ^+}^2+m_{\pi }^2 \left(m_{k^+}^2-u\right)\right]+\frac{16 L_8 m_k^2 m_{k^+}^2}{f_{\pi }^4}~\nonumber \\&&
-\frac{1}{360 t^2 f_{\pi }^4}\left\{120 \left(-m_{\pi }^2+m_{\pi ^+}^2+t\right) m_k^4+8 \left[59 t^2-30 m_{k^+}^2 \left(-m_{\pi }^2+m_{\pi ^+}^2+t\right)\right]m_k^2\right.~\nonumber \\&&
+120 t m_{k^+}^4+534 t^3+594 s t^2-15 t^2 m_{\eta }^2+472 t^2 m_{k^+}^2~\nonumber \\&&
+120 m_{k^+}^4 m_{\pi ^+}^2+4 t^2 m_{\pi ^+}^2+30 s t m_{\pi ^+}^2-30 t u m_{\pi ^+}^2-1056 t^2 u~\nonumber \\&&
\left.+3 m_{\pi }^2\left[-40 m_{k^+}^4-20 t m_{k^+}^2-20 t m_k^2+t (7 t+20 u)\right]\right\} \mu _{\pi }~\nonumber \\&&
-\frac{1}{180 t^2 f_{\pi }^4}\left[60 \left(m_{\pi }^2-m_{\pi ^+}^2\right) m_k^4-8 \left(8 t^2+15 m_{\pi }^2 m_{k^+}^2-15 m_{k^+}^2 m_{\pi ^+}^2\right)
m_k^2\right.~\nonumber \\&&
-48 t^3+12 s t^2+15 t^2 m_{\eta }^2-64 t^2 m_{k^+}^2-60 m_{k^+}^4 m_{\pi ^+}^2+5 t^2 m_{\pi ^+}^2~\nonumber \\&&
\left.-15 s t m_{\pi ^+}^2+15 t u m_{\pi ^+}^2+72 t^2 u+15 m_{\pi }^2 \left(4 m_{k^+}^4+s t-t u\right)\right]\mu _{\pi ^+}~\nonumber \\&&
+\frac{1}{60 t^2 f_{\pi }^4}\left\{10 m_k^6-10 \left(3 m_{k^+}^2+t\right) m_k^4+\left[30 m_{k^+}^4+10 t m_{k^+}^2+t (5 s+14 t)\right] m_k^2\right.~\nonumber \\&&
\left.-10 m_{k^+}^6+t (24 t-5 s) m_{k^+}^2+t^2 [3 s-2 (6 t+u)]\right\} \mu _k~\nonumber \\&&
-\frac{1}{12 t^2 f_{\pi }^4}\left\{2 m_k^6-6 m_{k^+}^2 m_k^4+\left[6 m_{k^+}^4-2 t m_{k^+}^2+(s-10 t) t\right]m_k^2\right.~\nonumber \\&&
\left.-2 m_{k^+}^6+2 t m_{k^+}^4-t (s+8 t) m_{k^+}^2+t^2 (3 t+5 u)\right\} \mu _{k^+}~\nonumber \\&&
+\frac{1}{360 f_{\pi }^4}\left(896 m_k^2+15 m_{\pi }^2-129 m_{\eta }^2+896 m_{k^+}^2-58 m_{\pi ^+}^2-810 u\right) \mu _{\eta }~\nonumber \\&&
-\frac{\left(2 m_k^2-2 m_{\pi }^2-2 m_{k^+}^2+2 m_{\pi ^+}^2+3 s\right) \left(-2 m_k^2-2 m_{\pi }^2+2 m_{k^+}^2+2 m_{\pi ^+}^2+3 s\right) J_{\pi
 \pi }(s)}{288 f_{\pi }^4}~\nonumber \\&&
-\frac{\left(s^2+m_k^2 s+m_{k^+}^2 s-t s+2 t m_{\pi ^+}^2-2 u m_{\pi ^+}^2\right) J_{\pi ^+ \pi ^+}(s)}{24 f_{\pi }^4}~\nonumber \\&&
-\frac{1}{72 t^2 f_{\pi }^4}\left\{4 \left[3 m_{\pi }^4-3 \left(2 m_{\pi ^+}^2+t\right) m_{\pi }^2+3 m_{\pi ^+}^4+t^2\right]m_k^4\right.~\nonumber \\&&
-8 m_{k^+}^2 \left[3 m_{\pi }^4-3 \left(2 m_{\pi ^+}^2+t\right) m_{\pi }^2+3 m_{\pi ^+}^4+t^2\right]m_k^2+4 t^2 m_{k^+}^4~\nonumber \\&&
+12 m_{k^+}^4 m_{\pi ^+}^4+3 s t m_{\pi ^+}^4-3 t u m_{\pi ^+}^4+3 s t^3-6 s t^2 m_{\pi ^+}^2~\nonumber \\&&
+6 t^2 u m_{\pi ^+}^2-3 t^3 u+3 m_{\pi }^4 \left[4 m_{k^+}^4+t (s-u)\right]~\nonumber \\&&
\left.-6 m_{\pi }^2\left[2 \left(2 m_{\pi ^+}^2+t\right) m_{k^+}^4+t (s-u) \left(m_{\pi ^+}^2+t\right)\right]\right\} J_{\pi  \pi ^+}(t)~\nonumber \\&&
-\frac{\left(2 m_k^4+\left(2 m_{k^+}^2-3 t+u\right) m_k^2+s \left(m_{k^+}^2+s-u\right)\right) J_{k k}(s)}{12 f_{\pi }^4}~\nonumber \\&&
-\frac{\left(s m_k^2+s (s-u)+m_{k^+}^2 \left(2 m_k^2+2 m_{k^+}^2-3 t+u\right)\right) J_{k^+ k^+}(s)}{12 f_{\pi }^4}~\nonumber \\&&
+\frac{\left(2 m_k^2+3 m_{\pi }^2+3 m_{\eta }^2+6 m_{k^+}^2-2 m_{\pi ^+}^2-9 s\right) \left(6 m_k^2+3 m_{\pi }^2+3 m_{\eta }^2+2 m_{k^+}^2-2
m_{\pi ^+}^2-9 s\right) J_{\pi  \eta }(s)}{432 f_{\pi }^4}~\nonumber \\&&
-\frac{\left(2 m_k^2-6 m_{\eta }^2-2 m_{k^+}^2-2 m_{\pi ^+}^2+9 s\right) \left(-2 m_k^2-6 m_{\eta }^2+2 m_{k^+}^2-2 m_{\pi ^+}^2+9 s\right)
J_{\eta  \eta }(s)}{288 f_{\pi }^4}~\nonumber \\&&
-\frac{1}{24 t^2 f_{\pi }^4}\left\{2 m_k^8-2 \left(4 m_{k^+}^2+t\right) m_k^6+\left(12 m_{k^+}^4+2 t m_{k^+}^2+s t\right) m_k^4\right.~\nonumber \\&&
-2\left[4 m_{k^+}^6-t m_{k^+}^4+(s-3 t) t m_{k^+}^2+s t^2\right] m_k^2~\nonumber \\&&
\left.+\left(t-m_{k^+}^2\right) \left[-2 m_{k^+}^6+t (2 t-s) m_{k^+}^2+t^2 (t-u)\right]\right\}J_{k k^+}(t)~\nonumber \\&&
-\frac{\left(m_k^2+m_{k^+}^2-u\right){}^2 J_{k k^+}(u)}{4 f_{\pi }^4}-\frac{\left(4 m_k^2+3 m_{\eta }^2+4 m_{k^+}^2+m_{\pi ^+}^2-9 t\right){}^2
J_{\eta  \pi ^+}(t)}{216 f_{\pi }^4} \ ;
\end{eqnarray}

\begin{eqnarray}
T_{K^{+}\pi^{0}\rightarrow K^{+}\eta}^{H}(s,t,u)&=&
\frac{ 9 t-2 m_k^2-3 m_{\pi }^2-3 m_{\eta }^2-6 m_{k^+}^2+2 m_{\pi ^+}^2 }{12 \sqrt{3}\text{  }f_{\pi }^2}+\frac{1}{1536 \sqrt{3}
\pi ^2 s u f_{\pi }^4}\left\{3 \left[-4 s m_{k^+}^6\right.\right.~\nonumber \\&&
-4 u m_{k^+}^6+s m_{\eta }^2 m_{k^+}^4+u m_{\eta }^2 m_{k^+}^4+4 s u m_{k^+}^4+3 s m_{\eta }^4 m_{k^+}^2~\nonumber \\&&
+3 u m_{\eta }^4 m_{k^+}^2+4 s u^2 m_{k^+}^2-4 s u m_{\eta }^2 m_{k^+}^2+4 s^2 u m_{k^+}^2-8 s t u m_{k^+}^2~\nonumber \\&&
+3 s u^2 m_{\eta }^2+3 s^2 u m_{\eta }^2-6 s t u m_{\eta }^2+2 \left(\left(m_{\eta }^2-m_{k^+}^2\right) \left(m_{\pi }^2+m_{\eta }^2+2 m_{k^+}^2-t\right)
m_{k^+}^2\right.~\nonumber \\&&
\left.+s u \left(m_{\pi }^2+m_{\eta }^2+2 m_{k^+}^2-3 t\right)\right] m_{\pi ^+}^2+2 s u [s (t-2 u)+t u]~\nonumber \\&&
-m_{\pi }^4 \left(m_{\eta }^2-m_{k^+}^2\right) \left(m_{\pi }^2+m_{\eta }^2+2 m_{k^+}^2-t\right)+2 m_k^2\left[s u \left(m_{\pi }^2+m_{\eta }^2+2
m_{k^+}^2-3 t\right)\right.~\nonumber \\&&
\left.+\left(m_{\pi }^2-m_{k^+}^2\right) \left(m_{k^+}^2-m_{\eta }^2\right) \left(m_{\pi }^2+m_{\eta }^2+2 m_{k^+}^2-t\right)\right]~\nonumber \\&&
+m_{\pi }^2 \left\{s u \left(m_{\pi }^2+m_{\eta }^2+2 m_{k^+}^2-3 t\right)-\left(m_{\eta }^2-m_{k^+}^2\right) \right.~\nonumber \\&&
\left.\left.\times \left[\left(m_{\pi }^2+m_{\eta }^2+2 m_{k^+}^2-t\right) \left(2 m_{\pi ^+}^2+3 m_{\eta }^2+3m_{k^+}^2\right)-4 s u\right]\right\}\right\}~\nonumber \\&&
+\frac{32 L_6 \left(m_{k^+}^2-m_k^2\right) m_{k^+}^2}{\sqrt{3} f_{\pi }^4}+\frac{16 L_8 \left(m_k^2-m_{\pi ^+}^2\right) \left(2 m_k^2-4 m_{k^+}^2-m_{\pi
^+}^2\right)}{3 \sqrt{3} f_{\pi }^4}~\nonumber \\&&
+\frac{32 L_7 \left(2 m_k^4-\left(m_{k^+}^2+3 m_{\pi ^+}^2\right) m_k^2-3 m_{k^+}^4+m_{\pi ^+}^4+4 m_{k^+}^2 m_{\pi ^+}^2\right)}{3 \sqrt{3}
f_{\pi }^4}~\nonumber \\&&
-\frac{L_3 \left[-m_{\pi }^4+3 \left(t-2 m_{k^+}^2\right) m_{\pi }^2-m_{\eta }^4-2 m_{k^+}^4+s^2-2 t^2+u^2+6 t m_{k^+}^2+3 m_{\eta }^2 \left(t-2
m_{k^+}^2\right)\right]}{\sqrt{3} f_{\pi }^4}~\nonumber \\&&
-\frac{1}{3 \sqrt{3} f_{\pi }^4}2 L_4\left[2 m_k^4+\left(5 m_{\pi }^2+3 m_{\eta }^2-20 m_{k^+}^2-4 m_{\pi ^+}^2+3 t\right) m_k^2+3 m_{\pi }^4+18
m_{k^+}^4\right.~\nonumber \\&&
+2 m_{\pi ^+}^4-3 m_{\eta }^2 m_{k^+}^2-3 t m_{k^+}^2-3 m_{\eta }^2 m_{\pi ^+}^2-4 m_{k^+}^2 m_{\pi ^+}^2+9 t m_{\pi ^+}^2~\nonumber \\&&
\left.+m_{\pi }^2 \left(3 m_{\eta }^2+3 m_{k^+}^2-5 m_{\pi ^+}^2-9 t\right)\right]+\frac{1}{9 \sqrt{3} f_{\pi }^4}L_5\left[4 m_k^4\right.~\nonumber \\&&
-2 \left(-6 m_{\pi }^2-12 m_{\eta }^2-50 m_{k^+}^2+6 m_{\pi ^+}^2+45 t\right) m_k^2-45 m_{\pi }^4+9 m_{\eta }^4~\nonumber \\&&
-24 m_{k^+}^4+8 m_{\pi ^+}^4-27 t m_{\eta }^2+6 m_{\eta }^2 m_{k^+}^2+36 t m_{k^+}^2+6 m_{\eta }^2 m_{\pi ^+}^2~\nonumber \\&&
\left.-40 m_{k^+}^2 m_{\pi ^+}^2+3 m_{\pi }^2 \left(-12 m_{\eta }^2-34 m_{k^+}^2+6 m_{\pi ^+}^2+45 t\right)\right]~\nonumber \\&&
+\frac{1}{480 \sqrt{3} s^2 u^2 f_{\pi }^4}\left\{-60 s^2 m_{k^+}^6-60 u^2 m_{k^+}^6+120 s u^2 m_{k^+}^4+60 s^2 m_{\eta }^2 m_{k^+}^4\right.~\nonumber \\&&
+60 u^2 m_{\eta }^2 m_{k^+}^4+120 s^2 u m_{k^+}^4+30 s u^3 m_{k^+}^2-140 s^2 u^2 m_{k^+}^2~\nonumber \\&&
-120 s u^2 m_{\eta }^2 m_{k^+}^2-90 s^2 u m_{\eta }^2 m_{k^+}^2+80 s u^2 m_{\pi ^+}^2 m_{k^+}^2~\nonumber \\&&
+80 s^2 u m_{\pi ^+}^2 m_{k^+}^2+30 s^3 u m_{k^+}^2+15 s^2 u m_{\eta }^4+1044 s^2 u^3~\nonumber \\&&
+1044 s^3 u^2-2016 s^2 t u^2+9 s^2 u^2 m_{\eta }^2-15 s^3 u m_{\eta }^2~\nonumber \\&&
-15 s^2 t u m_{\eta }^2-768 s^2 u^2 m_{\pi ^+}^2-60 s u^2 m_{\eta }^2 m_{\pi ^+}^2-60 s^2 u m_{\eta }^2 m_{\pi ^+}^2~\nonumber \\&&
+4 s u m_k^2 \left[5 m_{\pi }^4-5 \left(-4 m_{\eta }^2+2 m_{k^+}^2+t\right) m_{\pi }^2+15 m_{\eta }^4\right.~\nonumber \\&&
\left.-5 m_{\eta }^2 \left(3 t-2 m_{k^+}^2\right)+4 \left(-10 m_{k^+}^4+5 t m_{k^+}^2+43 s u\right)\right]~\nonumber \\&&
-5 m_{\pi }^4 \left[-12 \left(s^2-u s+u^2\right) m_{\eta }^2+12 \left(s^2+u^2\right) m_{k^+}^2+s u \left(4 m_{\pi ^+}^2+3 s\right)\right]~\nonumber \\&&
-5 m_{\pi }^2 \left\{12 s u m_{\eta }^4+4 \left[6 \left(s^2+u s+u^2\right) m_{k^+}^2+s u \left(m_{\pi ^+}^2-3 t\right)\right] m_{\eta }^2\right.~\nonumber \\&&
-24 \left(s^2+u^2\right) m_{k^+}^4-2 s u m_{k^+}^2 \left(-4 m_{\pi ^+}^2+3 s+6 u\right)~\nonumber \\&&
\left.\left.+s u \left(3 s^2-3 t s+u s+6 u^2-4 t m_{\pi ^+}^2\right)\right\}\right\} \mu _{\pi }~\nonumber \\&&
-\frac{1}{720 \sqrt{3} s^2 u^2 f_{\pi }^4}\left\{105 s u m_{\pi ^+}^2 m_{\pi }^4+3 \left\{5 \left[9 u s^2+\left(-12 s^2+7 u s-12 u^2\right)
m_{\pi ^+}^2\right] m_{\eta }^2\right.\right.~\nonumber \\&&
+s u \left(32 s u-35 t m_{\pi ^+}^2\right)+m_{k^+}^2\left[10 \left(6 s^2+7 u s+6 u^2\right) m_{\pi ^+}^2\right.~\nonumber \\&&
+5 s u (7 u-5 s)]\} m_{\pi }^2+135 s^2 u m_{\eta }^4+210 s u^2 m_{k^+}^4~\nonumber \\&&
-150 s^2 u m_{k^+}^4+36 s^2 u^3+36 s^3 u^2-504 s^2 t u^2+441 s^2 u^2 m_{\eta }^2~\nonumber \\&&
-135 s^3 u m_{\eta }^2-135 s^2 t u m_{\eta }^2-45 s u^3 m_{k^+}^2+370 s^2 u^2 m_{k^+}^2~\nonumber \\&&
-45 s t u^2 m_{k^+}^2-135 s u^2 m_{\eta }^2 m_{k^+}^2-45 s^2 u m_{\eta }^2 m_{k^+}^2+135 s^3 u m_{k^+}^2~\nonumber \\&&
+135 s^2 t u m_{k^+}^2-180 s^2 m_{k^+}^4 m_{\pi ^+}^2-180 u^2 m_{k^+}^4 m_{\pi ^+}^2+45 s u^3 m_{\pi ^+}^2~\nonumber \\&&
-318 s^2 u^2 m_{\pi ^+}^2-45 s t u^2 m_{\pi ^+}^2-135 s u^2 m_{\eta }^2 m_{\pi ^+}^2-135 s^2 u m_{\eta }^2 m_{\pi ^+}^2~\nonumber \\&&
+30 s u^2 m_{k^+}^2 m_{\pi ^+}^2+180 s^2 m_{\eta }^2 m_{k^+}^2 m_{\pi ^+}^2+180 u^2 m_{\eta }^2 m_{k^+}^2 m_{\pi ^+}^2~\nonumber \\&&
+30 s^2 u m_{k^+}^2 m_{\pi ^+}^2+45 s^3 u m_{\pi ^+}^2-45 s^2 t u m_{\pi ^+}^2+m_k^2 \left\{15 s u m_{\pi }^4\right.~\nonumber \\&&
+15 \left[2 \left(6 s^2-7 u s+6 u^2\right) m_{\eta }^2-2 \left(6 s^2-u s+6 u^2\right) m_{k^+}^2-s t u\right] m_{\pi }^2~\nonumber \\&&
-225 s u m_{\eta }^4+180 s^2 m_{k^+}^4+180 u^2 m_{k^+}^4-45 s u^3+62 s^2 u^2~\nonumber \\&&
+45 s t u^2+210 s u^2 m_{k^+}^2+210 s^2 u m_{k^+}^2-45 s^3 u+45 s^2 t u~\nonumber \\&&
\left.\left.-45 m_{\eta }^2\left[2 \left(2 s^2+5 u s+2 u^2\right) m_{k^+}^2-5 s t u\right]\right\}\right\}\mu _{\pi ^+}~\nonumber \\&&
-\frac{1}{144 \sqrt{3} s^2 u^2 f_{\pi }^4}\left\{-21 s u m_{\pi ^+}^2 m_{\pi }^4+3\left\{\left[\left(12 s^2-7 u s+12 u^2\right) m_{\pi ^+}^2\right.\right.\right.~\nonumber \\&&
+3 s u (s+4 u)] m_{\eta }^2+s u \left(7 t m_{\pi ^+}^2+24 s u\right)~\nonumber \\&&
\left.-m_{k^+}^2 \left[2 \left(6 s^2+7 u s+6 u^2\right) m_{\pi ^+}^2+s u (7 s+19 u)\right]\right\} m_{\pi }^2~\nonumber \\&&
-27 s^2 u m_{\eta }^4-6 s u^2 m_{k^+}^4+66 s^2 u m_{k^+}^4+36 s^2 u^3+36 s^3 u^2~\nonumber \\&&
-144 s^2 t u^2+3 s^2 u^2 m_{\eta }^2+27 s^3 u m_{\eta }^2+27 s^2 t u m_{\eta }^2+9 s u^3 m_{k^+}^2~\nonumber \\&&
+166 s^2 u^2 m_{k^+}^2+9 s t u^2 m_{k^+}^2-9 s u^2 m_{\eta }^2 m_{k^+}^2-27 s^2 u m_{\eta }^2 m_{k^+}^2~\nonumber \\&&
-27 s^3 u m_{k^+}^2-27 s^2 t u m_{k^+}^2+36 s^2 m_{k^+}^4 m_{\pi ^+}^2+36 u^2 m_{k^+}^4 m_{\pi ^+}^2~\nonumber \\&&
-9 s u^3 m_{\pi ^+}^2-82 s^2 u^2 m_{\pi ^+}^2+9 s t u^2 m_{\pi ^+}^2+27 s u^2 m_{\eta }^2 m_{\pi ^+}^2~\nonumber \\&&
+27 s^2 u m_{\eta }^2 m_{\pi ^+}^2-6 s u^2 m_{k^+}^2 m_{\pi ^+}^2-36 s^2 m_{\eta }^2 m_{k^+}^2 m_{\pi ^+}^2~\nonumber \\&&
-36 u^2 m_{\eta }^2 m_{k^+}^2 m_{\pi ^+}^2-6 s^2 u m_{k^+}^2 m_{\pi ^+}^2-9 s^3 u m_{\pi ^+}^2+9 s^2 t u m_{\pi ^+}^2~\nonumber \\&&
+3 m_k^2 \left\{-s u m_{\pi }^4+\left[-2 \left(6 s^2-7 u s+6 u^2\right) m_{\eta }^2+2 \left(6 s^2-u s+6 u^2\right) m_{k^+}^2\right.\right.~\nonumber \\&&
+s t u] m_{\pi }^2+15 s u m_{\eta }^4-12 s^2 m_{k^+}^4-12 u^2 m_{k^+}^4+3 s u^3~\nonumber \\&&
+6 s^2 u^2-3 s t u^2-14 s u^2 m_{k^+}^2-14 s^2 u m_{k^+}^2+3 s^3 u-3 s^2 t u~\nonumber \\&&
\left.\left.+3 m_{\eta }^2 \left[2 \left(2 s^2+5 u s+2 u^2\right) m_{k^+}^2-5 s t u\right]\right\}\right\} \mu _k~\nonumber \\&&
+\frac{1}{1440 \sqrt{3} s^2 u^2 f_{\pi }^4}\left\{720 s^2 m_{k^+}^6+720 u^2 m_{k^+}^6-1440 s u^2 m_{k^+}^4\right.~\nonumber \\&&
-1260 s^2 m_{\eta }^2 m_{k^+}^4-1260 u^2 m_{\eta }^2 m_{k^+}^4-1440 s^2 u m_{k^+}^4~\nonumber \\&&
+540 s^2 m_{\eta }^4 m_{k^+}^2+540 u^2 m_{\eta }^4 m_{k^+}^2-360 s u^3 m_{k^+}^2~\nonumber \\&&
-144 s^2 u^2 m_{k^+}^2+810 s u^2 m_{\eta }^2 m_{k^+}^2+450 s^2 u m_{\eta }^2 m_{k^+}^2~\nonumber \\&&
-240 s u^2 m_{\pi ^+}^2 m_{k^+}^2-240 s^2 u m_{\pi ^+}^2 m_{k^+}^2-360 s^3 u m_{k^+}^2~\nonumber \\&&
+135 s u^2 m_{\eta }^4-45 s^2 u m_{\eta }^4+144 s^2 u^3+144 s^3 u^2~\nonumber \\&&
+1584 s^2 t u^2+135 s u^3 m_{\eta }^2-726 s^2 u^2 m_{\eta }^2-135 s t u^2 m_{\eta }^2~\nonumber \\&&
+315 s^3 u m_{\eta }^2+45 s^2 t u m_{\eta }^2+128 s^2 u^2 m_{\pi ^+}^2+360 s u^2 m_{\eta }^2 m_{\pi ^+}^2~\nonumber \\&&
+360 s^2 u m_{\eta }^2 m_{\pi ^+}^2+45 m_{\pi }^4 \left[-4 \left(s^2+u^2\right) m_{\eta }^2+4 \left(s^2+u^2\right) m_{k^+}^2\right.~\nonumber \\&&
+s (s-3 u) u]-8 s u m_k^2 \left[-15 m_{\pi }^4+15 \left(2 m_{\eta }^2-2 m_{k^+}^2+t\right) m_{\pi }^2\right.~\nonumber \\&&
\left.+45 m_{\eta }^4-30 s m_{k^+}^2-30 u m_{k^+}^2+16 s u-45 m_{\eta }^2 \left(t-2 m_{k^+}^2\right)\right]~\nonumber \\&&
+3 m_{\pi }^2 \left\{-180 \left(s^2+u^2\right) m_{\eta }^4+480 \left(s^2+u^2\right) m_{k^+}^2 m_{\eta }^2\right.~\nonumber \\&&
-300 \left(s^2+u^2\right) m_{k^+}^4+30 s u (11 s+7 u) m_{k^+}^2~\nonumber \\&&
+s u \left\{3 \left[5 s^2-(5 t+74 u) s+5 u (3 t+5 u)\right]\right.~\nonumber \\&&
\left.\left.\left.-40 \left(m_{\pi }^2+m_{\eta }^2+2 m_{k^+}^2-t\right) m_{\pi ^+}^2\right\}\right\}\right\}\mu _{k^+}~\nonumber \\&&
-\frac{1}{480 \sqrt{3} s^2 u^2 f_{\pi }^4}\left\{180 s^2 m_{k^+}^6+180 u^2 m_{k^+}^6-120 s u^2 m_{k^+}^4\right.~\nonumber \\&&
-360 s^2 m_{\eta }^2 m_{k^+}^4-360 u^2 m_{\eta }^2 m_{k^+}^4-120 s^2 u m_{k^+}^4~\nonumber \\&&
+180 s^2 m_{\eta }^4 m_{k^+}^2+180 u^2 m_{\eta }^4 m_{k^+}^2-90 s u^3 m_{k^+}^2~\nonumber \\&&
+260 s^2 u^2 m_{k^+}^2-90 s u^2 m_{\eta }^2 m_{k^+}^2-180 s^2 u m_{\eta }^2 m_{k^+}^2~\nonumber \\&&
-90 s^3 u m_{k^+}^2+45 s u^2 m_{\eta }^4+108 s^2 u^3+108 s^3 u^2~\nonumber \\&&
-432 s^2 t u^2+45 s u^3 m_{\eta }^2+15 s^2 u^2 m_{\eta }^2-45 s t u^2 m_{\eta }^2~\nonumber \\&&
+90 s^3 u m_{\eta }^2-240 s^2 u^2 m_{\pi ^+}^2+60 s u^2 m_{\eta }^2 m_{\pi ^+}^2+60 s^2 u m_{\eta }^2 m_{\pi ^+}^2~\nonumber \\&&
+20 s u m_k^2 \left[3 m_{\pi }^4-3 \left(t-2 m_{k^+}^2\right) m_{\pi }^2-3 m_{\eta }^4+8 s u\right.~\nonumber \\&&
\left.+3 m_{\eta }^2 \left(t-2 m_{k^+}^2\right)\right]+60 s u m_{\pi }^4 \left(3 m_{\eta }^2+2 m_{k^+}^2-m_{\pi ^+}^2\right)~\nonumber \\&&
-3 m_{\pi }^2 \left\{60 \left(s^2-u s+u^2\right) m_{\eta }^4-5 \left[8 \left(3 s^2+4 u s+3 u^2\right) m_{k^+}^2\right.\right.~\nonumber \\&&
\left.+s u \left(-4 m_{\pi ^+}^2-12 t+3 u\right)\right] m_{\eta }^2+4 \left[5 \left(3 s^2-4 u s+3 u^2\right) m_{k^+}^4\right.~\nonumber \\&&
\left.\left.\left.+10 s u \left(m_{\pi ^+}^2+t\right) m_{k^+}^2+s u \left(6 s u-5 t m_{\pi ^+}^2\right)\right]\right\}\right\} \mu _{\eta }~\nonumber \\&&
-\frac{\left(m_k^2-m_{k^+}^2\right) \left(2 m_k^2+2 m_{\pi }^2-2 m_{k^+}^2-2 m_{\pi ^+}^2-3 t\right) J_{\pi  \pi }(t)}{24 \sqrt{3} f_{\pi }^4}~\nonumber \\&&
+\frac{t \left(-6 m_k^2-3 m_{\pi }^2-3 m_{\eta }^2-2 m_{k^+}^2+2 m_{\pi ^+}^2+9 t\right) J_{k k}(t)}{48 \sqrt{3} f_{\pi }^4}~\nonumber \\&&
+\frac{t \left(2 m_k^2-2 m_{\pi ^+}^2+3 \left(m_{\pi }^2+m_{\eta }^2+2 m_{k^+}^2-3 t\right)\right) J_{k^+ k^+}(t)}{24 \sqrt{3} f_{\pi }^4}~\nonumber \\&&
-\frac{m_{\pi ^+}^2 \left(-2 m_k^2-3 m_{\pi }^2-3 m_{\eta }^2-6 m_{k^+}^2+2 m_{\pi ^+}^2+9 t\right) J_{\pi  \eta }(t)}{36 \sqrt{3} f_{\pi }^4}~\nonumber \\&&
+\frac{\left(m_k^2-m_{k^+}^2\right) \left(2 m_k^2-6 m_{\eta }^2-2 m_{k^+}^2-2 m_{\pi ^+}^2+9 t\right) J_{\eta \eta }(t)}{72 \sqrt{3} f_{\pi
}^4}~\nonumber \\&&
+\frac{t \left(m_k^2-m_{k^+}^2\right) J_{\pi ^+ \pi ^+}(t)}{12 \sqrt{3} f_{\pi }^4}-\frac{1}{288 \sqrt{3} s^2 f_{\pi }^4}~\nonumber \\&&
\times \left\{\left\{\left[36 m_{k^+}^4+42 s m_{k^+}^2-5 s^2+9 s t-9 s u+3 m_{\pi }^2 \left(12 m_{\eta }^2-12 m_{k^+}^2+s\right)\right.\right.\right.~\nonumber \\&&
\left.-9 m_{\eta }^2 \left(4 m_{k^+}^2+5 s\right)\right]m_k^4+\left\{-72 m_{\pi ^+}^2 m_{k^+}^4+6 s m_{k^+}^4\right.~\nonumber \\&&
-39 s^2 m_{k^+}^2-36 s m_{\pi ^+}^2 m_{k^+}^2-9 s t m_{k^+}^2-9 s u m_{k^+}^2~\nonumber \\&&
+39 s^3-38 s^2 m_{\pi ^+}^2-18 s t m_{\pi ^+}^2+18 s u m_{\pi ^+}^2-33 s^2 t~\nonumber \\&&
+3 s^2 u+9 m_{\eta }^2 \left[\left(8 m_{\pi ^+}^2+s\right) m_{k^+}^2+s \left(2 m_{\pi ^+}^2+9 s\right)\right]~\nonumber \\&&
-3 m_{\pi }^2 \left[12 \left(2 m_{\pi ^+}^2+s\right) m_{\eta }^2+s \left(5 s-6 m_{\pi ^+}^2\right)\right.~\nonumber \\&&
\left.\left.-m_{k^+}^2 \left(24 m_{\pi ^+}^2+19 s\right)\right]\right\} m_k^2+2 \left\{18 m_{\pi ^+}^4 m_{k^+}^4\right.~\nonumber \\&&
+4 s^2 m_{k^+}^4-12 s m_{\pi ^+}^2 m_{k^+}^4-24 s m_{\pi ^+}^4 m_{k^+}^2-72 s^3 m_{k^+}^2~\nonumber \\&&
+40 s^2 m_{\pi ^+}^2 m_{k^+}^2+15 s t m_{\pi ^+}^4+6 s u m_{\pi ^+}^4~\nonumber \\&&
+36 s^3 m_{\pi ^+}^2-18 s^2 t m_{\pi ^+}^2+45 s^3 t+36 s^3 u~\nonumber \\&&
-3 m_{\eta }^2 \left[2 \left(3 m_{\pi ^+}^4-3 s m_{\pi ^+}^2+s^2\right) m_{k^+}^2\right.~\nonumber \\&&
\left.+s \left(-m_{\pi ^+}^4+12 s m_{\pi ^+}^2+9 s^2\right)\right]-3 m_{\pi }^2 \left[\left(7 s-6 m_{\eta }^2\right) m_{\pi ^+}^4\right.~\nonumber \\&&
\left.\left.\left.-4 s^2 m_{\pi ^+}^2+9 s^3+2 m_{k^+}^2 \left(3 m_{\pi ^+}^4+s m_{\pi ^+}^2+s^2\right)\right]\right\}\right\} J_{k \pi ^+}(s)~\nonumber \\&&
+\left\{18 m_{k^+}^8-18 m_{\eta }^2 m_{k^+}^6-54 s m_{k^+}^6+27 s^2 m_{k^+}^4+54 s m_{\eta }^2 m_{k^+}^4\right.~\nonumber \\&&
-24 s m_{\pi ^+}^2 m_{k^+}^4-9 s u m_{k^+}^4-30 s^3 m_{k^+}^2-18 s^2 m_{\eta }^2 m_{k^+}^2~\nonumber \\&&
+20 s^2 m_{\pi ^+}^2 m_{k^+}^2+18 s m_{\eta }^2 m_{\pi ^+}^2 m_{k^+}^2+6 s^2 t m_{k^+}^2~\nonumber \\&&
+24 s^2 u m_{k^+}^2+9 s^4+8 s^2 m_k^4+8 s^2 m_{\pi ^+}^4-18 s^3 m_{\pi ^+}^2~\nonumber \\&&
+6 s^2 t m_{\pi ^+}^2+6 s^2 u m_{\pi ^+}^2-9 s^3 u+18 m_{\pi }^6 \left(m_{\eta }^2-m_{k^+}^2\right)~\nonumber \\&&
+3 m_{\pi }^4 \left[18 m_{k^+}^4+6 s m_{k^+}^2-6 m_{\eta }^2 \left(3 m_{k^+}^2+s\right)+s \left(-2 m_{\pi ^+}^2+s-3 u\right)\right]~\nonumber \\&&
+2 s m_k^2 \left\{3 m_{\pi }^4+\left[9 m_{\eta }^2-5 \left(3 m_{k^+}^2+s\right)\right] m_{\pi }^2-3 m_{\eta }^2 \left(3 m_{k^+}^2+s\right)\right.~\nonumber \\&&
\left.+4 \left(3 m_{k^+}^4-4 s m_{k^+}^2+3 s^2-2 s m_{\pi ^+}^2\right)\right\}+2 m_{\pi }^2 \left\{-27 m_{k^+}^6\right.~\nonumber \\&&
+18 s m_{k^+}^4+3 s \left(5 m_{\pi ^+}^2-8 s+3 u\right) m_{k^+}^2+s^2 \left(2 m_{\pi ^+}^2+9 u\right)~\nonumber \\&&
\left.\left.+3 m_{\eta }^2 \left[9 m_{k^+}^4-6 s m_{k^+}^2+s \left(2 s-3 m_{\pi ^+}^2\right)\right]\right\}\right\} J_{\pi  k^+}(s)~\nonumber \\&&
+\frac{1}{2}\left\{9 \left(12 m_{\pi }^2-12 m_{k^+}^2+7 s\right) m_{\eta }^6-3 \left[-108 m_{k^+}^4-20 s m_{k^+}^2\right.\right.~\nonumber \\&&
\left.+3 m_{\pi }^2 \left(36 m_{k^+}^2+5 s\right)+3 s \left(4 m_{\pi ^+}^2+5 s+7 t+13 u\right)\right] m_{\eta }^4~\nonumber \\&&
+6 \left\{-54 m_{k^+}^6-46 s m_{k^+}^4+s \left(6 m_{\pi ^+}^2-21 s+29 t+47 u\right) m_{k^+}^2\right.~\nonumber \\&&
\left.+2 s^2 \left(m_{\pi ^+}^2+9 u\right)+m_{\pi }^2 \left[54 m_{k^+}^4-41 s m_{k^+}^2+6 s \left(m_{\pi ^+}^2+2 s\right)\right]\right\} m_{\eta
}^2~\nonumber \\&&
-16 s^2 m_k^4+4 s m_k^2 \left[9 m_{\eta }^4-3 \left(3 m_{k^+}^2+s\right) m_{\eta }^2+3 m_{\pi }^2 \left(-3 m_{\eta }^2+3 m_{k^+}^2+s\right)\right.~\nonumber \\&&
\left.+8 s \left(m_{\pi ^+}^2-m_{k^+}^2\right)\right]+2\left\{54 m_{k^+}^8-18 \left(3 m_{\pi }^2+5 s\right) m_{k^+}^6\right.~\nonumber \\&&
+3 s \left(30 m_{\pi }^2+23 s-9 u\right) m_{k^+}^4-2 s \left[\left(9 m_{\pi ^+}^2+30 s\right) m_{\pi }^2\right.~\nonumber \\&&
\left.+s \left(-8 m_{\pi ^+}^2+18 s-27 u\right)\right]m_{k^+}^2~\nonumber \\&&
\left.\left.\left.+s^2 \left[-8 m_{\pi ^+}^4-6 m_{\pi }^2 m_{\pi ^+}^2+27 s (s-u)\right]\right\}\right\}J_{\eta k^+}(s)+[s\longleftrightarrow
u]\right\} \ ;
\end{eqnarray}

\begin{eqnarray}
T_{K^{0}\pi^{0} \rightarrow K^{0}\eta}^{H}(s,t,u)&=&
\frac{2 m_{k^+}^2-2 m_{\pi ^+}^2+3 \left(2 m_k^2+m_{\pi }^2+m_{\eta }^2-3 t\right)}{12 \sqrt{3} f_{\pi }^2}~\nonumber \\&&
+\frac{1}{512 \sqrt{3} \pi ^2 s u f_{\pi }^4}\left\{4 \left(2 m_k^2+m_{\pi }^2+m_{\eta }^2-t\right) m_k^6\right.~\nonumber \\&&
-\left[4 s u+\left(2 m_k^2+m_{\pi }^2+m_{\eta }^2-t\right) \left(3 m_{\pi }^2+m_{\eta }^2-2 m_{k^+}^2-2m_{\pi ^+}^2\right)\right] m_k^4~\nonumber \\&&
-\left\{\left(2 m_k^2+m_{\pi }^2+m_{\eta }^2-t\right) m_{\pi }^4+2 \left[\left(2 m_k^2+m_{\pi }^2+m_{\eta }^2-t\right) \left(m_{k^+}^2+m_{\pi
^+}^2\right)\right.\right.~\nonumber \\&&
-2 s u] m_{\pi }^2+4 s u \left(2 m_k^2+m_{\pi }^2+m_{\eta }^2-3 t\right)~\nonumber \\&&
\left.+m_{\eta }^2 \left[\left(2 m_k^2+m_{\pi }^2+m_{\eta }^2-t\right)\left(3 m_{\eta }^2+2 m_{k^+}^2+2m_{\pi ^+}^2\right)-4 s u\right]\right\}
m_k^2~\nonumber \\&&
+m_{\pi }^4 m_{\eta }^2 \left(2 m_k^2+m_{\pi }^2+m_{\eta }^2-t\right)+s u \{-2 [s (t-2 u)+t u]~\nonumber \\&&
\left.-\left(2 m_k^2+m_{\pi }^2+m_{\eta }^2-3 t\right) \left(3 m_{\eta }^2+2 m_{k^+}^2+2m_{\pi ^+}^2\right)\right\}~\nonumber \\&&
+m_{\pi }^2 \left\{m_{\eta }^2 \left[\left(2 m_k^2+m_{\pi }^2+m_{\eta }^2-t\right) \left(3 m_{\eta }^2+2 m_{k^+}^2+2m_{\pi ^+}^2\right)-4 s
u\right]\right.~\nonumber \\&&
\left.\left.-s u \left(2 m_k^2+m_{\pi }^2+m_{\eta }^2-3 t\right)\right\}\right\}~\nonumber \\&&
+\frac{L_3\left[-2 m_k^4+6 \left(-m_{\pi }^2-m_{\eta }^2+t\right) m_k^2-m_{\pi }^4-m_{\eta }^4+s^2-2 t^2+u^2+3 t m_{\pi }^2+3 t m_{\eta }^2\right]}{\sqrt{3}
f_{\pi }^4}~\nonumber \\&&
+\frac{32 L_6 \left(m_{k^+}^2-m_k^2\right) m_k^2}{\sqrt{3} f_{\pi }^4}+\frac{16 L_8 \left(m_{k^+}^2-m_{\pi ^+}^2\right) \left(4 m_k^2-2 m_{k^+}^2+m_{\pi
^+}^2\right)}{3 \sqrt{3} f_{\pi }^4}~\nonumber \\&&
+\frac{32 L_7 \left[3 m_k^4+\left(m_{k^+}^2-4 m_{\pi ^+}^2\right) m_k^2-2 m_{k^+}^4-m_{\pi ^+}^4+3 m_{k^+}^2 m_{\pi ^+}^2\right]}{3 \sqrt{3}
f_{\pi }^4}~\nonumber \\&&
-\frac{1}{3 \sqrt{3} f_{\pi }^4}2 L_4\left[-30 m_k^4+\left(-9 m_{\pi }^2-3 m_{\eta }^2+28 m_{k^+}^2+8 m_{\pi ^+}^2+21 t\right) m_k^2\right.~\nonumber \\&&
-3 m_{\pi }^4+2 m_{k^+}^4-2 m_{\pi ^+}^4+3 m_{\eta }^2 m_{k^+}^2-21 t m_{k^+}^2+3 m_{\eta }^2 m_{\pi ^+}^2-9 t m_{\pi ^+}^2~\nonumber \\&&
\left.+m_{\pi }^2 \left(-3 m_{\eta }^2+m_{k^+}^2+5 m_{\pi ^+}^2+9 t\right)\right]+\frac{1}{9 \sqrt{3} f_{\pi }^4}L_5 \left[-12 m_k^4\right.~\nonumber \\&&
+2 \left(42 m_{\pi }^2-3 m_{\eta }^2-44 m_{k^+}^2+26 m_{\pi ^+}^2+9 t\right) m_k^2+45 m_{\pi }^4+8 m_{k^+}^4~\nonumber \\&&
-8 m_{\pi ^+}^4-9 s m_{\eta }^2+18 t m_{\eta }^2-9 u m_{\eta }^2-12 m_{\eta }^2 m_{k^+}^2+6 s m_{k^+}^2~\nonumber \\&&
\left.+42 t m_{k^+}^2+6 u m_{k^+}^2-6 m_{\eta }^2 m_{\pi ^+}^2-9 m_{\pi }^2 \left(-5 m_{\eta }^2+2 m_{\pi ^+}^2+15 t\right)\right]~\nonumber \\&&
+\frac{1}{480 \sqrt{3} s^2 u^2 f_{\pi }^4}\left\{60 \left(s^2-4 u s+u^2\right) m_k^6-20 \left[6 \left(s^2+u s+u^2\right) m_{\pi }^2\right.\right.~\nonumber \\&&
\left.+3 m_{\eta }^2 \left(2 m_k^2+m_{\pi }^2+m_{\eta }^2-t\right){}^2-2 s u \left(4 m_{k^+}^2+3 t\right)\right] m_k^4~\nonumber \\&&
-2 \left\{-30 \left(s^2+u^2\right) m_{\pi }^4+5 \left[s u \left(-4 m_{k^+}^2+3 s+6 u\right)-12 (s+u)^2 m_{\eta }^2\right] m_{\pi }^2\right.~\nonumber \\&&
+s u\left[-60 m_{\eta }^4+20 \left(-6 m_k^2+m_{k^+}^2-3 m_{\pi ^+}^2+3 t\right) m_{\eta }^2+15 s^2\right.~\nonumber \\&&
\left.\left.+15 u^2+40 t m_{k^+}^2+40 s m_{\pi ^+}^2+40 u m_{\pi ^+}^2+974 s u\right]\right\}m_k^2~\nonumber \\&&
+5 m_{\pi }^4 \left[s u \left(3 s-4 m_{k^+}^2\right)-12 \left(s^2-u s+u^2\right) m_{\eta }^2\right]~\nonumber \\&&
+4 s u \left[-15 \left(m_{k^+}^2-m_{\pi ^+}^2\right) m_{\eta }^4-3 \left(-5 t m_{k^+}^2+5 t m_{\pi ^+}^2+89 s u\right) m_{\eta }^2\right.~\nonumber \\&&
\left.+s u \left(-172 m_{k^+}^2+192 m_{\pi ^+}^2+765 t\right)\right]+s u m_{\pi }^2\left\{60 m_{\eta }^4\right.~\nonumber \\&&
+5 \left[3 \left(4 m_{\pi ^+}^2+s-4 t\right)-16 m_{k^+}^2\right] m_{\eta }^2+15 s^2+30 u^2~\nonumber \\&&
\left.\left.+20 t m_{k^+}^2+20 s m_{\pi ^+}^2+20 u m_{\pi ^+}^2-15 s t-1039 s u\right\}\right\}\mu _{\pi }~\nonumber \\&&
+\frac{1}{720 \sqrt{3} s^2 u^2 f_{\pi }^4}\left\{-30 \left[-2 \left(3 s^2+7 u s+3 u^2\right) m_{k^+}^2\right.\right.~\nonumber \\&&
\left.+\left(6 s^2-2 u s+6 u^2\right) m_{\pi ^+}^2+s (5 s-7 u) u\right] m_k^4~\nonumber \\&&
+5 \left\{3 \left[-4 \left(3 s^2-4 u s+3 u^2\right) m_{k^+}^2+4 \left(3 s^2+4 u s+3 u^2\right) m_{\pi ^+}^2\right.\right.~\nonumber \\&&
+s u (7 u-5 s)]m_{\pi }^2+s u \left(27 s^2+27 t s+74 u s-9 u^2\right.~\nonumber \\&&
\left.-42 t m_{k^+}^2-6 t m_{\pi ^+}^2-9 t u\right)-3 m_{\eta }^2 \left[4 \left(3 s^2+4 u s+3 u^2\right) m_{k^+}^2\right.~\nonumber \\&&
\left.\left.-4 \left(3 s^2-4 u s+3 u^2\right) m_{\pi ^+}^2+3 s u (7 s+3 u)\right]\right\}m_k^2~\nonumber \\&&
+15 s u m_{\pi }^4 \left(m_{k^+}^2+7 m_{\pi ^+}^2\right)+3 m_{\pi }^2 \left\{32 s^2 u^2+\left[10 \left(6 s^2-7 u s+6 u^2\right) m_{\eta }^2\right.\right.~\nonumber \\&&
\left.-5 s t u] m_{k^+}^2-5 \left[2 \left(6 s^2+u s+6 u^2\right) m_{\eta }^2+7 s t u\right] m_{\pi ^+}^2\right\}~\nonumber \\&&
+s u \left\{-45 \left(5 m_{k^+}^2+3 m_{\pi ^+}^2\right) m_{\eta }^4+9 \left(25 t m_{k^+}^2+15 t m_{\pi ^+}^2+64 s u\right) m_{\eta }^2\right.~\nonumber \\&&
+36 s u^2+\left[-45 s^2+(45 t+62 u) s+45 (t-u) u\right] m_{k^+}^2~\nonumber \\&&
+45 s^2 m_{\pi ^+}^2+45 u^2 m_{\pi ^+}^2-45 s t m_{\pi ^+}^2-318 s u m_{\pi ^+}^2~\nonumber \\&&
\left.\left.-45 t u m_{\pi ^+}^2+36 s^2 u-504 s t u\right\}\right\}\mu _{\pi ^+}~\nonumber \\&&
+\frac{1}{480 \sqrt{3} s^2 u^2 f_{\pi }^4}\left\{-240 \left(s^2-4 u s+u^2\right) m_k^6+20 \left[3 \left(5 s^2+8 u s+5 u^2\right) m_{\pi }^2\right.\right.~\nonumber \\&&
\left.+3 \left(7 s^2+8 u s+7 u^2\right) m_{\eta }^2-8 s u \left(m_{k^+}^2-m_{\pi ^+}^2+3 t\right)\right] m_k^4~\nonumber \\&&
+2 \left\{-30 \left(s^2+u^2\right) m_{\pi }^4-5 \left[48 \left(s^2+u^2\right) m_{\eta }^2+s u \left(16 m_{k^+}^2\right.\right.\right.~\nonumber \\&&
\left.\left.-16 m_{\pi ^+}^2+33 s+21 u\right)\right] m_{\pi }^2-90 \left(s^2+u^2\right) m_{\eta }^4~\nonumber \\&&
-5 s u m_{\eta }^2 \left(-16 m_{k^+}^2+16 m_{\pi ^+}^2+15 s+27 u\right)~\nonumber \\&&
\left.+4 s u \left(15 s^2+14 u s+15 u^2+10 t m_{k^+}^2-10 t m_{\pi ^+}^2\right)\right\}m_k^2~\nonumber \\&&
-5 m_{\pi }^4\left[s u \left(8 m_{k^+}^2-8 m_{\pi ^+}^2+3 s-9 u\right)-12 \left(s^2+u^2\right) m_{\eta }^2\right]~\nonumber \\&&
-s u \left[-15 \left(8 m_{k^+}^2-8 m_{\pi ^+}^2+s-3 u\right) m_{\eta }^4+\left(105 s^2+15 t s\right.\right.~\nonumber \\&&
\left.-274 u s+45 u^2+120 t m_{k^+}^2-120 t m_{\pi ^+}^2-45 t u\right) m_{\eta }^2~\nonumber \\&&
\left.+48 s u \left(-2 m_{k^+}^2+2 m_{\pi ^+}^2+15 t\right)\right]+m_{\pi }^2\left[180 \left(s^2+u^2\right) m_{\eta }^4\right.~\nonumber \\&&
+80 s u \left(m_{k^+}^2-m_{\pi ^+}^2\right) m_{\eta }^2+s u \left(-15 s^2+15 t s+254 u s\right.~\nonumber \\&&
\left.\left.\left.-75 u^2+40 t m_{k^+}^2-40 t m_{\pi ^+}^2-45 t u\right)\right]\right\}\mu _k~\nonumber \\&&
+\frac{1}{144 \sqrt{3} s^2 u^2 f_{\pi }^4}\left\{6 \left[-2 \left(3 s^2+7 u s+3 u^2\right) m_{k^+}^2+\left(6 s^2-2 u s+6 u^2\right) m_{\pi ^+}^2\right.\right.~\nonumber \\&&
+s (11 s-u) u]m_k^4+\left\{-3 \left[-4 \left(3 s^2-4 u s+3 u^2\right) m_{k^+}^2\right.\right.~\nonumber \\&&
\left.+4 \left(3 s^2+4 u s+3 u^2\right) m_{\pi ^+}^2+s u (7 s+19 u)\right] m_{\pi }^2~\nonumber \\&&
+s u \left(-27 s^2-27 t s+190 u s+9 u^2+42 t m_{k^+}^2+6 t m_{\pi ^+}^2+9 t u\right)~\nonumber \\&&
+3 m_{\eta }^2 \left[4 \left(3 s^2+4 u s+3 u^2\right) m_{k^+}^2-4 \left(3 s^2-4 u s+3 u^2\right) m_{\pi ^+}^2\right.~\nonumber \\&&
+3 s (3 s-u) u]\}m_k^2-3 s u m_{\pi }^4 \left(m_{k^+}^2+7 m_{\pi ^+}^2\right)~\nonumber \\&&
+s u \left\{9 \left(5 m_{k^+}^2+3 m_{\pi ^+}^2\right) m_{\eta }^4-3 \left(15 t m_{k^+}^2+9 t m_{\pi ^+}^2+4 s u\right) m_{\eta }^2\right.~\nonumber \\&&
+\left[9 s^2+(2 u-9 t) s+9 u (u-t)\right] m_{k^+}^2-3 \left[\left(3 s^2-3 t s\right.\right.~\nonumber \\&&
\left.\left.\left.+22 u s+3 u^2-3 t u\right) m_{\pi ^+}^2+36 s t u\right]\right\}~\nonumber \\&&
+3 m_{\pi }^2 \left\{2 \left[\left(-6 s^2+7 u s-6 u^2\right) m_{k^+}^2+\left(6 s^2+u s+6 u^2\right) m_{\pi ^+}^2\right.\right.~\nonumber \\&&
\left.\left.\left.+6 s u \left(2 m_k^2+m_{\pi }^2+m_{\eta }^2-t\right)\right] m_{\eta }^2+s u \left(t m_{k^+}^2+7 t m_{\pi ^+}^2+28 s u\right)\right\}\right\}\mu
_{k^+}~\nonumber \\&&
+\frac{1}{480 \sqrt{3} s^2 u^2 f_{\pi }^4}\left\{60 \left(3 s^2-4 u s+3 u^2\right) m_k^6-60 \left[\left(3 s^2+2 u s+3 u^2\right) m_{\pi }^2\right.\right.~\nonumber \\&&
\left.+2 \left(3 s^2+u s+3 u^2\right) m_{\eta }^2-2 s t u\right] m_k^4-2 \left\{-90 \left(s^2+u^2\right) m_{\eta }^4\right.~\nonumber \\&&
+15 s u \left(4 m_{k^+}^2-4 m_{\pi ^+}^2+6 s+3 u\right) m_{\eta }^2+s u \left(45 s^2-238 u s+45 u^2\right)~\nonumber \\&&
-60 m_{\pi }^2 \left[3 \left(s^2+u s+u^2\right) m_{\eta }^2+s u \left(2 m_k^2+m_{\pi }^2+m_{\eta }^2+m_{k^+}^2\right.\right.~\nonumber \\&&
\left.\left.\left.-m_{\pi ^+}^2-t\right)\right]\right\}m_k^2+60 s u m_{\pi }^4 \left(3 m_{\eta }^2+m_{k^+}^2-m_{\pi ^+}^2\right)~\nonumber \\&&
+s u \left[15 \left(-4 m_{k^+}^2+4 m_{\pi ^+}^2+3 u\right) m_{\eta }^4+3 \left(30 s^2+41 u s+15 u^2\right.\right.~\nonumber \\&&
\left.+20 t m_{k^+}^2-20 t m_{\pi ^+}^2-15 t u\right) m_{\eta }^2~\nonumber \\&&
\left.-20 s u \left(-8 m_{k^+}^2+12 m_{\pi ^+}^2+27 t\right)\right]+3 m_{\pi }^2 \left[-60 \left(s^2-u s+u^2\right) m_{\eta }^4\right.~\nonumber \\&&
\left.\left.+15 s u (u-4 t) m_{\eta }^2+4 s u \left(-5 t m_{k^+}^2+5 t m_{\pi ^+}^2+3 s u\right)\right]\right\}\mu _{\eta }~\nonumber \\&&
+\frac{\left(m_k^2-m_{k^+}^2\right) \left(2 m_k^2-2 m_{\pi }^2-2 m_{k^+}^2+2 m_{\pi ^+}^2+3 t\right) J_{\pi \pi }(t)}{24 \sqrt{3} f_{\pi }^4}~\nonumber \\&&
+\frac{t \left(-2 m_{k^+}^2+2 m_{\pi ^+}^2-3 \left(2 m_k^2+m_{\pi }^2+m_{\eta }^2-3 t\right)\right) J_{k k}(t)}{24 \sqrt{3} f_{\pi }^4}~\nonumber \\&&
-\frac{t \left(-2 m_k^2-3 m_{\pi }^2-3 m_{\eta }^2-6 m_{k^+}^2+2 m_{\pi ^+}^2+9 t\right) J_{k^+ k^+}(t)}{48 \sqrt{3} f_{\pi }^4}~\nonumber \\&&
+\frac{m_{\pi ^+}^2 \left(-6 m_k^2-3 m_{\pi }^2-3 m_{\eta }^2-2 m_{k^+}^2+2 m_{\pi ^+}^2+9 t\right) J_{\pi \eta }(t)}{36 \sqrt{3} f_{\pi }^4}~\nonumber \\&&
-\frac{\left(m_k^2-m_{k^+}^2\right) \left(2 m_k^2+6 m_{\eta }^2-2 m_{k^+}^2+2 m_{\pi ^+}^2-9 t\right) J_{\eta  \eta }(t)}{72 \sqrt{3} f_{\pi
}^4}~\nonumber \\&&
+\frac{t \left(m_k^2-m_{k^+}^2\right) J_{\pi ^+ \pi ^+}(t)}{12 \sqrt{3} f_{\pi }^4}~\nonumber \\&&
+\frac{1}{288 \sqrt{3} s^2 f_{\pi }^4}\left\{\left\{18 m_k^8-18 \left(3 m_{\pi }^2+m_{\eta }^2+3 s\right) m_k^6+\left[54 m_{\pi }^4+18 \left(3
m_{\eta }^2+2 s\right) m_{\pi }^2\right.\right.\right.~\nonumber \\&&
\left.-3 s \left(-18 m_{\eta }^2-8 m_{k^+}^2+8 m_{\pi ^+}^2+11 s+3 u\right)\right]m_k^4~\nonumber \\&&
+2 \left\{-9 m_{\pi }^6+9 \left(s-3 m_{\eta }^2\right) m_{\pi }^4-3 s \left(6 m_{\eta }^2+5 m_{k^+}^2-5 m_{\pi ^+}^2+13 s-3 u\right) m_{\pi
}^2\right.~\nonumber \\&&
+s \left[s \left(-16 m_{k^+}^2+16 m_{\pi ^+}^2+18 t+27 u\right)\right.~\nonumber \\&&
\left.\left.-3 m_{\eta }^2 \left(3 m_{k^+}^2-3 m_{\pi ^+}^2+8 s\right)\right]\right\} m_k^2+18 m_{\pi }^6 m_{\eta }^2~\nonumber \\&&
+3 s m_{\pi }^4 \left(-6 m_{\eta }^2+2 m_{k^+}^2-2 m_{\pi ^+}^2+s-3 u\right)+s^2 \left[8 m_{k^+}^4\right.~\nonumber \\&&
+2 \left(-3 m_{\eta }^2-8 m_{\pi ^+}^2+12 s\right) m_{k^+}^2+8 m_{\pi ^+}^4+6 \left(m_{\eta }^2-4 s\right) m_{\pi ^+}^2~\nonumber \\&&
+9 s (s-u)]+2 s m_{\pi }^2 \left[\left(9 m_{k^+}^2-9 m_{\pi ^+}^2+6 s\right) m_{\eta }^2\right.~\nonumber \\&&
\left.\left.+s \left(-5 m_{k^+}^2+5 m_{\pi ^+}^2+9 u\right)\right]\right\} J_{k \pi }(s)~\nonumber \\&&
+\left\{54 m_k^8-18 \left(3 m_{\pi }^2+9 m_{\eta }^2+5 s\right) m_k^6+3 \left[54 m_{\eta }^4-88 s m_{\eta }^2\right.\right.~\nonumber \\&&
\left.+s (23 s-9 u)+6 m_{\pi }^2 \left(9 m_{\eta }^2+5 s\right)\right] m_k^4-2 \left\{27 m_{\eta }^6+39 s m_{\eta }^4\right.~\nonumber \\&&
-3 s \left(-3 m_{k^+}^2+3 m_{\pi ^+}^2+25 t+34 u\right) m_{\eta }^2+s^2 \left(8 m_{k^+}^2-8 m_{\pi ^+}^2\right.~\nonumber \\&&
+18 s-27 u)+m_{\pi }^2\left[81 m_{\eta }^4+93 s m_{\eta }^2+3 s \left(-3 m_{k^+}^2+3 m_{\pi ^+}^2\right.\right.~\nonumber \\&&
+10 s)]\} m_k^2+9 \left(6 m_{\pi }^2+s\right) m_{\eta }^6-9 s m_{\eta }^4 \left(5 m_{\pi }^2-2 m_{k^+}^2\right.~\nonumber \\&&
\left.+2 m_{\pi ^+}^2+t+4 u\right)+s^2 \left[-8 m_{k^+}^4+16 m_{\pi ^+}^2 m_{k^+}^2-8 m_{\pi ^+}^4+27 s^2\right.~\nonumber \\&&
\left.-27 s u+6 m_{\pi }^2 \left(m_{k^+}^2-m_{\pi ^+}^2\right)\right]+6 s m_{\eta }^2 \left[3 \left(-m_{k^+}^2+m_{\pi ^+}^2+2 s\right) m_{\pi
}^2\right.~\nonumber \\&&
\left.\left.+s \left(-m_{k^+}^2+m_{\pi ^+}^2+9 u\right)\right]\right\} J_{k \eta }(s)-2\left\{-2 \left[9 m_{k^+}^4-3 \left(6 m_{\pi ^+}^2+s\right)
m_{k^+}^2\right.\right.~\nonumber \\&&
\left.+9 m_{\pi ^+}^4+2 s^2-6 s m_{\pi ^+}^2\right]m_k^4+2 \left\{3 \left[3 m_{k^+}^4-2 \left(3 m_{\pi ^+}^2+2 s\right) m_{k^+}^2\right.\right.~\nonumber \\&&
\left.+3 m_{\pi ^+}^4+s^2+s m_{\pi ^+}^2\right] m_{\pi }^2+3 m_{\eta }^2 \left(3 m_{k^+}^4-6 m_{\pi ^+}^2 m_{k^+}^2\right.~\nonumber \\&&
\left.+3 m_{\pi ^+}^4+s^2-3 s m_{\pi ^+}^2\right)+2 s\left[-4 m_{k^+}^4+\left(14 m_{\pi ^+}^2+3 s\right) m_{k^+}^2\right.~\nonumber \\&&
\left.\left.+6 m_{\pi ^+}^4-10 s m_{\pi ^+}^2\right]\right\} m_k^2+m_{\pi }^2 \left\{\left(s-18 m_{\eta }^2\right) m_{k^+}^4\right.~\nonumber \\&&
+2 \left[9 \left(2 m_{\pi ^+}^2+s\right) m_{\eta }^2+s \left(5 m_{\pi ^+}^2+3 s\right)\right]m_{k^+}^2-3 \left[\left(6 m_{\eta }^2-7 s\right)
m_{\pi ^+}^4\right.~\nonumber \\&&
\left.\left.+4 s^2 m_{\pi ^+}^2+3 s^3\right]\right\}+s \left\{(2 u-7 t) m_{k^+}^4-2 \left[(5 t+14 u) m_{\pi ^+}^2\right.\right.~\nonumber \\&&
+9 s (s-t)]m_{k^+}^2+3 \left[-(5 t+2 u) m_{\pi ^+}^4+6 s (t-2 s) m_{\pi ^+}^2\right.~\nonumber \\&&
\left.+3 s^2 (4 s-t)\right]+m_{\eta }^2 \left[25 m_{k^+}^4+\left(10 m_{\pi ^+}^2-42 s\right) m_{k^+}^2\right.~\nonumber \\&&
\left.\left.\left.\left.-3 \left(m_{\pi ^+}^4-12 s m_{\pi ^+}^2+3 s^2\right)\right]\right\}\right\}J_{k^+ \pi ^+}(s)+[s\longleftrightarrow u]\right\} \ ;
\end{eqnarray}

\begin{eqnarray}
T_{K^{+}\eta \rightarrow K^{+}\eta}^{H}(s,t,u)&=&
\frac{ 2 m_k^2-6 m_{\eta }^2-2 m_{k^+}^2-2 m_{\pi ^+}^2+9 t }{12f_{\pi }^2}+\frac{1}{512\pi ^2 s u f_{\pi }^4} \left\{-3 s m_{\eta
}^6-3 u m_{\eta }^6\right.~\nonumber \\&&
+2 s m_{k^+}^2 m_{\eta }^4+2 u m_{k^+}^2 m_{\eta }^4+4 s u m_{\eta }^4+5 s m_{k^+}^4 m_{\eta }^2+5 u m_{k^+}^4 m_{\eta }^2~\nonumber \\&&
+3 s u^2 m_{\eta }^2-8 s u m_{k^+}^2 m_{\eta }^2+3 s^2 u m_{\eta }^2-6 s t u m_{\eta }^2-4 s m_{k^+}^6~\nonumber \\&&
-4 u m_{k^+}^6+4 s u m_{k^+}^4+4 s u^2 m_{k^+}^2+4 s^2 u m_{k^+}^2-8 s t u m_{k^+}^2~\nonumber \\&&
+2 \left[s u \left(2 m_{\eta }^2+2 m_{k^+}^2-3 t\right)-\left(m_{\eta }^2-m_{k^+}^2\right){}^2 \left(2 m_{\eta }^2+2 m_{k^+}^2-t\right)\right]m_{\pi
^+}^2~\nonumber \\&&
+2 s u [s (t-2 u)+t u]+2 m_k^2 \left[s u \left(2 m_{\eta }^2+2 m_{k^+}^2-3 t\right)\right.~\nonumber \\&&
\left.-\left(m_{\eta }^2-m_{k^+}^2\right){}^2 \left(2 m_{\eta }^2+2 m_{k^+}^2-t\right)\right]+m_{\pi }^2 \left[s u \left(2 m_{\eta }^2+2 m_{k^+}^2-3
t\right)\right.~\nonumber \\&&
\left.\left.-\left(m_{\eta }^2-m_{k^+}^2\right){}^2 \left(2 m_{\eta }^2+2 m_{k^+}^2-t\right)\right]\right\}+\frac{8 L_1 \left(t-2 m_{\eta }^2\right)
\left(t-2 m_{k^+}^2\right)}{f_{\pi }^4}~\nonumber \\&&
+\frac{L_3 \left(-2 m_{\eta }^4-18 \left(t-2 m_{k^+}^2\right) m_{\eta }^2-2 m_{k^+}^4+s^2+10 t^2+u^2-18 t m_{k^+}^2\right)}{3 f_{\pi }^4}~\nonumber \\&&
+\frac{4 L_2 \left(\left(m_{\eta }^2+m_{k^+}^2-s\right){}^2+\left(m_{\eta }^2+m_{k^+}^2-u\right){}^2\right)}{f_{\pi }^4}+\frac{32 L_6 \left(2
\left(m_k^2+m_{k^+}^2\right)-m_{\pi ^+}^2\right) m_{k^+}^2}{3 f_{\pi }^4}~\nonumber \\&&
+\frac{32 L_7 \left(m_k^2+m_{k^+}^2-2 m_{\pi ^+}^2\right) \left(m_k^2+m_{k^+}^2-m_{\pi ^+}^2\right)}{3 f_{\pi }^4}~\nonumber \\&&
+\frac{16 L_8 \left(m_k^4+4 m_{k^+}^2 m_k^2+3 m_{k^+}^4+2 m_{\pi ^+}^4-\left(3 m_k^2+4 m_{k^+}^2\right) m_{\pi ^+}^2\right)}{3 f_{\pi }^4}~\nonumber \\&&
+\frac{1}{3 f_{\pi }^4}2 L_4 \left[2 m_k^4+\left(2 m_{\pi }^2-6 m_{\eta }^2-20 m_{k^+}^2-4 m_{\pi ^+}^2+17 t\right) m_k^2\right.~\nonumber \\&&
-14 m_{k^+}^4+2 m_{\pi ^+}^4-18 m_{\eta }^2 m_{k^+}^2+11 t m_{k^+}^2+6 m_{\eta }^2 m_{\pi ^+}^2~\nonumber \\&&
\left.+12 m_{k^+}^2 m_{\pi ^+}^2-13 t m_{\pi ^+}^2+m_{\pi }^2 \left(-6 m_{\eta }^2-2 m_{k^+}^2-2 m_{\pi ^+}^2+9 t\right)\right]~\nonumber \\&&
-\frac{1}{9 f_{\pi }^4}2 L_5 \left[4 m_k^4+6 \left(-3 m_{\pi }^2-3 m_{\eta }^2+3 m_{k^+}^2-2 m_{\pi ^+}^2+6 t\right) m_k^2+14 m_{k^+}^4\right.~\nonumber \\&&
+8 m_{\pi ^+}^4-9 s m_{\eta }^2+18 t m_{\eta }^2-9 u m_{\eta }^2+18 m_{\eta }^2 m_{k^+}^2~\nonumber \\&&
+6 s m_{k^+}^2-3 t m_{k^+}^2+6 u m_{k^+}^2-6 m_{\eta }^2 m_{\pi ^+}^2-26 m_{k^+}^2 m_{\pi ^+}^2~\nonumber \\&&
\left.-9 m_{\pi }^2 \left(-6 m_{\eta }^2-2 m_{k^+}^2-2 m_{\pi ^+}^2+9 t\right)\right]~\nonumber \\&&
+\frac{1}{480 s^2 u^2 f_{\pi }^4}\left\{-60 s^2 m_{k^+}^6-60 u^2 m_{k^+}^6+120 s^2 m_{\eta }^2 m_{k^+}^4\right.~\nonumber \\&&
+120 u^2 m_{\eta }^2 m_{k^+}^4-60 s^2 m_{\eta }^4 m_{k^+}^2-60 u^2 m_{\eta }^4 m_{k^+}^2~\nonumber \\&&
+30 s u^3 m_{k^+}^2-1740 s^2 u^2 m_{k^+}^2+30 s u^2 m_{\eta }^2 m_{k^+}^2~\nonumber \\&&
+30 s^2 u m_{\eta }^2 m_{k^+}^2+40 s u^2 m_{\pi ^+}^2 m_{k^+}^2+40 s^2 u m_{\pi ^+}^2 m_{k^+}^2~\nonumber \\&&
+30 s^3 u m_{k^+}^2-30 s u^2 m_{\eta }^4-30 s^2 u m_{\eta }^4+1284 s^2 u^3~\nonumber \\&&
+1284 s^3 u^2-2496 s^2 t u^2-15 s u^3 m_{\eta }^2+18 s^2 u^2 m_{\eta }^2~\nonumber \\&&
-15 s t u^2 m_{\eta }^2-15 s^3 u m_{\eta }^2-15 s^2 t u m_{\eta }^2+896 s^2 u^2 m_{\pi ^+}^2~\nonumber \\&&
-40 s u^2 m_{\eta }^2 m_{\pi ^+}^2-40 s^2 u m_{\eta }^2 m_{\pi ^+}^2-8 s u m_k^2 \left(-10 m_{\eta }^4+5 t m_{\eta }^2\right.~\nonumber \\&&
\left.+10 m_{k^+}^4-5 t m_{k^+}^2+102 s u\right)+m_{\pi }^2 \left\{60 \left(s^2-u s+u^2\right) m_{\eta }^4\right.~\nonumber \\&&
-30 \left[4 \left(s^2+u^2\right) m_{k^+}^2-s t u\right] m_{\eta }^2+60 \left(s^2+u s+u^2\right) m_{k^+}^4~\nonumber \\&&
\left.\left.-30 s t u m_{k^+}^2+s u \left[-15 s^2+(15 t-14 u) s+15 (t-u) u\right]\right\}\right\} \mu _{\pi }~\nonumber \\&&
+\frac{1}{720 s^2 u^2 f_{\pi }^4}\left\{-180 s u m_{\eta }^6-30 \left[-10 s u m_{k^+}^2-2 \left(3 s^2-7 u s+3 u^2\right) m_{\pi ^+}^2\right.\right.~\nonumber \\&&
+3 s u (s-t+u)] m_{\eta }^4+6 \left\{80 s u m_{k^+}^4-5 \left[2 \left(6 s^2+7 u s+6 u^2\right) m_{\pi ^+}^2\right.\right.~\nonumber \\&&
\left.+s u (3 s+8 t+3 u)]m_{k^+}^2+s u \left(35 t m_{\pi ^+}^2+24 s u\right)\right\}m_{\eta }^2~\nonumber \\&&
-150 s u^2 m_{k^+}^4-150 s^2 u m_{k^+}^4+72 s^2 u^3+72 s^3 u^2+72 s^2 t u^2~\nonumber \\&&
+45 s u^3 m_{k^+}^2-158 s^2 u^2 m_{k^+}^2+45 s t u^2 m_{k^+}^2+45 s^3 u m_{k^+}^2~\nonumber \\&&
+45 s^2 t u m_{k^+}^2+180 s^2 m_{k^+}^4 m_{\pi ^+}^2+180 u^2 m_{k^+}^4 m_{\pi ^+}^2-45 s u^3 m_{\pi ^+}^2~\nonumber \\&&
-114 s^2 u^2 m_{\pi ^+}^2+45 s t u^2 m_{\pi ^+}^2+210 s u^2 m_{k^+}^2 m_{\pi ^+}^2+210 s^2 u m_{k^+}^2 m_{\pi ^+}^2~\nonumber \\&&
-45 s^3 u m_{\pi ^+}^2+45 s^2 t u m_{\pi ^+}^2+5 m_k^2 \left\{-12 \left(3 s^2+u s+3 u^2\right) m_{\eta }^4\right.~\nonumber \\&&
+6 \left[12 \left(s^2+u^2\right) m_{k^+}^2+s t u\right]m_{\eta }^2-12 \left(3 s^2-u s+3 u^2\right) m_{k^+}^4~\nonumber \\&&
\left.\left.-6 s t u m_{k^+}^2+s u \left(9 s^2-9 t s+10 u s+9 u^2-9 t u\right)\right\}\right\} \mu _{\pi ^+}~\nonumber \\&&
-\frac{1}{720 s^2 u^2 f_{\pi }^4}\left\{60 \left[\left(3 s^2-7 u s+3 u^2\right) m_{\pi ^+}^2-4 s u m_{k^+}^2\right]m_{\eta }^4\right.~\nonumber \\&&
+6 \left\{-40 s u m_{k^+}^4-5\left[2 \left(6 s^2+7 u s+6 u^2\right) m_{\pi ^+}^2+s u (3 s-4 t+3 u)\right] m_{k^+}^2\right.~\nonumber \\&&
\left.+s u \left(35 t m_{\pi ^+}^2+108 s u\right)\right\} m_{\eta }^2+30 s u^2 m_{k^+}^4+30 s^2 u m_{k^+}^4~\nonumber \\&&
+504 s^2 u^3+504 s^3 u^2-1836 s^2 t u^2+45 s u^3 m_{k^+}^2-902 s^2 u^2 m_{k^+}^2~\nonumber \\&&
+45 s t u^2 m_{k^+}^2+45 s^3 u m_{k^+}^2+45 s^2 t u m_{k^+}^2+180 s^2 m_{k^+}^4 m_{\pi ^+}^2~\nonumber \\&&
+180 u^2 m_{k^+}^4 m_{\pi ^+}^2-45 s u^3 m_{\pi ^+}^2+622 s^2 u^2 m_{\pi ^+}^2+45 s t u^2 m_{\pi ^+}^2~\nonumber \\&&
+210 s u^2 m_{k^+}^2 m_{\pi ^+}^2+210 s^2 u m_{k^+}^2 m_{\pi ^+}^2-45 s^3 u m_{\pi ^+}^2+45 s^2 t u m_{\pi ^+}^2~\nonumber \\&&
+m_k^2 \left\{-60 \left(3 s^2+u s+3 u^2\right) m_{\eta }^4+30 \left[12 \left(s^2+u^2\right) m_{k^+}^2+s t u\right] m_{\eta }^2\right.~\nonumber \\&&
-60 \left(3 s^2-u s+3 u^2\right) m_{k^+}^4-30 s t u m_{k^+}^2~\nonumber \\&&
\left.\left.+s u \left[45 s^2-(45 t+494 u) s+45 u (u-t)\right]\right\}\right\} \mu _k~\nonumber \\&&
+\frac{1}{1440 s^2 u^2 f_{\pi }^4}\left\{-540 s^2 m_{\eta }^6-540 u^2 m_{\eta }^6-90 s u^2 m_{\eta }^4+1800 s^2 m_{k^+}^2 m_{\eta }^4\right.~\nonumber \\&&
+1800 u^2 m_{k^+}^2 m_{\eta }^4-90 s^2 u m_{\eta }^4-1980 s^2 m_{k^+}^4 m_{\eta }^2-1980 u^2 m_{k^+}^4 m_{\eta }^2~\nonumber \\&&
+315 s u^3 m_{\eta }^2-1602 s^2 u^2 m_{\eta }^2+45 s t u^2 m_{\eta }^2+1530 s u^2 m_{k^+}^2 m_{\eta }^2~\nonumber \\&&
+1530 s^2 u m_{k^+}^2 m_{\eta }^2-240 s u^2 m_{\pi ^+}^2 m_{\eta }^2-240 s^2 u m_{\pi ^+}^2 m_{\eta }^2~\nonumber \\&&
+315 s^3 u m_{\eta }^2+45 s^2 t u m_{\eta }^2+720 s^2 m_{k^+}^6+720 u^2 m_{k^+}^6~\nonumber \\&&
-1440 s u^2 m_{k^+}^4-1440 s^2 u m_{k^+}^4+144 s^2 u^3+144 s^3 u^2~\nonumber \\&&
+1944 s^2 t u^2-360 s u^3 m_{k^+}^2+96 s^2 u^2 m_{k^+}^2-360 s^3 u m_{k^+}^2~\nonumber \\&&
-208 s^2 u^2 m_{\pi ^+}^2+240 s u^2 m_{k^+}^2 m_{\pi ^+}^2+240 s^2 u m_{k^+}^2 m_{\pi ^+}^2~\nonumber \\&&
+16 s u m_k^2 \left(30 m_{\eta }^4-15 t m_{\eta }^2-30 m_{k^+}^4+15 t m_{k^+}^2+13 s u\right)~\nonumber \\&&
+15 m_{\pi }^2 \left\{-12 \left(s^2-u s+u^2\right) m_{\eta }^4+6 \left[4 \left(s^2+u^2\right) m_{k^+}^2-s t u\right]m_{\eta }^2\right.~\nonumber \\&&
-12 \left(s^2+u s+u^2\right) m_{k^+}^4+6 s t u m_{k^+}^2~\nonumber \\&&
\left.\left.+s u \left(3 s^2-3 t s-2 u s+3 u^2-3 t u\right)\right\}\right\} \mu _{k^+}~\nonumber \\&&
-\frac{1}{240 s^2 u^2 f_{\pi }^4}\left\{-90 (s-u)^2 m_{\eta }^6+30 \left[\left(9 s^2+8 u s+9 u^2\right) m_{k^+}^2\right.\right.~\nonumber \\&&
\left.-s u \left(4 m_{\pi ^+}^2+3 t\right)\right] m_{\eta }^4+3 \left[-10 \left(9 s^2-2 u s+9 u^2\right) m_{k^+}^4\right.~\nonumber \\&&
-10 s u \left(4 m_{\pi ^+}^2+t\right) m_{k^+}^2+s u \left(15 s^2-2 u s+15 u^2\right.~\nonumber \\&&
\left.\left.+20 t m_{\pi ^+}^2\right)\right] m_{\eta }^2+90 s^2 m_{k^+}^6+90 u^2 m_{k^+}^6~\nonumber \\&&
-120 s u^2 m_{k^+}^4-120 s^2 u m_{k^+}^4+54 s^2 u^3+54 s^3 u^2~\nonumber \\&&
-216 s^2 t u^2-45 s u^3 m_{k^+}^2+50 s^2 u^2 m_{k^+}^2-45 s^3 u m_{k^+}^2~\nonumber \\&&
+104 s^2 u^2 m_{\pi ^+}^2+60 s u^2 m_{k^+}^2 m_{\pi ^+}^2+60 s^2 u m_{k^+}^2 m_{\pi ^+}^2~\nonumber \\&&
\left.-4 s u m_k^2 \left(-30 m_{\eta }^4+15 t m_{\eta }^2+30 m_{k^+}^4-15 t m_{k^+}^2+16 s u\right)\right\} \mu _{\eta }~\nonumber \\&&
-\frac{m_{\pi ^+}^2 \left(-2 m_k^2-2 m_{\pi }^2+2 m_{k^+}^2+2 m_{\pi ^+}^2+3 t\right) J_{\pi  \pi }(t)}{72 f_{\pi }^4}~\nonumber \\&&
-\frac{t \left(-2 m_k^2-6 m_{\eta }^2+2 m_{k^+}^2-2 m_{\pi ^+}^2+9 t\right) J_{k k}(t)}{48 f_{\pi }^4}~\nonumber \\&&
-\frac{t m_{\pi ^+}^2 J_{\pi ^+ \pi ^+}(t)}{12 f_{\pi }^4}-\frac{t \left(2 m_k^2-6 m_{\eta }^2-2 m_{k^+}^2-2 m_{\pi ^+}^2+9 t\right) J_{k^+ k^+}(t)}{24
f_{\pi }^4}~\nonumber \\&&
-\frac{\left(m_k^2-m_{k^+}^2\right) \left(2 m_k^2+3 m_{\pi }^2+3 m_{\eta }^2+6 m_{k^+}^2-2 m_{\pi ^+}^2-9 t\right) J_{\pi  \eta }(t)}{108 f_{\pi
}^4}~\nonumber \\&&
-\frac{\left(8 m_k^2+8 m_{k^+}^2-7 m_{\pi ^+}^2\right) \left(2 m_k^2-6 m_{\eta }^2-2 m_{k^+}^2-2 m_{\pi ^+}^2+9 t\right) J_{\eta  \eta }(t)}{216
f_{\pi }^4}~\nonumber \\&&
-\frac{1}{288 s^2 f_{\pi }^4}\left\{\frac{1}{3}\left\{54 s^4+39 m_{k^+}^2 s^3-105 m_{\pi ^+}^2 s^3-54 u s^3+22 m_{k^+}^4 s^2\right.\right.~\nonumber \\&&
+49 m_{\pi ^+}^4 s^2+3 t m_{k^+}^2 s^2+3 u m_{k^+}^2 s^2-163 m_{k^+}^2 m_{\pi ^+}^2 s^2~\nonumber \\&&
-33 t m_{\pi ^+}^2 s^2+75 u m_{\pi ^+}^2 s^2+126 m_{k^+}^2 m_{\pi ^+}^4 s+27 t m_{\pi ^+}^4 s~\nonumber \\&&
-27 u m_{\pi ^+}^4 s-90 m_{k^+}^4 m_{\pi ^+}^2 s+27 t m_{k^+}^2 m_{\pi ^+}^2 s~\nonumber \\&&
+27 u m_{k^+}^2 m_{\pi ^+}^2 s+108 m_{k^+}^4 m_{\pi ^+}^4+2 m_k^4\left[54 m_{\eta }^4\right.~\nonumber \\&&
+2 \left(5 s-54 m_{k^+}^2\right) m_{\eta }^2+54 m_{k^+}^4-8 s m_{k^+}^2~\nonumber \\&&
+13 s t-14 s u]+36 m_{\eta }^4 \left(3 m_{\pi ^+}^4-3 s m_{\pi ^+}^2+s^2\right)~\nonumber \\&&
-18 m_{\eta }^2 \left[\left(12 m_{\pi ^+}^4-5 s m_{\pi ^+}^2+3 s^2\right) m_{k^+}^2+s m_{\pi ^+}^2 \left(7 m_{\pi ^+}^2-9 s\right)\right]~\nonumber \\&&
-4 m_k^2 \left\{54 m_{\pi ^+}^2 m_{\eta }^4+2 \left[s \left(24 s-17 m_{\pi ^+}^2\right)-9 m_{k^+}^2 \left(6 m_{\pi ^+}^2+s\right)\right] m_{\eta
}^2\right.~\nonumber \\&&
-4 s m_{k^+}^2 \left(8 s-5 m_{\pi ^+}^2\right)+18 m_{k^+}^4 \left(3 m_{\pi ^+}^2+s\right)~\nonumber \\&&
\left.\left.+s \left[(17 t-10 u) m_{\pi ^+}^2+9 s (t-2 u)\right]\right\}\right\} J_{k \pi ^+}(s)~\nonumber \\&&
+\frac{1}{6}\left\{108 m_{k^+}^8-216 m_{\eta }^2 m_{k^+}^6-108 s m_{k^+}^6+108 m_{\eta }^4 m_{k^+}^4+162 s^2 m_{k^+}^4\right.~\nonumber \\&&
+162 s m_{\eta }^2 m_{k^+}^4-72 s m_{\pi ^+}^2 m_{k^+}^4-54 s u m_{k^+}^4-54 s m_{\eta }^4 m_{k^+}^2~\nonumber \\&&
-36 s^3 m_{k^+}^2-207 s^2 m_{\eta }^2 m_{k^+}^2+27 s t m_{\eta }^2 m_{k^+}^2+27 s u m_{\eta }^2 m_{k^+}^2~\nonumber \\&&
+24 s^2 m_{\pi ^+}^2 m_{k^+}^2+72 s m_{\eta }^2 m_{\pi ^+}^2 m_{k^+}^2-36 s^2 t m_{k^+}^2+72 s^2 u m_{k^+}^2~\nonumber \\&&
+54 s^4+16 s^2 m_k^4+18 s^2 m_{\eta }^4+16 s^2 m_{\pi ^+}^4+9 s^3 m_{\eta }^2+9 s^2 t m_{\eta }^2~\nonumber \\&&
+9 s^2 u m_{\eta }^2-60 s^3 m_{\pi ^+}^2+12 s^2 t m_{\pi ^+}^2+12 s^2 u m_{\pi ^+}^2-54 s^3 u~\nonumber \\&&
+18 m_{\pi }^4\left[6 m_{\eta }^4-2 \left(6 m_{k^+}^2+s\right) m_{\eta }^2+6 m_{k^+}^4+4 s m_{k^+}^2+s (t-2 u)\right]~\nonumber \\&&
+8 s m_k^2 \left[9 m_{k^+}^4-6 s m_{k^+}^2+9 s^2-4 s m_{\pi ^+}^2-3 m_{\eta }^2 \left(3 m_{k^+}^2+s\right)\right.~\nonumber \\&&
\left.-3 m_{\pi }^2 \left(-3 m_{\eta }^2+3 m_{k^+}^2+s\right)\right]+12 m_{\pi }^2 \left\{-18 m_{k^+}^6-39 s m_{k^+}^4\right.~\nonumber \\&&
+3 s \left(2 m_{\pi ^+}^2+5 t+8 u\right) m_{k^+}^2-9 m_{\eta }^4 \left(2 m_{k^+}^2+s\right)~\nonumber \\&&
\left.\left.+s^2 \left(2 m_{\pi ^+}^2+9 u\right)+6 m_{\eta }^2 \left[6 m_{k^+}^4-5 s m_{k^+}^2+s \left(s-m_{\pi ^+}^2\right)\right]\right\}\right\}
J_{\pi  k^+}(s)~\nonumber \\&&
-\left\{-54 m_{\eta }^8+54 \left(4 m_{k^+}^2+s\right) m_{\eta }^6-9 \left[36 m_{k^+}^4-2 s m_{k^+}^2+s \left(4 m_{\pi ^+}^2+5 s-3 u\right)\right]
m_{\eta }^4\right.~\nonumber \\&&
-6 \left[-36 m_{k^+}^6+33 s m_{k^+}^4+s \left(-12 m_{\pi ^+}^2-29 s+9 u\right) m_{k^+}^2\right.~\nonumber \\&&
\left.+s^2 \left(9 u-2 m_{\pi ^+}^2\right)\right] m_{\eta }^2-54 m_{k^+}^8+126 s m_{k^+}^6-27 s^4-8 s^2 m_k^4~\nonumber \\&&
-137 s^2 m_{k^+}^4+27 s u m_{k^+}^4-8 s^2 m_{\pi ^+}^4+78 s^3 m_{k^+}^2+6 s^2 t m_{k^+}^2~\nonumber \\&&
-48 s^2 u m_{k^+}^2-36 s m_{k^+}^4 m_{\pi ^+}^2-30 s^3 m_{\pi ^+}^2+44 s^2 m_{k^+}^2 m_{\pi ^+}^2~\nonumber \\&&
+6 s^2 t m_{\pi ^+}^2+6 s^2 u m_{\pi ^+}^2+27 s^3 u+4 s m_k^2 \left[9 m_{\eta }^4-6 \left(3 m_{k^+}^2+s\right) m_{\eta }^2\right.~\nonumber \\&&
\left.\left.\left.+9 m_{k^+}^4+9 s^2-14 s m_{k^+}^2+4 s m_{\pi ^+}^2\right]\right\} J_{\eta k^+}(s)+[s\longleftrightarrow u]\right\} \ ;
\end{eqnarray}

\begin{eqnarray}
T_{K^{0}\eta \rightarrow K^{0}\eta}^{H}(s,t,u)&=&
\frac{-2 m_k^2-6 m_{\eta }^2+2 m_{k^+}^2-2 m_{\pi ^+}^2+9 t}{12 f_{\pi }^2}+\frac{1}{512 \pi ^2 s u f_{\pi }^4}\left\{-4 \left(2
m_k^2+2 m_{\eta }^2-t\right) m_k^6\right.~\nonumber \\&&
+\left[4 s u-\left(2 m_k^2+2 m_{\eta }^2-t\right) \left(m_{\pi }^2-5 m_{\eta }^2+2 m_{k^+}^2+2m_{\pi ^+}^2\right)\right] m_k^4~\nonumber \\&&
+2 \left\{\left[\left(2 m_k^2+2 m_{\eta }^2-t\right) \left(m_{\pi }^2+m_{\eta }^2+2 m_{k^+}^2+2m_{\pi ^+}^2\right)-4 s u\right] m_{\eta }^2\right.~\nonumber \\&&
\left.+2 s u \left(2 m_k^2+2 m_{\eta }^2-3 t\right)\right\}m_k^2-3 s m_{\eta }^6-3 u m_{\eta }^6+4 s u m_{\eta }^4~\nonumber \\&&
+3 s u^2 m_{\eta }^2+3 s^2 u m_{\eta }^2-6 s t u m_{\eta }^2-2 s m_{\eta }^4 m_{k^+}^2-2 u m_{\eta }^4 m_{k^+}^2~\nonumber \\&&
+2 s u^2 m_{k^+}^2+2 s^2 u m_{k^+}^2-4 s t u m_{k^+}^2+2\left[-2 m_{\eta }^6+t m_{\eta }^4+2 s u m_{\eta }^2\right.~\nonumber \\&&
\left.-3 s t u+2 m_k^2 \left(s u-m_{\eta }^4\right)\right] m_{\pi ^+}^2+2 s u [s (t-2 u)+t u]~\nonumber \\&&
\left.+m_{\pi }^2 \left[-2 m_{\eta }^6+t m_{\eta }^4+2 s u m_{\eta }^2-3 s t u+2 m_k^2 \left(s u-m_{\eta }^4\right)\right]\right\}~\nonumber \\&&
+\frac{8 L_1 \left(t-2 m_k^2\right) \left(t-2 m_{\eta }^2\right)}{f_{\pi }^4}+\frac{4 L_2 \left[\left(m_k^2+m_{\eta }^2-s\right){}^2+\left(m_k^2+m_{\eta
}^2-u\right){}^2\right]}{f_{\pi }^4}~\nonumber \\&&
+\frac{L_3 \left[-2 m_k^4-18 \left(t-2 m_{\eta }^2\right) m_k^2-2 m_{\eta }^4+s^2+10 t^2+u^2-18 t m_{\eta }^2\right]}{3 f_{\pi }^4}~\nonumber \\&&
+\frac{32 L_6 \left(2 m_k^2+2m_{k^+}^2-m_{\pi ^+}^2\right) m_k^2}{3 f_{\pi }^4}+\frac{32 L_7 \left(m_k^2+m_{k^+}^2-2 m_{\pi ^+}^2\right) \left(m_k^2+m_{k^+}^2-m_{\pi
^+}^2\right)}{3 f_{\pi }^4}~\nonumber \\&&
+\frac{16 L_8 \left[3 m_k^4+4 \left(m_{k^+}^2-m_{\pi ^+}^2\right) m_k^2+m_{k^+}^4+2 m_{\pi ^+}^4-3 m_{k^+}^2 m_{\pi ^+}^2\right]}{3 f_{\pi }^4}~\nonumber \\&&
-\frac{1}{3 f_{\pi }^4}2 L_4 \left[18 m_k^4+\left(2 m_{\pi }^2+30 m_{\eta }^2+12 m_{k^+}^2-8 m_{\pi ^+}^2-29 t\right) m_k^2+2 m_{k^+}^4-2 m_{\pi
^+}^4\right.~\nonumber \\&&
\left.-6 m_{\eta }^2 m_{k^+}^2+t m_{k^+}^2-6 m_{\eta }^2 m_{\pi ^+}^2+13 t m_{\pi ^+}^2+m_{\pi }^2 \left(6 m_{\eta }^2-2 m_{k^+}^2+2 m_{\pi
^+}^2-9 t\right)\right]~\nonumber \\&&
-\frac{1}{9 f_{\pi }^4}2 L_5 \left[20 m_k^4+2 \left(9 m_{\pi }^2+6 m_{\eta }^2+18 m_{k^+}^2-16 m_{\pi ^+}^2+9 t\right) m_k^2-2 m_{k^+}^4+8 m_{\pi
^+}^4\right.~\nonumber \\&&
-9 s m_{\eta }^2+18 t m_{\eta }^2-9 u m_{\eta }^2+6 m_{\eta }^2 m_{k^+}^2-3 s m_{k^+}^2+6 t m_{k^+}^2-3 u m_{k^+}^2~\nonumber \\&&
\left.-6 m_{\eta }^2 m_{\pi ^+}^2-6 m_{k^+}^2 m_{\pi ^+}^2-9 m_{\pi }^2 \left(-6 m_{\eta }^2+2 m_{k^+}^2-2 m_{\pi ^+}^2+9 t\right)\right]~\nonumber \\&&
+\frac{1}{480 s^2 u^2 f_{\pi }^4}\left\{-60 \left(s^2+u^2\right) m_k^6+20 \left[3 \left(s^2+u s+u^2\right) m_{\pi }^2\right.\right.~\nonumber \\&&
\left.+6 \left(s^2+u^2\right) m_{\eta }^2+4 s u \left(m_{\pi ^+}^2-m_{k^+}^2\right)\right] m_k^4~\nonumber \\&&
+2 \left\{-30 \left(s^2+u s+u^2\right) m_{\eta }^4-15 m_{\pi }^2 \left[4 \left(s^2+u^2\right) m_{\eta }^2+s t u\right]\right.~\nonumber \\&&
\left.+s u \left(15 s^2+414 u s+15 u^2+20 t m_{k^+}^2-20 t m_{\pi ^+}^2\right)\right\} m_k^2~\nonumber \\&&
+m_{\pi }^2 \left\{60 \left(s^2-u s+u^2\right) m_{\eta }^4+30 s t u m_{\eta }^2+s u \left[-15 s^2+(15 t-14 u) s\right.\right.~\nonumber \\&&
+15 (t-u) u]\}-2 s u \left\{30 m_{\eta }^6+5 \left[-8 m_{k^+}^2+8 m_{\pi ^+}^2+3 (s-t+u)\right] m_{\eta }^4\right.~\nonumber \\&&
\left.\left.-4 \left(-5 t m_{k^+}^2+5 t m_{\pi ^+}^2+327 s u\right) m_{\eta }^2+2 s u \left(204 m_{k^+}^2-224 m_{\pi ^+}^2+945 t\right)\right\}\right\}\mu
_{\pi }~\nonumber \\&&
+\frac{1}{720 s^2 u^2 f_{\pi }^4}\left\{-300 s u m_k^6-30 \left[-6 s u m_{\eta }^2+\left(6 s^2-2 u s+6 u^2\right) m_{k^+}^2\right.\right.~\nonumber \\&&
\left.-6 s^2 m_{\pi ^+}^2-6 u^2 m_{\pi ^+}^2-14 s u m_{\pi ^+}^2-5 s t u\right]m_k^4~\nonumber \\&&
+\left\{300 s u m_{\eta }^4-30 \left[-12 \left(s^2+u^2\right) m_{k^+}^2+12 \left(s^2+u^2\right) m_{\pi ^+}^2\right.\right.~\nonumber \\&&
+s u (3 s+8 t+3 u)] m_{\eta }^2+s u \left(45 s^2+45 t s-14 u s\right.~\nonumber \\&&
\left.\left.+45 u^2-30 t m_{k^+}^2-210 t m_{\pi ^+}^2+45 t u\right)\right\} m_k^2-180 s u m_{\eta }^6~\nonumber \\&&
+6 s u m_{\eta }^2 \left(5 t m_{k^+}^2+35 t m_{\pi ^+}^2+48 s u\right)-30 m_{\eta }^4 \left[2 \left(3 s^2+u s+3 u^2\right) m_{k^+}^2\right.~\nonumber \\&&
\left.-2 \left(3 s^2-7 u s+3 u^2\right) m_{\pi ^+}^2+3 s u (s-t+u)\right]~\nonumber \\&&
+s u \left\{5 \left[9 s^2+(10 u-9 t) s+9 u (u-t)\right] m_{k^+}^2\right.~\nonumber \\&&
\left.\left.-3 \left[15 s^2+(38 u-15 t) s+15 u (u-t)\right]m_{\pi ^+}^2\right\}\right\} \mu _{\pi ^+}~\nonumber \\&&
+\frac{1}{1440 s^2 u^2 f_{\pi }^4}\left\{720 \left(s^2-4 u s+u^2\right) m_k^6-60 \left[3 \left(s^2+u s+u^2\right) m_{\pi }^2\right.\right.~\nonumber \\&&
\left.+\left(33 s^2-3 u s+33 u^2\right) m_{\eta }^2-8 s u \left(-m_{k^+}^2+m_{\pi ^+}^2+3 t\right)\right] m_k^4~\nonumber \\&&
+2\left\{90 \left(10 s^2+17 u s+10 u^2\right) m_{\eta }^4-15 s u \left(16 m_{k^+}^2-16 m_{\pi ^+}^2+51 t\right) m_{\eta }^2\right.~\nonumber \\&&
+45 m_{\pi }^2 \left[4 \left(s^2+u^2\right) m_{\eta }^2+s t u\right]-4 s u \left(45 s^2-202 u s+45 u^2\right.~\nonumber \\&&
\left.\left.-30 t m_{k^+}^2+30 t m_{\pi ^+}^2\right)\right\} m_k^2-540 s^2 m_{\eta }^6-540 u^2 m_{\eta }^6~\nonumber \\&&
-90 s u^2 m_{\eta }^4-90 s^2 u m_{\eta }^4-936 s^2 u^3-936 s^3 u^2+3744 s^2 t u^2~\nonumber \\&&
+315 s u^3 m_{\eta }^2-1362 s^2 u^2 m_{\eta }^2+45 s t u^2 m_{\eta }^2+315 s^3 u m_{\eta }^2~\nonumber \\&&
+45 s^2 t u m_{\eta }^2+848 s^2 u^2 m_{k^+}^2+240 s u^2 m_{\eta }^2 m_{k^+}^2+240 s^2 u m_{\eta }^2 m_{k^+}^2~\nonumber \\&&
-848 s^2 u^2 m_{\pi ^+}^2-240 s u^2 m_{\eta }^2 m_{\pi ^+}^2-240 s^2 u m_{\eta }^2 m_{\pi ^+}^2~\nonumber \\&&
+15 m_{\pi }^2 \left\{-12 \left(s^2-u s+u^2\right) m_{\eta }^4-6 s t u m_{\eta }^2\right.~\nonumber \\&&
\left.\left.+s u \left[3 s^2-(3 t+2 u) s+3 u (u-t)\right]\right\}\right\} \mu _k~\nonumber \\&&
+\frac{1}{720 s^2 u^2 f_{\pi }^4}\left\{-60 s u m_k^6+30 \left[6 s u m_{\eta }^2+\left(6 s^2-2 u s+6 u^2\right) m_{k^+}^2\right.\right.~\nonumber \\&&
\left.-6 s^2 m_{\pi ^+}^2-6 u^2 m_{\pi ^+}^2-14 s u m_{\pi ^+}^2+s t u\right] m_k^4+\left\{60 s u m_{\eta }^4\right.~\nonumber \\&&
-120 \left[3 \left(s^2+u^2\right) m_{k^+}^2-3 \left(s^2+u^2\right) m_{\pi ^+}^2+s t u\right] m_{\eta }^2~\nonumber \\&&
+s u \left(-45 s^2-45 t s+214 u s-45 u^2+30 t m_{k^+}^2+210 t m_{\pi ^+}^2\right.~\nonumber \\&&
-45 t u)\}m_k^2-180 s u m_{\eta }^6+3 s u m_{\eta }^2 \left(15 s^2+15 t s\right.~\nonumber \\&&
\left.-202 u s+15 u^2-10 t m_{k^+}^2-70 t m_{\pi ^+}^2+15 t u\right)~\nonumber \\&&
+30 m_{\eta }^4\left[2 \left(3 s^2+u s+3 u^2\right) m_{k^+}^2-2 \left(3 s^2-7 u s+3 u^2\right) m_{\pi ^+}^2\right.~\nonumber \\&&
+3 s t u]+s u \left\{\left[-45 s^2+3 (15 t+58 u) s+45 (t-u) u\right] m_{k^+}^2\right.~\nonumber \\&&
\left.\left.+\left[45 s^2-(45 t+302 u) s+45 u (u-t)\right] m_{\pi ^+}^2+900 s t u\right\}\right\} \mu _{k^+}~\nonumber \\&&
+\frac{1}{240 s^2 u^2 f_{\pi }^4}\left\{-30 \left(3 s^2-8 u s+3 u^2\right) m_k^6+30 \left[\left(9 s^2+6 u s+9 u^2\right) m_{\eta }^2\right.\right.~\nonumber \\&&
\left.-4 s u \left(-m_{k^+}^2+m_{\pi ^+}^2+t\right)\right] m_k^4+\left[-30 \left(9 s^2+8 u s+9 u^2\right) m_{\eta }^4\right.~\nonumber \\&&
+30 s t u m_{\eta }^2+s u \left(45 s^2-158 u s+45 u^2-60 t m_{k^+}^2\right.~\nonumber \\&&
\left.\left.+60 t m_{\pi ^+}^2\right)\right] m_k^2+90 (s-u)^2 m_{\eta }^6+2 s^2 u^2 \left(32 m_{k^+}^2\right.~\nonumber \\&&
\left.-52 m_{\pi ^+}^2+135 t\right)+30 s u m_{\eta }^4 \left(-4 m_{k^+}^2+4 m_{\pi ^+}^2+3 t\right)~\nonumber \\&&
\left.-3 s u m_{\eta }^2 \left(15 s^2+34 u s+15 u^2-20 t m_{k^+}^2+20 t m_{\pi ^+}^2\right)\right\}\mu _{\eta }~\nonumber \\&&
-\frac{m_{\pi ^+}^2 \left(2 m_k^2-2 m_{\pi }^2-2 m_{k^+}^2+2 m_{\pi ^+}^2+3 t\right) J_{\pi  \pi }(t)}{72 f_{\pi }^4}~\nonumber \\&&
-\frac{t m_{\pi ^+}^2 J_{\pi ^+ \pi ^+}(t)}{12 f_{\pi }^4}-\frac{t \left(-2 m_k^2-6 m_{\eta }^2+2 m_{k^+}^2-2 m_{\pi ^+}^2+9 t\right) J_{k k}(t)}{24
f_{\pi }^4}~\nonumber \\&&
-\frac{t \left(2 m_k^2-6 m_{\eta }^2-2 m_{k^+}^2-2 m_{\pi ^+}^2+9 t\right) J_{k^+ k^+}(t)}{48 f_{\pi }^4}~\nonumber \\&&
+\frac{\left(m_k^2-m_{k^+}^2\right) \left(6 m_k^2+3 m_{\pi }^2+3 m_{\eta }^2+2 m_{k^+}^2-2 m_{\pi ^+}^2-9 t\right) J_{\pi  \eta }(t)}{108 f_{\pi
}^4}~\nonumber \\&&
+\frac{\left(8 m_k^2+8 m_{k^+}^2-7 m_{\pi ^+}^2\right) \left(2 m_k^2+6 m_{\eta }^2-2 m_{k^+}^2+2 m_{\pi ^+}^2-9 t\right) J_{\eta  \eta }(t)}{216
f_{\pi }^4}~\nonumber \\&&
-\frac{1}{864 s^2 f_{\pi }^4}\left\{\left\{54 m_k^8-54 \left(2 m_{\pi }^2+2 m_{\eta }^2+s\right) m_k^6+9 \left[6 m_{\pi }^4+\left(24 m_{\eta
}^2-26 s\right) m_{\pi }^2\right.\right.\right.~\nonumber \\&&
\left.+6 m_{\eta }^4+12 s m_{\eta }^2+s \left(4 m_{k^+}^2-4 m_{\pi ^+}^2+5 s-3 u\right)\right] m_k^4~\nonumber \\&&
+6 \left\{6 \left(s-3 m_{\eta }^2\right) m_{\pi }^4+3 \left[-6 m_{\eta }^4-10 s m_{\eta }^2+s \left(-2 m_{k^+}^2+2 m_{\pi ^+}^2\right.\right.\right.~\nonumber \\&&
+5 t+8 u)] m_{\pi }^2+s \left[s \left(-4 m_{k^+}^2+4 m_{\pi ^+}^2+9 u\right)\right.~\nonumber \\&&
\left.\left.-6 m_{\eta }^2 \left(m_{k^+}^2-m_{\pi ^+}^2+4 s\right)\right]\right\} m_k^2+9 m_{\pi }^4 \left[6 m_{\eta }^4-2 s m_{\eta }^2\right.~\nonumber \\&&
+s (t-2 u)]+s^2 \left[18 m_{\eta }^4-12 \left(m_{k^+}^2-m_{\pi ^+}^2\right) m_{\eta }^2+8 m_{k^+}^4\right.~\nonumber \\&&
\left.+8 m_{\pi ^+}^4+27 s^2-36 s m_{\pi ^+}^2-27 s u+4 m_{k^+}^2 \left(9 s-4 m_{\pi ^+}^2\right)\right]~\nonumber \\&&
+6 s m_{\pi }^2 \left[-9 m_{\eta }^4+6 \left(m_{k^+}^2-m_{\pi ^+}^2+s\right) m_{\eta }^2+s \left(-2 m_{k^+}^2+2 m_{\pi ^+}^2\right.\right.~\nonumber \\&&
+9 u)]\}J_{k \pi }(s)+3\left\{54 m_k^8-18 \left(12 m_{\eta }^2+7 s\right) m_k^6\right.~\nonumber \\&&
+\left[324 m_{\eta }^4+198 s m_{\eta }^2-s \left(36 m_{k^+}^2-36 m_{\pi ^+}^2+19 s+27 u\right)\right]m_k^4~\nonumber \\&&
+2 \left[-108 m_{\eta }^6-9 s m_{\eta }^4-3 s \left(-12 m_{k^+}^2+12 m_{\pi ^+}^2+55 s-9 u\right) m_{\eta }^2\right.~\nonumber \\&&
\left.+s^2 \left(28 m_{k^+}^2-28 m_{\pi ^+}^2+36 t+63 u\right)\right] m_k^2+54 m_{\eta }^8-54 s m_{\eta }^6~\nonumber \\&&
+6 s^2 m_{\eta }^2 \left(4 m_{k^+}^2-4 m_{\pi ^+}^2+9 u\right)+9 s m_{\eta }^4 \left(-4 m_{k^+}^2+4 m_{\pi ^+}^2+5 s\right.~\nonumber \\&&
-3 u)+s^2 \left[8 m_{k^+}^4-4 \left(4 m_{\pi ^+}^2+9 s\right) m_{k^+}^2+8 m_{\pi ^+}^4+36 s m_{\pi ^+}^2\right.~\nonumber \\&&
+27 s (s-u)]\}J_{k \eta }(s)+\left\{4\left[7 m_{k^+}^4-18 \left(3 m_{\pi ^+}^2+s\right) m_{k^+}^2+27 m_{\pi ^+}^4\right.\right.~\nonumber \\&&
\left.+7 s^2-9 s m_{\pi ^+}^2\right] m_k^4+2 \left\{s \left[-9 m_{k^+}^4+8 \left(8 s-5 m_{\pi ^+}^2\right) m_{k^+}^2\right.\right.~\nonumber \\&&
\left.+2 \left(56 m_{\pi ^+}^4-100 s m_{\pi ^+}^2+9 s^2\right)\right]-12 m_{\eta }^2 \left[9 m_{k^+}^4\right.~\nonumber \\&&
\left.\left.-3 \left(6 m_{\pi ^+}^2+s\right) m_{k^+}^2+9 m_{\pi ^+}^4+2 s^2-6 s m_{\pi ^+}^2\right]\right\} m_k^2~\nonumber \\&&
+36 m_{\eta }^4 \left(3 m_{k^+}^4-6 m_{\pi ^+}^2 m_{k^+}^2+3 m_{\pi ^+}^4+s^2-3 s m_{\pi ^+}^2\right)~\nonumber \\&&
-2 s m_{\eta }^2 \left[-9 m_{k^+}^4+\left(96 s-68 m_{\pi ^+}^2\right) m_{k^+}^2+14 m_{\pi ^+}^4+24 s m_{\pi ^+}^2\right]~\nonumber \\&&
+s\left\{[s+27 (t-u)]m_{k^+}^4+\left[(40 u-68 t) m_{\pi ^+}^2-36 s (t-2 u)\right] m_{k^+}^2\right.~\nonumber \\&&
-2 (11 t+38 u) m_{\pi ^+}^4+36 s (2 t+5 u) m_{\pi ^+}^2~\nonumber \\&&
\left.\left.\left.+54 s^2 (s-u)\right\}\right\}J_{k^+ \pi ^+}(s)+[s\longleftrightarrow u]\right\} \ .
\end{eqnarray}

\end{appendix}

\end{document}